\documentclass[12pt]{article}
\usepackage{a4wide}
\usepackage{amsmath,amssymb,bbm}
\usepackage{multirow}

\begin{document}
\setcounter{page}{1}
%


%

\def\pct#1{(see Fig. #1.)}

\begin{titlepage}
\hbox{\hskip 12cm KCL-MTH-09-02  \hfil}
\vskip 1.4cm
\begin{center}  {\Large  \bf  Local $E_{11}$}

\vspace{1.8cm}

{\large \large Fabio Riccioni \ and \ Peter West} \vspace{0.7cm}

{\sl Department of Mathematics\\
\vspace{0.3cm}
 King's College London  \\
\vspace{0.3cm}
Strand \ \  London \ \ WC2R 2LS \\
\vspace{0.3cm} UK}
\end{center}
\vskip 1.5cm

\abstract{We give a method of deriving the field-strengths of all
massless and massive maximal supergravity theories in any dimension
starting from the Kac-Moody algebra $E_{11}$. Considering the
subalgebra of $E_{11}$ that acts on the fields in the non-linear
realisation as a global symmetry, we show how this is promoted to a
gauge symmetry enlarging the algebra by the inclusion of additional
generators. We show how this works in eleven dimensions, and we call
the resulting enlarged algebra $E_{11}^{local}$. Torus reduction to
$D$ dimensions corresponds to taking a subalgebra of
$E_{11}^{local}$, called $E_{11,D}^{local}$, that encodes the full
gauge algebra of the corresponding $D$-dimensional massless
supergravity. We show that each massive maximal supergravity in $D$
dimensions is a non-linear realisation of an algebra
$\tilde{E}_{11,D}^{local}$. We show how this works in detail for the
case of Scherk-Schwarz reduction of IIB to nine dimensions, and in
particular we show how $\tilde{E}_{11,9}^{local}$ arises as a
subalgebra of the algebra $E_{11,10B}^{local}$ associated to the
ten-dimensional IIB theory. This subalgebra corresponds to taking a
combination of generators which is different to the massless case.
We then show that $\tilde{E}_{11,D}^{local}$ appears as a
deformation of the massless algebra $E_{11,D}^{local}$ in which the
commutation relations between the $E_{11}$ and the additional
generators are modified. We explicitly illustrate how the deformed
algebra is constructed in the case of massive IIA and of gauged
five-dimensional supergravity. These results prove the naturalness
and power of the method.}

\vfill
\end{titlepage}
\makeatletter \@addtoreset{equation}{section} \makeatother
\renewcommand{\theequation}{\thesection.\arabic{equation}}
\addtolength{\baselineskip}{0.3\baselineskip}

\section{Introduction}
It has been  conjectured in \cite{1} that eleven dimensional
supergravity could be extended so as to have a non-linearly realised
infinite-dimensional Kac-Moody symmetry called $E_{11}$, whose
Dynkin diagram is shown in Fig. \ref{fig1}. In a non-linear
realisation the algebra used to construct it is realised as a rigid
symmetry. However, in the eleven dimensional supergravity theory all
the symmetries are local. In this paper we will propose a non-linear
realisation in which  $E_{11}$ symmetries become local. To put this
work in context it will be useful to list some of the main
developments of the $E_{11}$ programme which are relevant for this
paper.

Eleven dimensional supergravity itself can be formulated as a
non-linear realisation  based on an algebra that includes generators
with non-trivial Lorentz character \cite{2}. To find the precise
dynamics one  takes the simultaneous non-linear realisation of this
algebra with the conformal group.  This naturally gives rise to both
a 3-form  and a 6-form fields and the resulting field equations are
first order duality relations, whose divergence reproduces the
3-form second-order field equations of 11-dimensional supergravity
provided on chooses one constant. The eleven-dimensional  gravity
field describes non-linearly $GL(11,\mathbb{R})$, which is a
subalgebra of this algebra. Indeed, gravity in $D$ dimensions can be
described as a non-linear realisation of the closure of the group
$GL(D,\mathbb{R})$ with the conformal group \cite{2}, as was
originally shown in the four dimensional case in \cite{3}.

$E_{11}$ first arose as  the smallest Kac-Moody algebra which
contains the algebra found in the non-linear realisation above. This
$E_{11}$ algebra is infinite-dimensional, and the $E_{11}$
non-linear realisation contains an infinite number of fields with
increasing number of indices. The first few fields are the graviton,
a three form, a six form and a field which has the right spacetime
indices to be interpreted as a dual graviton. This is the field
content of eleven dimensional supergravity, and keeping only the
first three of these fields one finds that the non-linear
realisation of $E_{11}$ reduces to the construction discussed in the
first point and so results in the dynamics of  this theory \cite{1}.
Theories in $D$ dimensions arise from the $E_{11}$ non-linear
realisation by choosing a suitable $GL(D,\mathbb{R})$ subalgebra,
which is  associated with $D$-dimensional gravity. The $A_{D-1}$
Dynkin diagram of this subalgebra, called the gravity line, must
include the node labelled 1 in the Dynkin diagram of Fig.
\ref{fig1}. In ten dimensions there are two possible ways of
constructing this subalgebra, and the corresponding non-linear
realisations give rise to two theories that contain the fields of
the IIA and IIB supergravity theories and their electromagnetic
duals \cite{1,4}. Below ten dimensions, there is a unique choice for
this subalgebra, and this corresponds to the fact that massless
maximal supergravity theories in dimensions below ten are unique.
Again, the non-linear realisation in each case describes, among an
infinite set of other fields, the fields of the corresponding
supergravity and their electromagnetic duals. In each dimension, the
part of the $E_{11}$ Dynkin diagram which is not connected to the
gravity line corresponds to the internal hidden symmetry of the $D$
dimensional theory. This not only reproduces all the hidden
symmetries found long ago in the dimensionally reduced theories, but
it also gives an eleven-dimensional origin to these symmetries.

All the maximal supergravity theories mentioned so far are massless
in the sense that no other dimensional parameter other than the
Planck scale is present. In fact, even this parameter can be
absorbed into the fields such that it is absent from the equations
of motion. There are however other theories that are also maximal,
{\it i.e.} invariant under 32 supersymmetries, but are massive in
the sense that they possess additional dimensionful parameters.
These can be viewed as deformations of the massless maximal
theories. However, unlike the massless maximal supergravity theories
they can not in general be obtained by a process of dimensional
reduction and in each dimension they have been determined by
analysing the deformations that the corresponding massless maximal
supergravity admits. With the exception of the one deformation
allowed for type IIA supergravity in ten dimensions, called Roman's
theory, all the massive maximal supergravities possess a local gauge
symmetry carried by vector fields that is a subgroup of the symmetry
group $G$ of the corresponding maximal supergravity theory, and are
therefore called gauged supergravities. In general these theories
also have potentials for the scalars fields which contain the
dimensionful parameters as well as a cosmological constant. In
recent years there have been a number of systematic searches for
gauged maximal supergravity theories and in particular in nine
dimensions and in dimension from seven to three all such theories
have been classified \cite{5,6,7}.

It will be useful to recall how  $E_{11}$ has from a very different
perspective lead to the classification of gauged supergravities that
agrees with these results and how the $E_{11}$ formulation of the
gauged supergravity theories has lead to new work in these theories.
The cosmological constant of ten-dimensional Romans IIA theory
\cite{8} can be described as the dual of a 10-form field-strength
\cite{9}, and the supersymmetry algebra closes on the corresponding
9-form potential \cite{10}. The Romans theory was found to be a
non-linear realisation \cite{11} which includes all form fields up
to and including a 9-form with a corresponding set of generators.
This 9-form is automatically encoded in the non-linear realisation
of $E_{11}$ \cite{12}. From the eleven dimensional $E_{11}$ theory
it arises as the dimensional reduction of the eleven-dimensional
field $A_{a_1 \dots a_{10} , (bc)}$  in the irreducible
representation of $GL(11,\mathbb{R})$ with ten antisymmetric indices
$a_1 \dots a_{10}$ and two symmetric indices $b$ and $c$. Therefore
$E_{11}$ not only contains Romans IIA, but it also provides it for
the first time with an eleven-dimensional origin \cite{13}.

By studying the eleven-dimensional fields of the $E_{11}$ non-linear
realisation, one can determine all the forms, {\it i.e.} fields with
completely antisymmetric indices, that arise from dimensional
reduction to any dimension \cite{14}. In particular, in addition  to
all the lower rank forms, this analysis gives all the $D-1$-forms
and the $D$-forms in $D$ dimensions. The list of all form fields
obtained in this way for all supergravity theories is given in table
\ref{Table1}. The $D-1$ and $D$-forms predicted by $E_{11}$ can also
be derived in each dimension separately \cite{15}. The $D-1$-forms
have $D$-form field strengths, that are related by duality to the
mass deformations of gauged maximal supergravities, and the $E_{11}$
analysis shows perfect agreement with the complete classification of
gauged supergravities performed in \cite{6,7}. Therefore $E_{11}$
not only contains all the possible massive deformations of maximal
supergravities in a unified framework, but it also provides an
eleven-dimensional origin to all of them. Indeed, while some gauged
supergravities were known to be obtainable using dimensional
reduction of ten or eleven dimensional supergravities, this was not
generically the case. As a result the gauged supergravities were
outside the framework of M-theory as it is usually understood.

One striking feature of the $E_{11}$ formulation of massless or
massive supergravity theories  is that it includes fields together
with all their dual fields. The presence of the dual forms is
essential to formulate the field equations as duality relations.
Some  dual forms have been introduced in the past in an ad-hoc way
beginning with \cite{16}, but it is only with $E_{11}$ that they
have arisen from an underlying principle. Indeed, the forms of table
\ref{Table1} were proposed in \cite{14,15} to play a crucial role in
gauged supergravities, the $D-1$ forms classifying the gauged
supergravities and the lower forms providing a chain of form fields
that occur in the duality relations. This is compatible with the
structure of the gauge algebra arising in gauged supergravities, in
which one is forced to introduce a $p+1$ form to close the gauge
algebra of a $p$ form, thus determining a hierarchy of forms
\cite{17}. For the cases in which this latter method has been
subsequently used to compute the hierarchy of forms, the results are
precisely in agreement with $E_{11}$ \cite{18}, and indeed the
presence of the forms given in table \ref{Table1} has now been
systematically adopted by those studying gauged supergravities.
\begin{table}
\begin{tiny}
\begin{center}
\begin{tabular}{|c|c||c|c|c|c|c|c|c|c|c|c|}
\hline \rule[-1mm]{0mm}{6mm}
D & G & 1-forms & 2-forms & 3-forms & 4-forms & 5-forms & 6-forms & 7-forms & 8-forms & 9-forms & 10-forms\\
\hline \rule[-1mm]{0mm}{6mm} \multirow{2}{*}{10A} &
\multirow{2}{*}{$\mathbb{R}^+$} & \multirow{2}{*}{${\bf 1}$} &
\multirow{2}{*}{${\bf 1}$} & \multirow{2}{*}{${\bf 1}$} &  &
\multirow{2}{*}{${\bf 1}$} & \multirow{2}{*}{${\bf 1}$} &
\multirow{2}{*}{${\bf 1}$} & \multirow{2}{*}{${\bf 1}$} &
\multirow{2}{*}{${\bf 1}$} & ${\bf 1}$ \\
& & & & & & & & & & & ${\bf 1}$ \\
\hline \rule[-1mm]{0mm}{6mm} \multirow{2}{*}{10B} &
\multirow{2}{*}{$SL(2,\mathbb{R})$} & & \multirow{2}{*}{${\bf 2}$} &
&
\multirow{2}{*}{${\bf 1}$} & & \multirow{2}{*}{${\bf 2}$} & & \multirow{2}{*}{${\bf 3}$} & & ${\bf 4}$ \\
& & & & & & & & & & & ${\bf 2}$ \\
\cline{1-12} \rule[-1mm]{0mm}{6mm} \multirow{3}{*}{ 9} &
\multirow{3}{*}{ $SL(2,\mathbb{R})\times \mathbb{R}^+$} & ${\bf 2}$
& \multirow{3}{*}{${\bf 2 }$} & \multirow{3}{*}{ ${\bf 1}$} &
\multirow{3}{*}{ ${\bf 1}$} & \multirow{3}{*}{ ${\bf
2 }$} & ${\bf 2}$ & ${\bf 3}$ & ${\bf 3}$ & ${\bf 4}$  \\
& & & & & & & & & & ${\bf 2}$ \\
& & ${\bf 1}$ & & & & & ${\bf 1}$ & ${\bf 1}$ & ${\bf 2}$ & ${\bf 2 }$  \\
\cline{1-11} \rule[-1mm]{0mm}{6mm} \multirow{4}{*}{8} &
\multirow{4}{*}{$SL(3,\mathbb{R}) \times SL(2,\mathbb{R})$} &
\multirow{4}{*}{${\bf (\overline{3}, 2)}$} & \multirow{4}{*}{${\bf
(3,1) }$} & \multirow{4}{*}{${\bf (1,2)}$} & \multirow{4}{*}{${\bf
(\overline{3},1)}$} & \multirow{4}{*}{${\bf (3,2) }$} &  &  & ${\bf (15,1)}$  \\
& & & & & & & ${\bf (8,1)}$ & ${\bf (6,2)}$ & ${\bf (3,3)}$  \\
& & & & & & & ${\bf (1,3)}$ & ${\bf (\overline{3},2)}$ & ${\bf (3,1)}$  \\
& & & & & & &  &  & ${\bf (3,1)}$ \\
 \cline{1-10} \rule[-1mm]{0mm}{6mm} \multirow{3}{*}{7} & \multirow{3}{*}{$SL(5,\mathbb{R})$} & \multirow{3}{*}{${\bf
\overline{10}}$} & \multirow{3}{*}{${\bf 5 }$} &
\multirow{3}{*}{${\bf \overline{5}}$} & \multirow{3}{*}{${\bf 10}$}
&
\multirow{3}{*}{${\bf 24 }$} & ${\bf \overline{40}}$ & ${\bf 70}$  \\
& & & & & & & & ${\bf  45}$  \\
& & & & & & & ${\bf \overline{15}}$ & ${\bf 5}$  \\
 \cline{1-9} \rule[-1mm]{0mm}{6mm}\multirow{3}{*}{6} & \multirow{3}{*}{$SO(5,5)$} & \multirow{3}{*}{${\bf
16}$} & \multirow{3}{*}{${\bf 10 }$} & \multirow{3}{*}{${\bf
\overline{16} }$} & \multirow{3}{*}{${\bf 45}$} &
\multirow{3}{*}{${\bf 144 }$} & ${\bf 320}$  \\
& & & & & & & ${\bf \overline{126}}$ \\
& & & & & & & ${\bf 10}$ \\
\cline{1-8} \rule[-1mm]{0mm}{6mm} \multirow{2}{*}{5} &
\multirow{2}{*}{$E_{6(+6)}$} & \multirow{2}{*}{${\bf 27}$} &
\multirow{2}{*}{${\bf \overline{27} }$} & \multirow{2}{*}{${\bf 78
}$} & \multirow{2}{*}{${\bf 351}$} &
${\bf \overline{1728}}$  \\
& & & & & &  ${\bf \overline{27}}$  \\
 \cline{1-7} \rule[-1mm]{0mm}{6mm} \multirow{2}{*}{4} & \multirow{2}{*}{$E_{7(+7)}$} & \multirow{2}{*}{${\bf
56}$} & \multirow{2}{*}{${\bf 133 }$} & \multirow{2}{*}{${\bf 912 }$} & ${\bf 8645}$ \\
 & & & & & ${\bf 133}$ \\
 \cline{1-6} \rule[-1mm]{0mm}{6mm}
 \multirow{3}{*}{3} &  \multirow{2}{*}{$E_{8(+8)}$} &  \multirow{2}{*}{${\bf 248}$} & ${\bf 3875}$
 & {\bf 147250}  \\
 & & & & {\bf 3875} \\
 & & & ${\bf 1}$ & {\bf 248} \\
 \cline{1-5}
\end{tabular}
\end{center}
\end{tiny}
\caption{\sl Table giving the representations of the symmetry group
$G$ of all the forms of maximal supergravities in any dimension
\cite{14}. The 3-forms in three dimensions were determined in
\cite{15}. \label{Table1}}
\end{table}

All in all there is considerable evidence for an $E_{11}$ symmetry
in the low energy limit of what is often called M theory. The above
evidence concerns the adjoint representation of $E_{11}$, or the
part of the non-linear realisation that involves the fields
associated with the $E_{11}$ generators. However, there is also the
question of how space-time is encoded in the theory. In the
non-linear realisations mentioned above the  generator of space-time
translations $P_a$ was introduced by hand in order to encode the
coordinates of space-time. From the beginning it was understood that
this was an ad-hoc step that did not respect the $E_{11}$ symmetry.
It  was subsequently proposed \cite{19} that one could include an
$E_{11}$ multiplet of generators which had as its lowest component
the generator of space-time translations. This is just the
fundamental representation of $E_{11}$ associated with the node
labelled 1 in the Dynkin diagram of Fig. \ref{fig1} and it is
denoted by $l$. A method of constructing the gauged supergravities
was given in reference \cite{20} using $E_{11}$ and the $l$
multiplet of generators. Indeed as an example all the gauged
supergravity generators in five dimensions were derived from this
viewpoint. This reference also contains a review of the evidence for
the $l$ multiplet as the multiplet of brane charges and a table of
its low level content in dimensions three and above. In this context
there has been a recent interesting paper \cite{21} which keeps the
scalar charges in the $l$ multiplet for the seven-dimensional
maximal supergravity and still finds diffeomorphism invariance in
seven dimensions.

What was not clear from this method was how the global $E_{11}$
symmetries  would become local and this is the subject of this
paper. In the context of purely gravity this was achieved long ago
in reference \cite{3} by taking the simultaneous non-linear
realisation of $IGL(4,\mathbb{R} )$ with the conformal group in four
dimensions. As mentioned above, if one took the non-linear
realisation of $E_{11}$ at low levels, that is to include the six
form generator, took only the Lorentz group as the local subgroup
and the simultaneous non-linear realisation with the conformal
group, then the dynamics predicted by the non-linear realisation is
just the maximal supergravity theory in eleven dimensions. This can
be seen by realising that the low level $E_{11}$ algebra \cite{1} is
just that used in reference \cite{2} to construct the eleven
dimensional supergravity theory as a non-linear realisation once one
includes the conformal group. The effect of latter is that it makes
not only the space-time translations a local symmetry but also turns
the shifts associated with form fields into gauge transformations
\cite{2}. However, it is not clear how to combine the conformal
group with the algebra formed from $E_{11}$ and the $l$ multiplet.
In particular how to extend the action of the conformal group on the
usual coordinates of space-time to include the other coordinates
encoded in the $l$ multiplet.

In this paper we will not use the conformal group, but rather add
the generators that the closure of this algebra with $E_{11}$ would
generate. We will also not use the generators from the $l$
multiplet, but only the space-time translations $P_a$. The prototype
example of this mechanism was given long ago for the case of
Yang-Mills theory \cite{22}. Essentially one takes an algebra that
contains the generators $P_a$ and the Yang Mills generators
$Q^\alpha$, as well as the generators $R^{a,\alpha}$ for which the
gauge fields are Goldstone bosons and an infinite number of
generators $K^{a_1\ldots a_n,\alpha}$, symmetric in their spacetime
indices, which do not commute with $P_a$ and whose role is to make
the rigid symmetry generated by $R^{a,\alpha}$ local. We will review
this construction later on in this introduction.

We will first show the analogous mechanism for pure gravity. In
particular, we will show how to construct Einstein's theory of
gravity using a non-linear realisation which takes as its underlying
algebra one that consists of  $IGL(D,\mathbb{R} )$ and an infinite
set of additional generators whose effect will  to promote the rigid
$IGL(D,\mathbb{R})$ to be local. The generators $P_a$  lead in the
non-linear realisation to the coordinates of space-time while the
Goldstone boson for $GL(D, \mathbb{R})$ is the vierbein which is
subject to local Lorentz transformations. The infinite number of
additional generators lead to local translations, that is general
coordinate transformations, but to no new fields in the final theory
as  their Goldstone fields are solved in terms of the graviton field
using a set of invariant constraints placed on the Cartan forms.
This is an example of what has been called the inverse Higgs effect
\cite{23}. The unique theory resulting from this non-linear
realisation with only two space-time derivatives is Einstein's
theory up to a possible cosmological term. In this case one can see
that the additional generators we have added are just those found by
taking the closure of $IGL(D,\mathbb{R})$ with the conformal group.

We will then generalise this procedure to $E_{11}$ at low levels. We
take the algebra, called $E_{11}^{local}$  consisting of
non-negative level $E_{11}$ generators, the generators $P_a$  and an
infinite number of additional generators.  While the latter lead in
the final result to no new Goldstone fields they do result in all
the low level $E_{11}$ symmetries becoming local, thus we find
general coordinate transformations and gauge transformations for all
the form fields. For the eleven dimensional theory, space-time
arises in the group element due to the $P_a$ generators, however,
for lower dimensional theories we will take space-time to be not
only the translation operator $P_a$ for that dimension but also
certain other Lorentz scalar charges that include the translation
operators for the dimensionally reduced generators, in effect we
take only the Lorentz scalar part of the $l$ multiplet. As we add
just the spacetime translations rather than the whole $l$ multiplet
we will take $P_a$ to commute with the non-negative  level
generators of $E_{11}$. The price for  proceeding in this way is
that we are working with only the non-negative level generators of
$E_{11}$ and we have essentially thrown out the  negative level
generators. We show that the non-linear realisation of the algebra
$E_{11}^{local}$ describes at low levels in eleven dimensions the
3-form and the 6-form of the eleven dimensional supergravity theory
with all their gauge symmetries. This can be thought of as
equivalent to taking the non-linear realisation of $E_{11}$ at low
levels and taking the simultaneous non-linear realisation with the
conformal group as was discussed  earlier \cite{2,1}, but here the
procedure is more transparent.

We then  consider the formulation of lower dimensional maximal
gauged supergravity theories from the viewpoint of the enlarged
algebra $E_{11,D}^{local}$. The $D$ refers to the fact that although
we take the same non-negative level $E_{11}$ generators and
generators $P_a$, the infinite number of additional generators we
take vary from dimension to dimension. We first consider as a toy
model the Scherk-Schwarz dimensional reduction of the IIB
supergravity theory from this viewpoint. We begin with an algebra
consisting of $E_{11,10B}^{local}$ and take the ten dimensional
space-time to arise from an operator  $\tilde Q$ which is
constructed from $Q=P_{9}$ and part of the $SL(2,\mathbb{R})$
symmetry of the theory. This means that the 10th direction of
space-time is twisted to contain a part in the $SL(2,\mathbb{R})$
coset symmetry of the theory. This non-linear realisation gives a
nine dimensional gauged supergravity. We observe that not all of the
algebra $E_{11,10B}^{local}$ is essential for the construction of
the gauged supergravity in nine dimensions, but only an algebra
which we call $\tilde E_{11, 9}^{local}$ which is the subalgebra of
$E_{11,10B}^{local}$ that commutes with $\tilde Q$. Its generators
are non-trivial combinations of $E_{11}$ generators and the
additional generators and in general the generators of $\tilde
E_{11, 9}^{local}$ have non-trivial commutation relations with nine
dimensional  space-time translations. Although the subalgebra
$\tilde E_{11, 9}^{local}$ appears to be a deformation of the
original $E_{11}$ algebra and the space-time translations we have
not changed the original commutators,  but rather the new algebra
arises due to the presence of the additional generators which are
added to the $E_{11}$ generators.

However, we then show that one can find the algebra $\tilde E_{11,
9}^{local}$ without carrying out all the above steps. Given the
non-trivial relation between the lowest non-trivial positive level
generator of $\tilde E_{11,9}^{local}$ and the nine dimensional
space-time translations one can derive the rest of the algebra
$\tilde E_{11, 9}^{local}$ simply using Jacobi identities. This
algebra determines uniquely all the field strengths of the theory,
and thus one finds a very quick way of deriving the gauged
supergravity theory.

This picture applies to all gauged supergravity theories, as one can
easily find the algebra  $\tilde E_{11, D}^{local}$  without using
its derivation from $E_{11}^{local}$ and this provides a very
efficient method of constructing all gauged supergravities. We
illustrate how this works by constructing the massive IIA theory as
well as all the gauged maximal supergravities in five dimensions.

Finally, we consider how this construction generalises to the fields
with mixed symmetry, {\it i.e.} not completely antisymmetric, of
$E_{11}$ and in general of any non-linear realisation of a
very-extended Kac-Moody algebra. We will consider as a prototype of
such fields the dual graviton in four dimensions, which is a field
$A_{ab}$ symmetric in its two spacetime indices. We will show that
if one tries to promote the global shift symmetry of the dual
graviton field to a gauge symmetry, one finds that this is not
compatible with the $E_{11}$ algebra. The solution of this problem
is that actually $E_{11}$ forces to include additional generators,
whose role is to enlarge the gauge symmetry of the dual graviton so
that one can gauge away the field completely. We show this first for
the simpler case of the non-linear realisation of the Kac-Moody
algebra $A_1^{+++}$ in four dimensions. We then consider the case of
$E_{11}$ in four dimensions. For simplicity in this case we neglect
the gravity generators, and we still find that even considering
consistency conditions involving only the generators associated to
the form fields and those associated to the dual graviton, one is
forced to include additional generators for the dual graviton that
generate a local symmetry that gauges away the dual graviton
completely. We claim that this picture generalises to all mixed
symmetry fields in any dimension. It is important to stress that the
dynamics is compatible with this result. Indeed, while the field
strengths of the antisymmetric fields are first order in
derivatives, and therefore one needs fields and dual fields to
construct duality relations which are first order equations for
these fields, the gravity Riemann tensor is at second order in
derivatives and thus there is no need of a dual field to construct
its equation of motion.

It will be helpful to recall some facts about non-linear
realisations. A non-linear realisation of a group $G$ with respect
to a subgroup $H$ is by definition a theory invariant under the two
separate transformations
  \begin{equation}
  g (x) \to g_0g(x) ,\ \ \ g (x) \to g(x) h(x) \label{1.1}
  \end{equation}
where $g\in G$, $g_0\in G$ while  $h\in H$. The dependence on the
generic symbol $x$ signifies which group elements dependence on the
coordinates of the space-time. For the case of an internal symmetry
the space-time dependence is incorporated by hand. However, in this
paper space-time will arise naturally in that its associated
generators are part of the Lie algebra of the group $G$. Indeed, a
part of the group element is just space-time viewed as a coset. We
note that the $h$ transformations  depend on the space-time
coordinates so can be said to be local, unlike the rigid $g_0$
transformations. Working with the most general group element $g$ we
must then find a theory that is invariant under both $g_0$ and local
$h$ transformations.

It is often more transparent to  use the $h$ transformations to
choose $g(x)$ to be of a particular form, that is choose coset
representatives. If one does this then when making a rigid $g_0$
transformation  one finds a group element $ g_0 g$ which is in
general not one of the chosen coset representatives. To rectify this
one must make a compensating $h_c$ that depends on $g_0$ and the
original coset representative $g(x)$. That is $g\to g'=g_0gh_c^{-1}$
where both $g$ and $g'$ are  chosen coset representatives.

The  problem of finding the invariant dynamics is most often solved
by using the Cartan forms ${\cal V}=g^{-1}dg$.  This is obviously
invariant under rigid $g_0$ transformations and transforms as
  \begin{equation}
  {\cal V}\to  h^{-1}{\cal V} h+h^{-1}dh
  \label{1.2}
  \end{equation}
under local $h$ transformations. We note that $g^{-1}dg= dx\cdot
g^{-1}\partial g$ is invariant but $g^{-1}\partial g$ is not as the
coordinates of space-time $x$ transform under $g_0$ transformations.
To be more explicit we consider a group that contains the generators
$L_N$ and we denote the remaining generators by the generic symbol
$T^*$. We will assume that the generators $L_N$ from a
representation of the $T^*$'s. The general group element is of the
form
  \begin{equation}
  g=e^{x\cdot L}e^{\phi(x)\cdot T} \quad . \label{1.3}
  \end{equation}
We recognise $x$ as the coordinates and $\phi$ as the fields. The
local subgroup can be used to set some of the fields $\phi$ to zero.
The discussion below holds if one makes this choice or work with the
general group element. The Cartan forms can be  written as
  \begin{equation}
 {\cal V}= g^{-1}d g= dx^\Pi E_\Pi{}^N L_N+ dx^\Pi G_{\Pi,*} T^*
 \quad . \label{1.4}
 \end{equation}
Since $ {\cal V}$ is invariant under $g\to g_0g$ it follows that
each of the coefficients of the above generators is invariant, that
is $dx^\Pi E_\Pi{}^N$ and $dx^\Pi G_{\Pi, *}$ are invariant.
However, $dx^\Pi$ does transform under $g_0$ and so $E_\Pi{}^N$ and
$G_{\Pi, *}$ are not invariant. To find quantities that only
transform under the local subalgebra we can rewrite $ {\cal V}$ as
   \begin{equation}
   {\cal V}= g^{-1}d g= dx^\Pi E_\Pi{}^N( L_N+  G_{N,*} T^*) \quad ,
   \label{1.5}
   \end{equation}
where we recognise that $G_{N,*}=(E^{-1})_N{}^\Pi G_{\Pi,*}$. It
follows that $G_{N,*}$ are inert under $g_0$ transformations and
just transform under local transformations. As such they are useful
quantities with which to construct the dynamics as one must now only
solve the problem of finding objects which are invariant under the
local symmetry. We may think of $G_{N,*}$  as covariant derivatives
of the fields $\phi$.

There  is one subtle point that is sometimes worth remembering if
one chooses coset representatives. Although $G_{N,*}$ is naively
invariant under $g_0$ transformations it is not invariant under the
required  compensating $h_c$ transformation under which $G_{N,*}$
transforms as in eq. (\ref{1.2}) with $h$ replaced by $h_c^{-1}$.
However, having found a set of dynamics that is invariant under $h$
transformations it is of course also invariant under the
compensating transformations.

Realising Yang-Mills theory as a non-linear realisation was first
given by Ivanov and Ogievetsky \cite{22} and we now summarise this
approach as it will serve as a prototype model for the later
sections of this paper. We begin with the algebra
  \begin{equation}
  P_a \ , \quad J_{ab} \ , \quad   Q^\alpha \ , \quad
  R^{a, \alpha} \ , \quad  K^{a_1 a_2 , \alpha} \ ,\quad  K^{a_1 a_2 a_3 ,\alpha} \
  , \ ...  \ K^{a_1 ... a_n ,\alpha} \
  \ldots \label{1.6}
  \end{equation}
which will generate the group $G$ of the non-linear realisation. The
generators $P_a$ and $J_{ab}$ are those of the Poincare group while
the $Q^{\alpha}$'s will become identified with those of the gauge
group. The generator $R^{a, \alpha}$ is the generator associated to
the gauge vector in the non-linear realisation, while the generators
$K^{a_1 ... a_n ,\alpha}$ are symmetric in the spacetime indices and
will be responsible for the symmetry of the vectors to be promoted
to a gauge symmetry. The $Q^\alpha$ generators obey the commutators
  \begin{equation}
  [Q^{\alpha}, Q^{\beta}]=g  f^{\alpha\beta}{}_\gamma Q^\gamma \quad
  , \label{1.7}
  \end{equation}
where $g$ is the coupling constant. The remaining commutation
relations are given by
  \begin{equation}
  [K^{a_1\ldots a_n, \alpha}, P_b]=n  \delta_b^{(a_1} K^{a_2\ldots a_n),
  \alpha} \ ,\  \ \ [K^{a_1\ldots a_n, \alpha}, K^{b_1\ldots b_m,
  \beta}] = g f^{\alpha\beta}{}_\gamma K^{a_1\ldots a_n b_1\ldots b_m ,
  \gamma} \quad . \label{1.8}
  \end{equation}
Although the $K^{a_1 \dots a_n, \alpha}$ generators have at least
two indices, the commutation relations of $Q^\alpha$ and $R^{a ,
\alpha}$ with all the generators are encoded in the equation above
making the identification $K^{a ,\alpha} = R^{a,\alpha}$ and
$K^{\alpha} = Q^\alpha$. The Lorentz generators $J_{ab}$ have the
usual commutators with the above generators.  The local sub-group
$H$ is generated by the $Q^\alpha$ and the $J_{ab}$. As a result we
may choose the group element to be of the form
  \begin{equation}
  g=e^{x^a P_a} \dots e^{\Phi_{a_1 a_2 a_3,\alpha}(x) K^{a_1 a_2 a_3
  ,\alpha}}  e^{\Phi_{a_1 a_2,\alpha}(x) K^{a_1
  a_2,\alpha}} e^{A_{a,\alpha}(x) R^{a,\alpha}} \quad .
  \label{1.9}
  \end{equation}

Computing the Cartan forms we find that
  \begin{eqnarray}
  g^{-1} d g & =& d x^a [ P_a + G_{a,b ,\alpha} R^{b ,\alpha} +
  G_{a,bc , \alpha} K^{bc , \alpha} - A_{a ,\alpha} Q^\alpha + ... ]
  \nonumber \\
  & =& d x^a [ P_a + ( \partial_a A_{b , \alpha} - {1
  \over 2} g A_{a , \beta} A_{b , \gamma} f^{\beta \gamma}{}_\alpha -
  2 \Phi_{ab , \alpha} ) R^{b , \alpha} \nonumber \\
  & + & ( \partial_a \Phi_{bc , \alpha} - {1 \over 6} g^2  A_{a ,
  \epsilon} A_{b , \beta} A_{c, \gamma} f^{\epsilon \beta}{}_\delta
  f^{\delta \gamma}{}_\alpha - 2 g\Phi_{ab ,\beta} A_{c , \gamma}
  f^{\beta \gamma}{}_\alpha \nonumber \\
  & +& {1 \over 2} g \partial_a A_{b , \beta} A_{c , \gamma} f^{\beta
  \gamma}{}_\alpha -3 \Phi_{abc , \alpha} ) K^{bc ,\alpha }- A_{a
  ,\alpha} Q^\alpha + ... ] \quad , \label{1.10}
  \end{eqnarray}
where the dots denote $K$ generators with more than two spacetime
indices. Only the last term in eq. (\ref{1.10}) is in the local
sub-algebra and as such we can identify $A_{a , \alpha}$ as the
connection, { \it i.e.} the gauge field, for the gauge group
generated by $Q^{\alpha}$. Each of the other terms separately
transform covariantly under the local subgroup and so we can place
constraints on them and still preserve all the symmetries. In
particular we can set
  \begin{equation}
  G_{(a,b), \alpha}=0 \quad , \label{1.11}
  \end{equation}
which implies
  \begin{equation}
  2\Phi_{ab , \alpha}=\partial_{(a} A_{b) , \alpha} \quad ,
  \label{1.12}
  \end{equation}
and also
  \begin{equation}
  G_{(a,bc)\alpha}=0\quad , \label{1.13}
  \end{equation}
which implies
  \begin{equation}
  3\Phi_{abc, \alpha}=\partial_{(a} \Phi_{bc), \alpha} -2 g \Phi_{(ab, \beta}
  A _{c),\gamma} f^{\beta\gamma}{}_\alpha + {1 \over 2} g
  \partial_{(a} A_{b , \beta} A_{c),\gamma} f^{\beta \gamma}{}_\alpha
  \quad . \label{1.14}
  \end{equation}
Indeed one can solve in this way for all the $\Phi$ fields leaving
only with the field $A_{a, \alpha}$. The elimination of some fields
using constraints on the Cartan forms that preserve the symmetries
is sometimes called the inverse Higgs mechanism \cite{23}.

Substituting the above solutions for the $\Phi$ fields into the
Cartan forms one finds expressions that contain $A_{a, \alpha}$
alone which are given by
 \begin{equation}
  g^{-1}d g= dx^a [ P_a + F_{ab , \alpha} R^{b , \alpha} +{2\over 3}
  D_b F_{ac, \alpha} K^{bc, \alpha} + ... - A_{a , \alpha}Q^\alpha ]
  \quad , \label{1.15}
  \end{equation}
where $F_{ab,\alpha}= \partial _{[a}A_{b]\alpha} - {1\over 2} g
A_{a,\beta} A_{b,\gamma} f^{\beta\gamma}{}_\alpha$ and $D_a$ is the
expected covariant derivative. We recognise this as the Yang-Mills
field strength and the higher Cartan form as its covariant
derivatives. The object invariant under the symmetries of the
non-linear realisation, which is lowest order in derivatives, is
just  the usual Yang-Mills action.

In fact only the lowest order Cartan form $G_{a,b,\alpha}$ was
evaluated in reference \cite{22}, but it is interesting to realise
that the Cartan forms do contain all the gauge covariant derivatives
of the field strength.

One way to arrive at the above set of generators of eq. (\ref{1.6})
is to write the Yang-Mills gauge parameter as a Taylor expansion
  \begin{equation}
  \lambda_\alpha(x)= a_\alpha +
  a_{a, \alpha}x^a+a_{ab , \alpha}x^ax^b+\ldots
  \label{1.16}
  \end{equation}
where the parameters $a$ do not depend on space-time. The usual
Yang-Mills transformation can then be interpreted as a an infinite
set of rigid transformations whose generators are just those of eq.
(\ref{1.6}) with the commutation relations of eqs. (\ref{1.7}) and
(\ref{1.8}). Indeed carrying rigid transformations $e^{a\cdot R}$
and $e^{a \cdot K}$ on the group element of eq. (\ref{1.9}) one
finds the same result that a Yang-Mills transformation would produce
if the gauge parameter were expanded as in eq. (\ref{1.16}).

In a tribute to Ogievetsky's important contributions to the theory
of non-linear realisations we will call the additional generators
$K^{a_1 \ldots a_n, \alpha}$ Ogievetsky generators (Og for short)
and similarly for their associated fields. They will be used
throughout this paper and they are the generators that make the
original symmetry, in this case that of the $R^{a,\alpha}$, local.
We can systematically assign a grade to the generators, in
particular $Q^\alpha$ and $P_a$ have grade -1, $R^{a,\alpha}$ has
grade 0 and $K^{a_1\ldots a_{n+1}, \alpha}$ have grade $n$. The
coupling constant $g$ has grade -1. We denote the Og generator of
grade $n$ as Og $n$. The algebra of eq. (\ref{1.8}) can then
schematically be written as
  \begin{equation}
  [G, {\rm Og} \ n ]= g \ {\rm Og} \ n \quad  [{ \rm Og}\  n ,P_a ]={\rm
  Og} \ (n-1) \quad  [{\rm Og}\  n ,{\rm Og}\  m ]= g \ {\rm Og}\  (m+n+1) \
  . \label{1.17}
  \end{equation}

It will be instructive to consider the dimensional reduction of the
above non-linear realisation in $D$ dimensions on a circle with
coordinate $y$. For simplicity we will just consider the abelian
case here, and we will therefore drop the index $\alpha$. After
dimensional reduction, the vector field becomes $A_a,
A_\star=\varphi$, while the Og 1 field becomes $\Phi_{ab},
\Phi_{a\star}, \Phi_{\star\star}$ and similarly for the higher grade
Og fields. Here $\star$ denotes the $y$th, {\it i.e.} circle,
components and $a,b=0,\ldots , D-2$. Neglecting for simplicity the
contribution along the Og 1 generator, the Cartan form of eq.
(\ref{1.10}) becomes
  \begin{eqnarray}
  g^{-1}d g & = & dx^a P_a + dy P_\star + dx^a (\partial_a
  A_{b}-2\phi_{ab} )R^{b} + dx^a(\partial_a \varphi -2\Phi_{a \star
  })R^{\star} \nonumber \\
  & + & dy(\partial_\star A_{a}-2\Phi_ {\star a})R^{a} + dy
  (\partial_\star \varphi -2\Phi_{\star \star})R^{\star}
  -dx^a A_{a} Q-dy \varphi Q \ . \label{1.18}
  \end{eqnarray}
We now take all  the fields to be independent of $y$. Imposing that
the Cartan form in the $dy$ direction vanishes, apart from the term
in the local subalgebra, we find that $\Phi_{\star
a}=\Phi_{\star\star}=0$. This generalises to all the Og fields of
any grade having at least one index in the internal direction.
Solving for the remaining Cartan forms as above one finds that one
is then left with the fields $A_{a}$ and $\varphi$ with the expected
dynamics. The net effect of these steps is that from the original
set of generators in the higher dimension we take only those that
commute with $Q$, the generator of $y$ transformations, and
construct the non-linear realisation from the sub-algebra formed by
these generators. Since the $\Phi$ fields are related to the
derivatives of the usual fields it is to be expected that some of
the Ogievetsky fields will vanish in dimensional reduction on a
circle.

The non-linear realisation of the Yang-Mills theory will be the
prototype example of all the analysis that we will perform
throughout this paper. The paper is organised as follows. Section 2
discusses the non-linear realisation of gravity, while section 3 is
devoted to the analysis of the 3-form and the 6-form of
eleven-dimensional supergravity from $E_{11}$. In section 4 we show
how to derive from $E_{11}$ the Scherk-Schwarz reduction of the IIB
theory to nine dimensions. Sections 5 and 6 are devoted to the
$E_{11}$ derivation of the massive IIA theory of Romans and of
gauged five-dimensional maximal supergravities respectively. In
section 7 we discuss the dual graviton in four dimensions,
considering first the algebra of the dual graviton alone, and then
the cases of gravity and dual gravity in $A_1^{+++}$ in four
dimensions and of dual graviton coupled to vectors in $E_{11}$ in
four dimensions. Finally, section 8 contains the conclusions.

\section{Gravity as a non-linear realisation}
It was shown long ago by Borisov and Ogievetsky that
four-dimensional gravity could be formulated as a non-linear
realisation \cite{3}. These authors showed that gravity in four
dimensions could be formulated as the  non-linear realisation of
$IGL(4,\mathbb{R})$ with local subgroup $SO(4)$ if taken together
with the simultaneous realisation of the four dimensional conformal
group $SO(2,4)$ with local subgroup $SO(4)$. The first non-linear
realisation possesses coset representatives $g=e^{x\cdot P}
e^{h\cdot K}$ that contain the coordinates of spacetime $x^\mu$ as
coefficients of the spacetime translation generator $P_a$ and the
field $h_a{}^b$, which was taken to depend on $x^\mu$, and are
associated with the generators $K^a{}_b$ of $GL(4,\mathbb{R})$. The
non-linear realisation of the conformal group has coset
representatives $g=e^{x\cdot P} e^{\phi D} e^{\phi_a K^a}$ that are
labelled by the coordinates of space-time $x^\mu$ and the fields
$\phi$ and $\phi_a$ associated with the dilation generator $D$ and
special conformal generator $K^a$. The field $\phi_a$ can be
eliminated using the inverse Higgs mechanism, that is by setting
constraints on the Cartan forms that preserve all the symmetries.
The simultaneous non-linear realisation of the two groups is
achieved by constructing the dynamics  from only the Cartan forms of
$IGL(4,\mathbb{R})$ which also transform covariantly under the
conformal group. The transformations of the two groups are linked in
that the dilation generator $D$ and the trace of the
$GL(4,\mathbb{R})$ generators $K^a{}_a$ generate the same scaling of
the coordinates $x^\mu$ and so their corresponding Goldstone fields
$\phi$ and $h^a{}_a$ must be identified with an appropriate
proportionality constant. Although a little complicated the result
of this procedure is Einstein's theory if one restricts one's
attention to terms that are second order in spacetime derivatives.
Taking only the non-linear realisation of $IGL(4,\mathbb{R})$ one
can also find Einstein's theory from the Cartan forms provided one
fixes a number of coefficients in a way not determined by the
symmetries of $IGL(4,\mathbb{R})$ alone.  The results can  be
generalised to $D$ dimensions \cite{2}. However, this latter
reference did not use the Lorentz group to make a particular choice
of coset representative and introduces a vierbein rather than a
metric.

The derivation of gravity as a non-linear realisation  was
anticipated by an earlier paper of Ogievetsky's \cite{24} that
showed that the closure of $IGL(4,\mathbb{R})$ and the conformal
group as realised on the coordinates of space-time $x^\mu$ in the
well known way  is equivalent to just considering all infinitesimal
general coordinates transformations $x^\mu\to x^\mu +f^\mu (x)$
where $f^\mu (x)$ is an arbitrary function of $x^\mu$. Thus the
closure of the two groups is an infinite dimensional group that is
just the group of general coordinate transformations. We note that
the starting point {\it i.e.} the  well known transformations on $x^
\mu$ are just those found by taking space-time to be  a coset or
equivalently a non-linear realisation in which the fields are
absent.

As such an equivalent more straightforward approach would be take
the non-linear realisation of the infinite group which is the
closure of the two groups, that is the algebra of general coordinate
transformations. This calculation is the subject of this section.
Such an approach was adopted by Pashnev \cite{25}, however, although
we will begin from the same starting point our method will depart in
some important ways, some of which are discussed in
\cite{kirsch,boulanger}, that are explained below.

Let us begin with the infinite dimensional algebra that contains the
generators
  \begin{equation}
  P_a, K^a{}_b, K^{ab}{}_c,\ldots , K^{a_1\ldots a_n}{}_c,\ldots
  \label{2.1}
  \end{equation}
where $K^{a_1\ldots a_n}{}_c=K^{(a_1\ldots a_n )}{}_c$. These
generators obey the relations
  \begin{equation}
  [ K^{a_1\ldots a_n}{}_c, P_b]= (n-1) \delta_b^{(a_1}K^{a_2\ldots
  a_n)}{}_c \label{2.2}
  \end{equation}
and
  \begin{equation}
  [K^{a_1\ldots a_n}{}_c, K^{b_1\ldots b_m}{}_d]= (n+m-1 )[{1 \over n}
  \delta_c^{(b_1|}
  K^{a_1\ldots a_n|b_2\ldots b_m)}{}_d - {1 \over m}
  \delta_d^{(a_1}K^{a_2\ldots
  a_n)b_1\ldots b_m}{}{}_c ] \ . \label{2.3}
  \end{equation}
The generators $P_a, K^a{}_b$ are those of $IGL(D,\mathbb{R})$ while
the special conformal transformations are contained in $K^{ab}{}_c$.
Indeed  the entire algebra can be generated by $P_a, K^a{}_b$ and $
K^{ab}{}_c$. We note that one can assign grade to the generators;
$K^{a_1\ldots a_{n+1}}{}_c$ has grade $n$,  $P_a$ has grade $-1$ and
$K^a{}_b$ has grade zero. This notion of grade is preserved by the
above commutation relations. In terms of our previous notation we
call the additional generators Ogievetsky, or Og, generators. In
particular, $K^{a_1\ldots a_{n+1}} {}_c$ is an Og $n$ generator. The
commutators of eqs. (\ref{2.2}) and (\ref{2.3}) can thus
schematically be written as
  \begin{equation}
  [ {\rm Og} \ n, {\rm Og } \ m ] = {\rm Og } \ (n+m) \quad ,
  \label{2.4}
  \end{equation}
which includes all possible commutators provided that we denote with
Og (-1) the momentum operator and with Og 0 the $GL(D,\mathbb{R})$
generators.

We now carry out the non-linear realisation of the group based on
the algebra of eqs. (\ref{2.2}) and (\ref{2.3}) taking  as our local
subgroup the Lorentz group which has the generators $J_{ab}= \eta_{[
a \vert c \vert }K^{c}{}_{b]}$. As such we may choose our group
element, or coset representative to be given by
  \begin{equation}
  g=e^{x^aP_a}\ldots
  e^{\Phi_{a_1\ldots a_n}^b (x) K^{a_1\ldots a_n}{}_b}\ldots e^{\Phi _{a_1 a_2}^b (x)
  K^{a_1 a_2}{}_b} e^{
  h_a{}^b (x) K^a{}_b}\equiv e^{x^aP_a}g_\phi g_h \quad .\label{2.5}
  \end{equation}
In fact this is the most general group element as we have not used
the Lorentz group to make any choice. The Cartan forms are given by
  \begin{eqnarray}
  g^{-1}d g& =& g_h^{-1} g_\phi^{-1}dx^a P_a g_\phi g_h +
  g_h^{-1}(g_\phi^{-1}d g_\phi) g_h +g_h^{-1}d g_h
  \nonumber \\
  & =& dx^\mu (e_\mu{}^a P_a+ G_{\mu , b}{}^c K^b{}_c +G_{\mu ,a b}{}^c
  K^{ab}{}_c+\ldots ) \quad .\label{2.6}
  \end{eqnarray}
A straightforward calculation gives
  \begin{equation}
  e_\mu{}^a= (e^h)_\mu{}^a,\ \ G_{\mu , b}{}^c = (e^{-1}\partial _\mu
  e)_b{}^c -\Phi _{\mu \rho}{}^\kappa (e^{-1})_b{}^\rho
  e_\kappa{}^c,\ \ \label{2.7}
  \end{equation}
  \begin{equation}
  G_{\mu , a b}{}^c= (\partial_\mu \Phi _{\rho\kappa}^\lambda -2
  \Phi_{\mu\rho\kappa}^\lambda - \Phi_{\mu
  (\rho}^\tau \Phi_{\kappa ) \tau}^\lambda + {1 \over 2} \Phi_{\rho \kappa}^\tau
  \Phi_{\mu \tau}^\lambda ) (e^{-1})_a{}^\rho (e^{-1})_b{}^\kappa
  e_\lambda{}^c, \ldots \label{2.8}
  \end{equation}
In deriving these expressions no conversion of indices on the
objects has taken place but the indices have been relabelled with
curved or flat indices suitable for their latter interpretation. The
factors of $e$ come from the final factor of $g_h$ in the group
element. Indeed, carrying out a local Lorentz transformation the
Cartan forms transform as in eq. (\ref{1.2}) and one sees that the
$e_\mu{}^a$ are rotated on their $a$ index by a local Lorentz
rotation allowing us to interpreted $e_\mu{}^a$ as the vierbein.

The part of the Cartan form involving the local subalgebra is
contained in the second term of eq. (\ref{2.7}) which we may write
as
  \begin{equation}
  G_{\mu ,a }{}^b K^{a}{}_b= G_{\mu , (a }{}^{b)} K^{(a}{}_{b)}
  +\omega_{\mu a}{}^b J^{a}{}_b \label{2.9}
  \end{equation}
where
  \begin{equation}
  \omega_{\mu a}{}^b= (e^{-1}\partial _\mu e)_{[a}{}^{b]} -\Phi _{\mu
  \rho}{}^\kappa (e^{-1})_{[a}{}^\rho e_\kappa{}^{b]} \quad .\label{2.10}
  \end{equation}
We note that although the algebra of eqs. (\ref{2.2}) and
(\ref{2.3}) is formulated in terms of the generators of
$GL(D,\mathbb{R} )$ and other generators that are representations of
$GL(D,\mathbb{R})$ the choice of the local sub-algebra to be
$SO(1,D-1)$ allows us to introduce the tangent space metric
$\eta_{ab}$ with which we may raise and lower indices to achieve the
above (anti-)symmetrisations.

Thus far we agree with the paper of Pashnev \cite{25}. However, in
this reference it was proposed that the Maurer Cartan equations $d
{\cal V}+{\cal V}\wedge {\cal V}=0$, which are identities, would
place constraints on the fields. Imposing inverse Higgs conditions
to find the Christoffel symbol in terms of the metric was correctly
carried out in \cite{kirsch,boulanger}.

From now on we follow a different path. The dynamics are constructed
in the way explained in the introduction with the Cartan forms
transforming as in eq. (\ref{1.2}). Recalling our discussion in the
introduction we conclude that $G_{a,\star}\equiv (e^{-1})_a{}^\mu
G_{\mu\star}$, where $\star$ stands for  any form except those lying
in the Poincare algebra, transform under the Lorentz group as its
indices suggest.  As such we can place constraints on these Cartan
forms and preserve all the symmetries, that is use the inverse Higgs
mechanism. Indeed, we can set
  \begin{equation}
  G_{c,(a }{}^{b)}=(e^{-1})_c{}^\mu G_{\mu (a }{}^{b)}
  =(e^{-1})_c{}^\mu (e^{-1}\partial _\mu e)_{(a}{}^{b)} - \Phi _{\mu
  \rho}{}^\kappa (e^{-1})_{(a}{}^\rho e_\kappa{}^{b)}=0 \quad .\label{2.11}
  \end{equation}
The effect of this is to solve for $\Phi _{\mu \rho}{}^\kappa=\Phi _
{(\mu \rho)}{}^\kappa$ in terms of the $e_\mu{}^a$. The result is
\cite{kirsch}
  \begin{equation}
  \Phi _{\mu \nu}{}^\kappa= \Gamma _{\mu \nu}{}^\kappa \equiv {1\over
  2}g^{\kappa \tau}(\partial_\nu g_{\tau\mu}+
  \partial_\mu g_{\tau\nu}-\partial_\tau g_{\mu\nu}) \quad .
  \label{2.12}
  \end{equation}
We define $g_{\mu\nu}=e_\mu{}^ae_\nu{}^b\eta_{ab}$ and recognise $
\Gamma _{\mu \nu}{}^\kappa$ as the usual Christoffel connection of
general relativity. A quick check of this result is to verify that
eq. (\ref{2.11}) implies that $2 g_{\lambda\kappa} \Phi _{\mu
\nu}{}^\kappa = \partial_\mu g_{\lambda \nu}$. Substituting into eq.
(\ref{2.10}) we find that
  \begin{eqnarray}
  \omega_{\mu a}{}^b & = & (e^{-1}\partial _\mu e)_{[a}{}^{b]} -
  \Gamma_{\mu \rho}{}^\kappa (e^{-1})_{[a}{}^\rho e_\kappa{}^{b]} ={1\over
  2} e_a{}^\tau(\partial_\mu e_\tau{}^b -\partial_\tau e_\mu{} ^b )
  \nonumber \\
  & -& {1\over 2} \eta^{bc}e_c{}^\tau(\partial_\mu e_\tau{}^a
  -\partial_\tau e_\mu{}^a ) -{1\over 2} e_a{}^\tau
  \eta^{bc}e_c{}^\sigma (\partial_\tau e_\sigma{}^d -\partial_\sigma
  e_\tau{}^d )e_\mu{}^d  \quad , \label{2.13}
  \end{eqnarray}
which is the well known formula for the spin connection.

At the next level we can covariantly set
  \begin{equation}
  G_{(d,ab)}{}^c=0  \quad , \label{2.14}
  \end{equation}
where $G_{d,ab}{}^c\equiv e_d{}^\mu G_{\mu,ab}^c\equiv e_d{}^\mu
e_a{}^\rho e_b{}^\kappa G_{\mu,\rho\kappa}{}^\lambda e_\lambda{}^c$.
This solves for the field $\Phi_{\mu\nu\rho}{}^\lambda=
\Phi_{(\mu\nu\rho )}{}^\lambda$ in terms of
$\Phi_{\mu\nu}{}^\lambda$ by imposing symmetrisation in the obvious
way. Substituting the solution into the part of this Cartan form
that remains we find that
  \begin{equation}
  2G_{\mu,\rho\kappa}{}^\lambda= R_{\mu\rho}{}^\lambda{}_\kappa\equiv
  \partial_\mu \Gamma _{\rho\kappa}{}^\lambda-\partial_\rho
  \Gamma_{\mu\kappa}{}^\lambda +\Gamma_{\mu\tau}{}^\lambda \Gamma_{\rho
  \kappa}{}^\tau - \Gamma_{\rho\tau}{}^\lambda \Gamma_{\mu
  \kappa}{}^\tau \label{2.15}
  \quad ,
  \end{equation}
which we recognise as the well known expression of the Riemann
tensor.

At higher orders one imposes covariant constraints on the Cartan
forms so as to solve for all the Og fields $\Phi$ to leave only the
field $h_a{}^b$ or equivalently $ e_\mu{}^a= (e^h)_\mu{}^a$.
Substituting the solutions back into the Cartan forms we find that
  \begin{equation}
  g^{-1}dg= dx^\mu (e_\mu{}^a P_a+ \omega_{\mu a}{}^b
  J^a{}_b+{1\over 2}R_{\mu\rho}{}^\lambda{}_\kappa e_a{}^\rho
  e_b{}^\kappa e_c {}^\lambda K_c^{ab}+\dots )
  \quad , \label{2.16}
  \end{equation}
where $+\dots$ denotes terms which contain covariant derivatives of
the Riemann tensor.

The Og generators play the role of turning $GL(D,\mathbb{R})$ into a
local symmetry and one can verify that carrying out a general rigid
group transformation $g\to g_0g$ on the group element of eq.
(\ref{2.5}) we recover the usual general coordinate  transformations
of general relativity on the vierbein $e_\mu{}^a$.

We will now summarise the above discussion. We started with the
group $GL(D,\mathbb{R})$, generators $K^a{}_b$ and the translations
$P_a$ to which we assigned grades $0$ and $-1$ respectively. To
these we added an infinite number of Ogievetsky generators
$K^{a_1\ldots a_{n+1}}{}_c$ each with grade $n$. These obey the Lie
algebra of eqs. (\ref{2.2}) and (\ref{2.3}). We then placed
covariant constraints on the Cartan forms solving for all the
Ogievetsky fields whereupon the remaining parts of the Cartan form
contain the spin connection at lowest grade and then the Riemann
tensor and its covariant derivatives. The introduction of the
Ogievetsky generators leads in the non-linear realisation to general
coordinate invariance. As such we find Einstein's theory in a
completely systematic way from the viewpoint of non-linear
realisations.

We now consider the dimensional reduction of this non-linear
realisation that is equivalent to the usual dimensional reduction on
a circle. Let us denote by $y$ the coordinate of the circle, $*$ the
components in this direction and let $Q=P_*$. Dimensionally reducing
the Cartan forms of eq. (\ref{2.6}) we find
  \begin{eqnarray}
  g^{-1}d g& =& dx^\mu (e_\mu{}^a P_a+e_\mu{}^*Q+ G_{\mu ,b}{}^c K^b{}_c
  +G_{\mu , *}{}^c K^*{}_c+G_{\mu , b}{}^* K^b{}_*+G_{\mu ,*}{}^*
  K^*{}_* \nonumber \\
  & +&G_{\mu , a b}{}^c K^{ab}{}_c+2G_{\mu , * b}{}^c K^{*b}{}_c +\ldots
  ) \nonumber \\
  &+&dy (e_*{}^a P_a+e_*{}^*Q+ G_{*, b}{}^c K^b{}_c+G_{*, *}{}^c K^*{}_c
  +G_{*, b}{}^* K^b{}_*+G_{*, *}{}^* K^*{}_* \nonumber \\
  &+&G_{* ,a b}{}^c K^{ab}{}_c+2G_{* ,* b}{}^c K^{*b}{}_c +G_{* ,a
  b}{}^* K^{ab}{}_*+\ldots ) \quad .\label{2.17}
  \end{eqnarray}
The coefficients $G$ can be read off from eqs. (\ref{2.7}) and
(\ref{2.8}). We now take all the fields not to depend on $y$ and
imposing the inverse Higgs constraint on all $G_{*,\bullet}$ where
${\bullet }$ is any index, that is set the part of the Cartan form
in the $dy$ direction to zero. We find that all the Ogievetsky
fields that contain a lower $*$  index  vanish. Thus all the
Ogievetsky generators that do not commute with $Q$ disappear from
the group element and so the Cartan form.  The only Ogievetsky
fields left are $\Phi_{ab}{}^c$ and $\Phi_{ab}^*$. The later field
occurs in the Cartan form in the term
  \begin{equation}
  dx^\mu G_{\mu , b}{}^* K^b{}_*=((e^{-1}\partial _\mu e)_b{}^*
  -\Phi_{\mu \rho}{}^* (e^{-1})_b{}^\rho e_*{}^* ) K^b{}_*
  \label{2.18} \quad .
  \end{equation}
In the dimensionally reduced theory we set the coefficient of
$dx^\mu$ lying in $K^{(a}{}_{b)}$ of the Cartan form to zero and
solve for  $\Phi_{ab}{}^c$ which just plays the role of the
Ogievetsky field of gravity in the lower dimension. Setting the part
of the Cartan form of eq. (\ref{2.18})  $(e^{-1})_{(a}{}^\mu G_{\mu
, b )}{}^*=0$ we solve for $\Phi_{ab}^*$ in terms of $e_\mu{}^*$.
The latter field is just the vector field that arises in this
dimensional reduction and so this step is as we found for the case
of the vector studied earlier. In the dimensionally reduced theory
we have as our local symmetry only the Local Lorentz group in the
lower dimension. Substituting for $\Phi_{ab}{}^c$ in the part of the
Cartan form in this part of the algebra we find the spin connection
for the lower dimensional theory. There remains, however, the term
containing $(e^{-1})_{[a}{}^\mu  G_{\mu , b ]}{}^*$, but this we
recognise as  just the field strength for the vector.

\section{$E_{11}$ and eleven-dimensional supergravity}
\begin{figure}[h]
\begin{center}
\begin{picture}(380,70)
\multiput(10,10)(40,0){6}{\circle{10}}
\multiput(250,10)(40,0){3}{\circle{10}} \put(370,10){\circle{10}}
\multiput(15,10)(40,0){9}{\line(1,0){30}} \put(290,50){\circle{10}}
\put(290,15){\line(0,1){30}} \put(8,-8){$1$} \put(48,-8){$2$}
\put(88,-8){$3$} \put(128,-8){$4$} \put(168,-8){$5$}
\put(208,-8){$6$} \put(248,-8){$7$} \put(288,-8){$8$}
\put(328,-8){$9$} \put(365,-8){$10$} \put(300,47){$11$}
\end{picture}
\caption{\sl The $E_8^{+++}$, or $E_{11}$, Dynkin diagram.
\label{fig1}}
\end{center}
\end{figure}
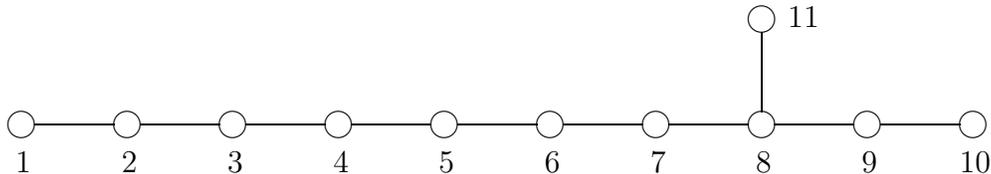

In this section we want to repeat the analysis of the previous
section for the non-linear realisation based on the very-extended
Kac-Moody algebra $E_{11}$, whose Dynkin diagram is shown in fig.
\ref{fig1}.

The decomposition of the adjoint representation of $E_{11}$ with
respect to the subalgebra $GL(11, \mathbb{R})$ corresponding to
nodes from 1 to 10 in the diagram leads to the generators $K^a{}_b$
of $GL(11, \mathbb{R})$ and $R^{abc}$ and $R_{abc}$ in the
completely antisymmetric representations of $GL(11, \mathbb{R})$,
together with an infinite set of generators which can be obtained by
multiple commutators of the generators $R^{abc}$ and $R_{abc}$
subject to the Serre relations. Defining the level $l$ as the number
of times the generator $R^{abc}$ occurs in such multiple
commutators, one obtains for instance at level 2 the generator
$R^{a_1 \dots a_6}$ with completely antisymmetric indices and at
level 3 the generator $R^{a,b_1 \dots b_8}$ antisymmetric in the
indices $b_1 \dots b_8$ and with $R^{[a,b_1 \dots b_8]} =0$. The
generator $R^{abc}$ itself has level 1, while the generator
$R_{abc}$ has level -1 and correspondingly multiple commutators of
this generator have negative level \cite{1}.

In the last section we have shown how spacetime arises in the
nonlinear realisation based on the algebra $GL(D, \mathbb{R})$ in
$D$ dimensions. This corresponds to introducing the momentum
operator $P_a$, together with an infinite set of Og $n$ operators
$K^{a_1 ... a_{n+1}}{}_b$. In $E_{11}$ the momentum operator arises
as the lowest component of the $E_{11}$ representation corresponding
to $\lambda_1 =1$, where $\lambda_1$ is the Dynkin index associated
to node 1 in fig. \ref{fig1}, and called the $l$ multiplet
\cite{19}. In this paper we will consider a different approach, that
is we will consider the momentum operator as commuting with all the
positive level generators. This approach has the advantage that one
can naturally introduce the Og operators for each positive level
generator of $E_{11}$, although it has the disadvantage of breaking
$E_{11}$ to Borel $E_{11}$, or more precisely to the subgroup of
$E_{11}$ generated by $GL(11, \mathbb{R})$ and all the positive
level generators. The corresponding local subalgebra is $SO(11)$, or
$SO(10,1)$ in Minkowski signature. We denote with $E_{11}^{local}$
the algebra generated by the momentum operator, the non-negative
level $E_{11}$ operators and the Og operators.

In the non-linear realisation, the fields associated to $R^{abc}$
and $R^{a_1 \dots a_6}$ correspond to the 3-form and its dual 6-form
of eleven dimensional supergravity. The field associated to the
generator $R^{a,b_1 \dots b_8}$ has the right indices to be
associated to the dual graviton, and we will call it the dual
graviton for short. In this section we will concentrate on the
3-form and 6-form, while section 7 will be devoted to the dual
graviton, although not in eleven dimensions but in the simpler four
dimensional case.

Following the analysis of the previous section, we take Og operators
for the 3-form and the 6-form in the representations obtained adding
symmetrised indices to the set of 3 or 6 antisymmetric indices
respectively. The Young Tableaux corresponding to the first three Og
operators is shown in fig. \ref{fig2}. In particular, the Og 1
operators $K_1^{a, b_1 b_2 b_3}$ and $K_1^{a, b_1 \dots b_6}$ belong
to the $GL(11,\mathbb{R})$ representations defined by
   \begin{eqnarray}
   & & K_1^{a, b_1 b_2 b_3} = K_1^{a,[ b_1 b_2 b_3 ]} \qquad K_1^{[a, b_1 b_2 b_3
   ]}=0 \nonumber \\
   & & K_1^{a, b_1 \dots  b_6} = K_1^{a,[ b_1 \dots  b_6 ]} \qquad K_1^{[a, b_1 \dots
   b_6 ]}=0 \label{3.1}
   \end{eqnarray}
and we take their commutation relation with $P_a$ to be
   \begin{eqnarray}
   & & [ K_1^{a, b_1 b_2 b_3} , P_c] = \delta^a_c R^{b_1 b_2 b_3} - \delta^{  [ a}_c
   R^{b_1 b_2 b_3 ]}
   \nonumber \\
   & & [ K_1^{a, b_1 \dots b_6} , P_c] = \delta^a_c R^{b_1 \dots b_6} - \delta^{  [ a}_c
   R^{b_1 \dots b_6 ]} \quad . \label{3.2}
   \end{eqnarray}
The Og 2 operator for the 3-form $K_2^{a,b,c_1 c_2 c_3}$ belongs to
the representation defined by
   \begin{equation}
   K_2^{a,b,c_1 c_2 c_3} =  K_2^{( a,b ),c_1 c_2 c_3} =  K_2^{a,b,[ c_1 c_2 c_3
   ]} \qquad  K_2^{a,[b,c_1 c_2 c_3]} = 0 \label{3.3}
   \end{equation}
and we take its commutation relation with $P_a$ to be
  \begin{equation}
  [  K_2^{a,b,c_1 c_2 c_3}, P_d ] = \delta^a_d K_1^{b, c_1 c_2 c_3}
  + \delta^b_d K_1^{a, c_1 c_2 c_3} + {3 \over 4} \delta^{ [c_1 }_d K_1^{ \vert a , b \vert  c_2 c_3] }
  + { 3 \over 4 }  \delta^{ [c_1 }_d K_1^{ \vert b , a \vert  c_2 c_3] } \quad
  , \label{3.4}
  \end{equation}
and similarly for the Og 2 operator for the 6-form $K_2^{a,b,c_1
\dots c_6}$. Proceeding this way one can write down the
representation and the commutation relation with $P_a$ of the next
Og operators. Denoting with $n$ the grade of the Og operators, {\it
i.e.} the Og 1 operators have grade 1, then the commutator of an Og
$n$ operator with $P_a$ gives an Og $(n-1)$ operator.
\begin{figure}[h]
\centering
\begin{picture}(380,140)

\put(10,115){$E_{11}$} \put(100,115){Og 1} \put(190,115){Og 2}
\put(310,115){Og 3}

\put(-10,100){\line(1,0){390}}

\put(-10,80){$R^{a_1 a_2 a_3}$:}

\put(40,80){\line(1,0){10}} \put(40,90){\line(1,0){10}}
\put(40,70){\line(1,0){10}} \put(40,60){\line(1,0){10}}
\put(40,60){\line(0,1){30}} \put(50,60){\line(0,1){30}}

\put(-10,40){$R^{a_1 \dots a_6}$:}

\put(40,50){\line(1,0){10}} \put(40,40){\line(1,0){10}}
\put(40,30){\line(1,0){10}} \put(40,20){\line(1,0){10}}
\put(40,10){\line(1,0){10}} \put(40,0){\line(1,0){10}}
\put(40,-10){\line(1,0){10}} \put(40,-10){\line(0,1){60}}
\put(50,-10){\line(0,1){60}}

\put(60,-10){\line(0,1){140}}

\put(70,80){$K_1^{a,b_1 \dots b_3}$:}

\put(120,80){\line(1,0){20}} \put(120,90){\line(1,0){20}}
\put(120,70){\line(1,0){10}} \put(120,60){\line(1,0){10}}
\put(120,60){\line(0,1){30}} \put(130,60){\line(0,1){30}}
\put(140,80){\line(0,1){10}}

\put(70,40){$K_1^{a,b_1 \dots b_6}$:}

\put(120,50){\line(1,0){20}} \put(120,40){\line(1,0){20}}
\put(120,30){\line(1,0){10}} \put(120,20){\line(1,0){10}}
\put(120,10){\line(1,0){10}} \put(120,0){\line(1,0){10}}
\put(120,-10){\line(1,0){10}} \put(120,-10){\line(0,1){60}}
\put(130,-10){\line(0,1){60}} \put(140,40){\line(0,1){10}}

\put(150,-10){\line(0,1){140}}

\put(160,80){$K_2^{a,b,c_1 \dots c_3}$:}

\put(220,80){\line(1,0){30}} \put(220,90){\line(1,0){30}}
\put(220,70){\line(1,0){10}} \put(220,60){\line(1,0){10}}
\put(220,60){\line(0,1){30}} \put(230,60){\line(0,1){30}}
\put(240,80){\line(0,1){10}} \put(250,80){\line(0,1){10}}

\put(160,40){$K_2^{a,b,c_1\dots c_6}$:}

\put(220,50){\line(1,0){30}} \put(220,40){\line(1,0){30}}
\put(220,30){\line(1,0){10}} \put(220,20){\line(1,0){10}}
\put(220,10){\line(1,0){10}} \put(220,0){\line(1,0){10}}
\put(220,-10){\line(1,0){10}} \put(220,-10){\line(0,1){60}}
\put(230,-10){\line(0,1){60}} \put(240,40){\line(0,1){10}}
\put(250,40){\line(0,1){10}}

\put(260,-10){\line(0,1){140}}

\put(270,80){$K_3^{a,b,c,d_1 \dots d_3}$:}

\put(335,80){\line(1,0){40}} \put(335,90){\line(1,0){40}}
\put(335,70){\line(1,0){10}} \put(335,60){\line(1,0){10}}
\put(335,60){\line(0,1){30}} \put(345,60){\line(0,1){30}}
\put(355,80){\line(0,1){10}} \put(365,80){\line(0,1){10}}
\put(375,80){\line(0,1){10}}

\put(270,40){$K_3^{a,b,c,d_1 \dots d_6}$:}

\put(335,50){\line(1,0){40}} \put(335,40){\line(1,0){40}}
\put(335,30){\line(1,0){10}} \put(335,20){\line(1,0){10}}
\put(335,10){\line(1,0){10}} \put(335,0){\line(1,0){10}}
\put(335,-10){\line(1,0){10}} \put(335,-10){\line(0,1){60}}
\put(345,-10){\line(0,1){60}} \put(355,40){\line(0,1){10}}
\put(365,40){\line(0,1){10}} \put(375,40){\line(0,1){10}}

\end{picture}
\caption{\label{fig2}\sl The Young Tableaux of the ${\rm Og}$
generators associated to the eleven-dimensional $E_{11}$ generators
$R^{abc}$ and $R^{a_1 \dots a_6}$.}
\end{figure}

The 6-form generator occurs in the commutator
   \begin{equation}
   [ R^{a_1 a_2 a_3} , R^{a_4 a_5 a_6} ] = 2 R^{a_1 \dots a_6} \quad
   . \label{3.5}
   \end{equation}
Using this relation and eq. (\ref{3.2}) one can then determine the
commutation relations between the Og 1 generators and $R^{abc}$
requiring that the Jacobi identities are satisfied. This gives
   \begin{equation}
   [ K_1^{a, b_1 b_2 b_3 } , R^{c_1 c_2 c_3} ] = 2 K_1^{a , b_1 b_2 b_3
   c_1 c_2 c_3 } -  2 K_1^{[ a , b_1 b_2 b_3 ] c_1 c_2 c_3 } \quad .
   \label{3.6}
   \end{equation}
Neglecting higher level generators and the gravity contribution, as
well as higher Og generators, we can thus write down the group
element as
  \begin{equation}
  g = e^{x \cdot P} e^{\Phi_{\rm Og} K_1^{\rm Og}} e^{A_{a_1 \dots
  a_6 } R^{a_1 \dots a_6}} e^{A_{a_1 \dots a_3} R^{a_1 \dots a_3}}
  \quad , \label{3.7}
  \end{equation}
where we have denoted with $\Phi_{\rm Og}$ the Og 1 fields for both
the 3-form and the 6-form, and similarly $K_1^{\rm Og}$ denotes
collectively the Og 1 operators for the 3-form and the 6-form. One
can then compute the Maurer Cartan form, which is
   \begin{eqnarray}
   g^{-1} \partial_\mu g & = & P_\mu + ( \partial_\mu A_{a_1 a_2 a_3} -
   \Phi_{\mu , a_1 a_2 a_3 } ) R^{a_1 a_2 a_3 }
   + ( \partial_\mu
   A_{a_1 \dots a_6} +  \partial_\mu A_{a_1 a_2 a_3 } A_{a_4 a_5
   a_6} \nonumber \\
   & - & \Phi_{\mu , a_1 \dots a_6} - 2 \Phi_{\mu , a_1 a_2 a_3} A_{a_4 a_5 a_6}
   ) R^{a_1 \dots a_6} + ... \label{3.8}
   \end{eqnarray}
The inverse Higgs mechanism allows one to express the Og 1 fields in
terms of the 3-form and the 6-form in such a way that only the
completely antisymmetric expressions are left in (\ref{3.8}). This
corresponds to
  \begin{eqnarray}
  \Phi_{\mu , a_1 a_2 a_3} &=& \partial_\mu A_{a_1 a_2 a_3} -
  \partial_{[\mu} A_{a_1 a_2 a_3 ]}
  \nonumber \\
  \Phi_{\mu ,a_1 \dots a_6} &=&  \partial_\mu A_{a_1 \dots a_6} -
  \partial_{[\mu} A_{a_1 \dots a_6 ]} - \partial_\mu A_{a_1 a_2 a_3 }
  A_{a_4 a_5 a_6} \nonumber \\
  &-& \partial_{[ \mu} A_{a_1 a_2 a_3 } A_{a_4 a_5
   a_6]} + 2 \partial_{[\mu} A_{a_1 a_2 a_3 ]} A_{a_4 a_5
   a_6} \quad , \label{3.9}
  \end{eqnarray}
where antisymmetry in the $a$ indices is understood. These relations
are all invariant with respect to the local subalgebra. Plugging
this into the Maurer-Cartan form one gets
  \begin{equation}
  g^{-1} \partial_\mu g  = P_\mu + F_{\mu a_1 a_2 a_3 } R^{a_1 a_2
  a_3} + F_{\mu a_1 \dots a_6} R^{a_1 \dots a_6 } + ... \quad ,
  \label{3.10}
  \end{equation}
where
  \begin{eqnarray}
  & & F_{a_1 a_2 a_3 a_4 } = \partial_{[ a_1} A_{a_2 a_3 a_4 ]}
  \nonumber \\
  & & F_{a_1 \dots a_7} = \partial_{[a_1} A_{a_2 \dots a_7 ]} +
  F_{[a_1 \dots a_4 } A_{a_5 a_6 a_7]}
  \label{3.11}
  \end{eqnarray}
are the field strengths of the 3-form and its dual 6-form of
11-dimensional supergravity.

The Maurer-Cartan form is invariant under transformations in the
Borel subalgebra, and we use this to derive transformations for the
fields. In the particular case discussed in this section, where we
have restricted the group element to be as in eq. (\ref{3.7}), we
consider the action of
  \begin{equation}
  g_0 = e^{a_{a_1 a_2 a_3} R^{a_1 a_2 a_3 } } e^{a_{a_1 \dots a_6}
  R^{a_1 \dots  a_6 } } e^{b_{a, b_1 b_2 b_3} K_1^{a, b_1 b_2 b_3 } }
  e^{b_{a, b_1 \dots  b_6} K_1^{a, b_1 \dots  b_6 } } \label{3.12}
  \end{equation}
from the left. Taking the parameters $a$ and $b$ to be
infinitesimal, we derive the transformations of the fields to be
  \begin{eqnarray}
  & & \delta A_{a_1 a_2 a_3} = a_{a_1 a_2 a_3} + x^b b_{b ,a_1 a_2 a_3
  } \nonumber \\
  & & \delta A_{a_1 \dots a_6} = a_{a_1 \dots a_6} + a_{a_1 a_2
  a_3}A_{a_4 a_5 a_6} + x^b b_{b , a_1 \dots a_6} + x^b b_{b , a_1
  a_2 a_3} A_{a_4 a_5 a_6} \nonumber \\
  & & \delta  \Phi_{b , a_1 a_2 a_3} = b_{b , a_1 a_2 a_3} \nonumber \\
  & & \delta \Phi_{b , a_1 \dots a_6} = b_{b , a_1 \dots a_6} -2 \Phi_{b ,
  a_1 a_2 a_3} a_{a_4 a_5 a_6} -2 \Phi_{ b ,
  a_1 a_2 a_3} x^c b_{c , a_4 a_5 a_6} \quad . \label{3.13}
  \end{eqnarray}
The eqs. (\ref{3.9}) and the field strengths of eqs. (\ref{3.11})
are separately invariant under these transformations, and in
particular the transformations of the 3-form and the 6-form can be
written as
  \begin{eqnarray}
  & & \delta A_{a_1 a_2 a_3} = \partial_{[ a_1 } \Lambda_{a_2 a_3 ]}
  \nonumber \\
  & & \delta A_{a_1 \dots a_6} = \partial_{[ a_1 } \Lambda_{a_2 \dots  a_6 ]}
  + \partial_{[ a_1 } \Lambda_{a_2 a_3 } A_{a_4 a_5 a_6 ]}
  \label{3.14}
  \end{eqnarray}
with gauge parameters
  \begin{eqnarray}
  & & \Lambda_{a_1 a_2} = x^b a_{b a_1 a_2} + {3 \over 4} x^b x^c
  b_{ b ,c a_1 a_2} \nonumber \\
  & & \Lambda_{a_1 \dots a_5} = x^b a_{b a_1 \dots a_5} + {6 \over
  7} x^b x^c b_{b,c a_1 \dots a_5} \quad . \label{3.15}
  \end{eqnarray}
Including higher order Og generators corresponds to higher powers of
$x$ in the equations above. The full gauge invariance is obtained
including all the Og generators.

It is worth mentioning that the normalisation used here is different
from the one used in the original $E_{11}$ paper \cite{1}. This is
for consistency with the normalisation used in the rest of this
paper. Going from this normalisation to the original one in \cite{1}
corresponds to making the field redefinitions
  \begin{eqnarray}
  & & A_{a_1 a_2 a_3 } \rightarrow {1 \over 3!} A_{a_1 a_2 a_3}
  \nonumber \\
  & & A_{a_1 \dots a_6} \rightarrow {1 \over 6!} A_{a_1 \dots a_6}
  \quad , \label{3.16}
  \end{eqnarray}
as can be deduced from eq. (2.6) in \cite{1}.

It is also instructive to consider the Maurer-Cartan form at the
next order in the Og generators. For simplicity we will now perform
this analysis only for the 3-form, so that we can neglect the
contributions coming from commutators of Og generators among
themselves. The generalisation to include the 6-form generators is
straightforward, although technically more complicated. We thus
consider the group element as only containing the 3-form generators
in the $E_{11}$ sector and including the Og 1 operator of eq.
(\ref{3.1}) and the Og 2 operator of eq. (\ref{3.3}), that is
  \begin{equation}
  g = e^{x \cdot P} e^{\Phi_{a,b,c_1 c_2 c_3} K_2^{a,b,c_1 c_2 c_3}}
  e^{\Phi_{a,b_1 b_2 b_3} K_1^{a, b_1 b_2 b_3}} e^{A_{a_1 a_2 a_3 }
  R^{a_1 a_2 a_3 }} \quad . \label{3.17}
  \end{equation}
Using eq. (\ref{3.4}), as well as eq. (\ref{3.2}), one gets
  \begin{eqnarray}
  g^{-1} \partial_\mu g  & = & P_\mu + ( \partial_\mu A_{a_1 a_2 a_3} -
   \Phi_{\mu , a_1 a_2 a_3 } ) R^{a_1 a_2 a_3 } \nonumber \\
   & + &   (\partial_\mu \Phi_{a , b_1 b_2 b_3 } - {5 \over 2}
   \Phi_{\mu , a , b_1 b_2 b_3}) K_1^{a , b_1 b_2 b_3}+ \dots  \quad
   . \label{3.18}
  \end{eqnarray}
Using the inverse Higgs mechanism one solves for the Og 1 field in
terms of the derivative of the 3-form, and the Og 2 field in terms
of the derivative of the Og 1 field. Plugging this into the group
element leads to
  \begin{equation}
  g^{-1} \partial_\mu g  = P_\mu  + F_{\mu a_1 a_2 a_3 } R^{a_1 a_2
  a_3 } + \partial_a F_{\mu b_1 b_2 b_3 } K_1^{a , b_1 b_2 b_3}+ ...
  \quad . \label{3.19}
  \end{equation}

This is an example of the general picture, in which after applying
the inverse Higgs mechanism one is left with the field strengths of
the 3-form and the 6-form together with infinitely many derivatives
of those, without breaking any of the original symmetries. These
fields are the only forms that arise in the decomposition of
$E_{11}$ with respect to $GL(11, \mathbb{R} )$. Indeed, in \cite{26}
it was shown that all the positive level 11-dimensional generators
of $E_{11}$ can be cast in generators of the form $R^{9,9,...,9,3}$,
$R^{9,9,...,9,6}$ and $R^{9,9,...,9,8,1}$, together with generators
with at least one set of 10 or 11 completely antisymmetric indices
(here we are using a shortcut notation, in which each number
corresponds to the number of antisymmetric indices; for example the
Og 2 generator for the 6-form is written as $K_2^{6,1,1}$ in this
notation). The fields associated to the former generators (the ones
with sets of 9 antisymmetric indices) were interpreted in \cite{26}
as being all the possible dual formulations of the 3-form and the
graviton, while the latter were interpreted as giving rise to
non-propagating fields. In section 7 we will consider the case of
$E_{11}$ fields with mixed symmetries, focusing in particular on the
case of the dual graviton in four dimensions, while in the next
section we will show that the introduction of the Og generators is
crucial to understand and derive the algebra that describes gauged
supergravity theories.

\section{Scherk-Schwarz reduction of IIB supergravity from $E_{11}$}

In this section we will show how to dimensionally reduce maximal
supergravities in the context of their $E_{11}$ formulation
including the Og extension. We will in particular focus on the case
of ten-dimensional IIB reduced to nine dimensions and study both the
dimensional reduction on a circle and the Scherk-Schwarz reduction.

We will first introduce the Og generators required to encode the
gauge symmetries of the ten-dimensional theory. This gives rise to
the algebra $E_{11,10B}^{local}$. We will then express the $E_{11}$
and Og generators of the IIB theory in a nine-dimensional set-up.
The consistency of the truncation from ten-dimensional IIB
supergravity to maximal supergravity in nine dimensions corresponds
to the fact that within the algebra $E_{11,10B}^{local}$ of the
ten-dimensional $E_{11}$ and Og generators one can find a
sub-algebra $E_{11,9}^{local}$ appropriate to the nine-dimensional
theory. This indeed corresponds to a maximal supergravity theory in
nine dimensions, which is a compactification of the ten-dimensional
IIB theory on a coordinate $y$. If one takes the ten-dimensional
group element not to depend on $y$ apart from the momentum
contribution $e^{y Q}$, where $Q$ is the internal momentum, then
this corresponds to standard, {\it i.e.} massless, dimensional
reduction on a circle parametrised by $y$, and the form of the group
element is preserved by the sub-algebra $E_{11,9}^{local}$ of the
ten-dimensional algebra $E_{11,10B}^{local}$ of $E_{11}$ plus Og
generators appropriate to massless dimensional reduction. One can
also consider a ten-dimensional group element with a suitable $y$
dependence, which we show to give rise in nine dimensions to the
massive theory corresponding to the Scherk-Schwarz reduction of the
IIB theory \cite{27}. This different form of the group element is
preserved by a different sub-algebra of the ten-dimensional algebra
$E_{11,10B}^{local}$ of $E_{11}$ and Og generators that we call
$\tilde{E}_{11,9}^{local}$. We show how to construct this subalgebra
corresponding to Scherk-Schwarz reduction. The mass parameter mixes
$E_{11}$ and Og generators, and from the nine-dimensional
perspective this corresponds to a deformation of the massless
$E_{11}$ algebra. The occurrence of a deformed $E_{11}$ algebra
associated to massive theories was shown for the first time in
\cite{11} for the case of the ten-dimensional massive IIA theory. In
that case the occurrence of a mass parameter for the 2-form was
shown to arise from requiring that the commutator of the 2-form
generator with momentum does not vanish, but is instead equal to the
vector generator times the Romans mass parameter.

We now consider the decomposition of the $E_{11}$ generators
appropriate to the IIB theory, that arises from deleting node 9 in
the Dynkin diagram of fig. \ref{fig1}. The $GL(10, \mathbb{R})$
subalgebra associated to the non-linear realisation of gravity
corresponds to nodes from 1 to 8 and node 11, while node 10
corresponds to the internal $SL(2 , \mathbb{R} )$ symmetry of the
IIB theory. We denote tangent spacetime indices in ten dimensions
with $\hat{a}, \hat{b}, ...$ and curved spacetime indices with
$\hat{\mu}, \hat{\nu},...$,  where the indices go from 1 to 10. One
constructs the positive level generators as multiple commutators of
the 2-form generator $R^{\hat{a} \hat{b} ,\alpha}$, $\alpha = 1,2$,
which is a doublet of $SL(2 , \mathbb{R} )$. Together with the
$GL(10, \mathbb{R} )$ generators $K^{\hat{a}}{}_{\hat{b}}$ and the
$SL(2, \mathbb{R} )$ generators $R^i$, $i=1,2,3$ at level zero, one
has the doublet of 2-form generators at level 1, a 4-form generator
$R^{\hat{a}\hat{b}\hat{c}\hat{d}}$ at level 2, and then a doublet of
6-forms at level 3, a triplet of 8-forms at level 4 and a doublet
and a quadruplet of 10-forms at level 5, together with an infinite
set of generators with mixed, {\it i.e.} not completely
antisymmetric, indices. We consider the positive level generators as
commuting with the momentum operator $P_{\hat{a}}$.

We now want to write down the relevant algebra in ten dimensions.
For simplicity, we consider a level truncation  and we therefore
only consider in ten dimensions the 2-form generators
$R^{\hat{a}\hat{b} , \alpha}$, together with the $GL(10,\mathbb{R})$
generators $K^{\hat{a}}{}_{\hat{b}}$ and the $SL(2, \mathbb{R})$
generators $R^i$. We have the commutation relations
  \begin{eqnarray}
  & & [ R^i , R^j] = f^{ij}{}_k R^k \nonumber \\
  & & [ R^i , R^{\hat{a}\hat{b},\alpha } ]= D^i_\beta{}^\alpha R^{\hat{a}\hat{b} ,\beta}
  \label{4.1}
  \end{eqnarray}
where $D^i_\beta{}^\alpha$ are the generators of $SL(2, \mathbb{R})$
satisfying
  \begin{equation}
  [ D^i , D^j ]_\beta{}^\alpha = f^{ij}{}_k D^k_\beta{}^\alpha
  \label{4.2}
  \end{equation}
and $f^{ij}{}_k$ are the structure constants of $SL(2,\mathbb{R})$.
In terms of Pauli matrices, a choice of $D^i_\beta{}^\alpha$ is
  \begin{equation}
  D_1 = {\sigma_1 \over 2} \qquad D_2 = {i \sigma_2 \over 2} \qquad D_3 = {\sigma_3 \over
  2} \quad . \label{4.3}
  \end{equation}

We now add the Og generators to the $E_{11}$ formulation of
ten-dimensional IIB. In this way we encode all the local gauge
symmetries of the ten-dimensional IIB theory. The procedure is much
like the one discussed in the previous sections for other cases. The
Og 1 operator for the 2-form is a doublet of generators
$K^{\hat{a},\hat{b}\hat{c} , \alpha}$, satisfying
  \begin{equation}
  K^{\hat{a},\hat{b}\hat{c} , \alpha} =K^{\hat{a},[\hat{b}\hat{c}] ,
  \alpha} \qquad K^{[\hat{a},\hat{b}\hat{c}] , \alpha} =
  0 \quad , \label{4.4}
  \end{equation}
and whose commutation relation with the momentum operator
$P_{\hat{a}}$ is
  \begin{equation}
  [ K^{\hat{a},\hat{b}\hat{c} , \alpha} , P_{\hat{d}} ] = \delta^{\hat{a}}_{\hat{d}} R^{\hat{b}\hat{c} , \alpha} -
  \delta^{[\hat{a}}_{\hat{d}} R^{\hat{b}\hat{c} ] , \alpha} \quad .
  \label{4.5}
  \end{equation}
Ignoring for simplicity the gravity contribution, the non-linear
realisation can be constructed from the group element
  \begin{equation}
  g = e^{x \cdot P} e^{\Phi_{\hat{a},\hat{b}\hat{c} , \alpha} K^{\hat{a},\hat{b}\hat{c} , \alpha}}
  e^{A_{\hat{a}\hat{b} ,\alpha} R^{\hat{a}\hat{b} ,\alpha}} e^{\phi_i R^i} \quad
  , \label{4.6}
  \end{equation}
and the corresponding Maurer-Cartan form gives
  \begin{equation}
  g^{-1} d g = dx^{\hat{\mu}} [ P_{\hat{\mu}} + ( \partial_{\hat{\mu}} A_{\hat{a}\hat{b} , \alpha} -
  \Phi_{{\hat{\mu}} , \hat{a}\hat{b} , \alpha})e^{- \phi_i R^i}  R^{\hat{a}\hat{b} , \alpha} e^{\phi_i R^i} +
  e^{-\phi_i R^i} \partial_{\hat{\mu}} e^{\phi_i R^i} +  ...] \quad
  . \label{4.7}
  \end{equation}
The inverse Higgs mechanism then fixes $\Phi_{{\hat{\mu}} ,
\hat{a}\hat{b} , \alpha}$ in terms of $\partial_{\hat{\mu}}
A_{\hat{a}\hat{b} , \alpha}$ so that the $R^{\hat{a}\hat{b},\alpha}$
term becomes proportional to
  \begin{equation}
  F_{\hat{a}\hat{b}\hat{c} , \alpha} = \partial_{[ \hat{a}} A_{\hat{b}\hat{c} ] ,\alpha} \quad
  , \label{4.8}
  \end{equation}
which is the field strength for the 2-form. This procedure is
completely consistent because the inverse Higgs mechanism preserves
entirely the local subalgebra, which is $SO(9,1) \times SO(2)$.

We now consider a generic compactification of the IIB theory in the
above $E_{11}$ formulation to nine dimensions. This will include the
derivation of both the massless theory and the Scherk-Schwarz
reduction, which both have maximal supersymmetry. We thus split the
ten-dimensional coordinates in $x^\mu$, $\mu = 1, ..., 9$, and the
10th coordinate $y$. Correspondingly, the momentum operator splits
in $P_a$ and $Q$, where $Q=P_y$. As we did in ten-dimensions, we
consider a level truncation and thus we are only interested in
1-forms and 2-forms in nine dimensions. The doublet of 2-form
generators in ten dimensions gives a doublet of 2-forms $R^{ab ,
\alpha}$ and a doublet of 1-forms $R^{a , \alpha} = R^{ay ,
\alpha}$. One also obtains a 1-form from the $GL(10 , \mathbb{R})$
generators, namely $R^a = K^a{}_y$, whose commutator with $P_a$ is
  \begin{equation}
  [ R^a , P_b ] =  - \delta^a_b Q \quad .
  \label{4.9}
  \end{equation}
One also has the $SL(2,\mathbb{R})$ triplet of scalar generators
$R^i$, as well as the singlet scalar generator $R = K^y{}_y$,
satisfying
  \begin{equation}
  [ R, Q ] = -Q \quad .
  \label{4.10}
  \end{equation}
The commutator between $R^a$ and $R^{a ,\alpha}$ is
  \begin{equation}
  [ R^a , R^{b , \alpha}] = - R^{ab ,\alpha} \quad ,
  \label{4.11}
  \end{equation}
while the non-vanishing commutators with the scalars are
  \begin{eqnarray}
  & & [ R, R^a ] = -R^a \qquad \qquad \ \ [ R , R^{a , \alpha} ] = R^{a ,
  \alpha} \nonumber \\
  & & [ R^i , R^{a , \alpha} ] = D^i_\beta{}^\alpha R^{a ,\beta}
  \qquad [ R^i , R^{ab , \alpha} ] = D^i_\beta{}^\alpha R^{ab
  ,\beta} \quad . \label{4.12}
  \end{eqnarray}
The commutator of $R^a$ with itself and the commutator of $R^{a
,\alpha}$ with itself vanish,
  \begin{equation}
  [ R^a , R^b ] =0 \qquad [ R^{a , \alpha} , R^{b , \beta} ] = 0
  \quad . \label{4.13}
  \end{equation}
These are all the $E_{11}$ commutators we need consider at the level
we are analysing. At the end of this section we will also consider
the 3-form generator $R^{abc}$ and 4-form generator $R^{abcd}$. The
first arises from the 4-form generator of IIB with one index in the
internal direction, $R^{abcy}$, while the latter is just the 4-form
of IIB with all indices along the nine-dimensional spacetime.

Just as for the $E_{11}$ generators,  we also rewrite the
ten-dimensional Og generators as decomposed in $GL(9 , \mathbb{R})$
representations. The generator $K^{\hat{a},\hat{b} \hat{c} ,
\alpha}$ thus gives rise to $K^{a,bc , \alpha}$, $K^{[ab],\alpha}$,
$K^{(ab), \alpha}$ and $K^{a , \alpha}$, where
  \begin{equation}
  K^{[ab], \alpha} = K^{y, ab , \alpha} - K^{[a,b]y , \alpha} \qquad
  K^{(ab), \alpha} = K^{(a,b)y , \alpha} \qquad K^{a , \alpha}
  = K^{y, a y , \alpha} \quad . \label{4.14}
  \end{equation}
The commutation relations of these operators with $P_a$ and $Q$ are
  \begin{eqnarray}
  &  [ K^{a,bc , \alpha} , P_d ]  = \delta^a_d R^{bc , \alpha} -
  \delta^{[a}_d R^{bc ] , \alpha} \qquad & [ K^{a,bc , \alpha} , Q ] =
  0 \nonumber \\
  & \! \! \! \! \! \! \! \! \! \! \! \! \! \! \! \! \! \! \! \! \! \! \! \! \! \! \! \!
  \! \! \! \! \! \! \! [ K^{[ab], \alpha} , P_c ]  = - \delta^{[a}_c R^{b],\alpha}
  & [ K^{[ab], \alpha} , Q ] = R^{ab , \alpha} \nonumber \\
  & \! \! \! \! \! \! \! \!\! \! \! \! \! \! \! \! \! \! \! \! \! \!\! \! \! \! \! \! \!
  \! \! \! \! \! \! \![ K^{(ab), \alpha} , P_c ] = \delta^{(a}_c R^{b),\alpha}
  & [ K^{(ab), \alpha} , Q ] = 0 \nonumber \\
  & \! \! \! \! \! \! \! \!\! \! \! \! \! \! \!\! \! \! \! \! \! \!\! \! \! \! \! \! \!
  \! \! \! \! \! \! \!\! \! \! \! \! \! \!\! \! \! \! \! \! \!
  \! \! \! \! \! \! \![ K^{a ,\alpha} , P_b ] = 0 & [ K^{a ,\alpha} , Q ] =
  R^{a , \alpha} \quad . \label{4.15}
  \end{eqnarray}
Similarly, the dimensional reduction of the gravity Og 1 operator
gives the Og operators $K^{(ab)}$, $K^a$ and $K$, that satisfy
  \begin{eqnarray}
  & & [K^{(ab)} , P_c ] = \delta^{(a}_c R^{b )} \qquad [  K^{(ab)} , Q ]
  =0 \nonumber \\
  & & [ K^a , P_b ] = \delta^a_b R \qquad \quad  \ \ [ K^a , Q ] = R^a
  \nonumber \\
  & & [ K , P_a ] = 0 \qquad \qquad \quad \  [ K , Q ] = R \quad .
  \label{4.16}
  \end{eqnarray}

We now write down the group element. For simplicity we will neglect
Og 2 contributions, and therefore we will consider the Og 1
generators as commuting among themselves. We will denote with
$\Phi_{\rm Og}$ and $K^{\rm Og}$ the whole set of Og 1 fields and
generators. Thus the group element is
  \begin{equation}
  g = e^{x \cdot P} e^{y Q} e^{\Phi_{\rm Og} K^{\rm Og}}
  e^{A_{ab ,\alpha} R^{ab ,\alpha}} e^{A_{a , \alpha} R^{a , \alpha}}
  e^{A_a R^a} e^{\phi R} e^{\phi_i R^i} \quad , \label{4.17}
  \end{equation}
where all the fields are taken to depend on $x$ and $y$. We now
compute the Maurer-Cartan form. The result is
  \begin{eqnarray}
  g^{-1} d g & = & dx^\mu [ P_\mu + A_\mu e^\phi Q +
  ( \partial_\mu A_{ab , \alpha} - \partial_\mu A_{a , \alpha} A_b -
  \Phi_{\mu , ab , \alpha} + \Phi_{(\mu a ) , \alpha} A_b -
  \Phi_{[\mu a ] , \alpha} A_b \nonumber \\
  & +& \Phi_{(\mu a)} A_{b , \alpha} + \Phi_\mu A_{a , \alpha} A_b
  )e^{- \phi_i R^i}  R^{ab , \alpha} e^{\phi_i R^i}
  + (\partial_\mu A_{a , \alpha} - \Phi_{(\mu a ) , \alpha} +
  \Phi_{[\mu a ] , \alpha} \nonumber \\
  &-& \Phi_\mu A_{a , \alpha}) e^{-\phi} e^{- \phi_i R^i}  R^{a , \alpha} e^{\phi_i
  R^i}
  +  (\partial_\mu A_a - \Phi_{(\mu a )} + \Phi_\mu A_a  ) e^\phi R^a
  + ( \partial_\mu \phi - \Phi_\mu ) R \nonumber \\
  &+&
  e^{-\phi_i R^i} \partial_\mu e^{\phi_i R^i} +  ...]
  + d y [ e^\phi Q +
  (\partial_y A_{ab , \alpha}  -  \partial_y A_{a , \alpha} A_b
  -  \Phi_{[ab ] , \alpha}  +
  \Phi_{ a , \alpha} A_b \nonumber \\
  &+& \Phi A_{a ,\alpha} A_b )
  e^{- \phi_i R^i}  R^{ab , \alpha} e^{\phi_i R^i} + ( \partial_y
  A_{a , \alpha}- \Phi_{a , \alpha} - \Phi A_{a ,\alpha} ) e^{-\phi} e^{- \phi_i R^i}  R^{a , \alpha}
  e^{\phi_i R^i}\nonumber \\
  & +& ( \partial_y A_a  - \Phi_a + \Phi A_a ) e^\phi R^a
  + ( \partial_y \phi - \Phi ) R
  + e^{-\phi_i R^i} \partial_y e^{\phi_i
  R^i} + \dots ]
  \quad , \label{4.18}
  \end{eqnarray}
where the dots denote contributions from higher level $E_{11}$
generators and Og generators.

Before discussing the Scherk-Schwarz reduction of the IIB theory, we
first consider the derivation of the massless nine-dimensional
supergravity. We take all the fields in the group element of eq.
(\ref{4.17}) not to depend on $y$, and using the inverse Higgs
mechanism we set the part of the Maurer-Cartan form of eq.
(\ref{4.18}) in the $dy$ direction and proportional to the $E_{11}$
and Og generators to zero. This imposes
  \begin{equation}
  \Phi_{[ab],\alpha} = \Phi_{a ,\alpha} = \Phi_a = \Phi = 0 \quad . \label{4.19}
  \end{equation}
Considering the $d x^\mu$ part and imposing the inverse Higgs
mechanism on the remaining Og fields one finds that the
Maurer-Cartan form gives the field strengths
  \begin{eqnarray}
  & & F_{abc ,\alpha} = \partial_{[a} A_{bc ] , \alpha} -
  \partial_{[a} A_{b , \alpha} A_{c ]}  \nonumber \\
  & & F_{ab , \alpha} = \partial_{[a} A_{b ] , \alpha} \nonumber \\
  & & F_{ab} = \partial_{[a} A_{b ]} \quad , \label{4.20}
  \end{eqnarray}
which are invariant under the gauge transformations
  \begin{equation}
  \delta A_{ab , \alpha} = \partial_{[a}  \Lambda_{b ] , \alpha} -
  \partial_{[a} \Lambda A_{b ] , \alpha} \qquad \delta A_{a ,
  \alpha} = \partial_{a} \Lambda_{\alpha} \qquad \delta A_a  =
  \partial_{a} \Lambda \quad . \label{.4.21}
  \end{equation}
This construction is consistent because the relations that the
inverse Higgs mechanism imposes are invariant under the local
subalgebra, which is $SO(1,8) \otimes SO(2)$. The field-strengths
and gauge transformations we have derived are those of the 1-forms
and 2-forms of massless maximal nine-dimensional supergravity.

In the above, we have set to zero the Og fields corresponding to the
generators $K^{[ab],\alpha}$, $K^{a , \alpha}$, $K^a$ and $K$.
Implementing this in the group element, and so the Cartan form, we
find that these generators in fact play no role. These generators
are indeed the only ones in eqs. (\ref{4.15}) and (\ref{4.16}) that
do not commute with $Q$. As such, one is left with the original
$E_{11}$ generators and a subset of the Og generators, all of which
commute with $Q$ apart from the scalar generator $R$, and all fields
which do not depend on $y$. We note that the operator $Q$ appears in
the commutation relations of eq. (\ref{4.9}) and (\ref{4.10}).
However, $Q$ commutes with every operator in the theory other that
$R$, and the commutator of $R$ with $Q$ is proportional to $Q$, and
so one can consistently set the commutator of $R^a$ with $P_a$ to
zero and ignore $Q$ in the algebra. Correspondingly, one can ignore
the presence of $Q$ in the group element of eq. (\ref{4.17}), which
corresponds to no $y$ dependence at all. Thus one finds a non-linear
realisation that is the one that arises if one constructs the
massless nine-dimensional theory using the formulation of $E_{11}$
appropriate to nine dimensions, which corresponds to deleting nodes
9 and 11 in the Dynkin diagram in fig. \ref{fig1} and decomposing
$E_{11}$ in terms of the $GL(9, \mathbb{R})$ subalgebra. The Og
generators that are left are the Og generators that encode the gauge
symmetries of the nine-dimensional theory, and they form with the
non-negative level $E_{11}$ generators the algebra
$E_{11,9}^{local}$. To summarise, the massless nine-dimensional
theory arises from taking the subset of Og generators that commute
with $Q$. This implies that one can consistently remove $Q$ from the
algebra, which can be used to construct the non-linear realisation.
This in the nine-dimensional $E_{11}$ formulation of massless
maximal supergravity.

We now describe the Scherk-Schwarz dimensional reduction of the
ten-dimensional IIB supergravity theory to nine dimensions in an
analogous way. We take the same starting point, namely the $E_{11}$
formulation of the IIB theory in ten dimensions together with the Og
generators and corresponding fields. We consider the group element
  \begin{equation}
  g  = e^{x \cdot P} e^{y ( Q +  m_i R^i ) } e^{\Phi_{\rm Og} (x) K^{\rm Og}}
  e^{A_{ab ,\alpha}(x) R^{ab ,\alpha}} e^{A_{a , \alpha} (x) R^{a , \alpha}}
  e^{A_a (x) R^a} e^{\phi(x) R } e^{\phi_i (x) R^i} \quad . \label{4.22}
  \end{equation}
We thus take the dependence on the coordinate $y$ in the group
element to be in the form $e^{y (Q + m_i R^i )}$. This is equivalent
to taking the theory to be defined on the conventional
nine-dimensional spacetime tensored with a manifold that is a circle
constructed form the usual ten-dimensional circle of spacetime and a
circle, or one parameter subgroup of $SL(2,\mathbb{R})$, which is
specified by the mass parameter $m_i$. The factor $e^{y (Q + m_i R^i
)}$ occurs at the beginning of the group element in the usual place
for the introduction of spacetime, and the fields are taken to not
depend on $y$, however we can rearrange the group element by taking
the $e^{ y m_i R^i}$ factor to the right whereupon the fields
acquire a $y$ dependence, that is
  \begin{equation}
  g = e^{x \cdot P} e^{y Q} e^{\Phi_{\rm Og} (x,y) K^{\rm Og}}
  e^{A_{ab ,\alpha} (x,y) R^{ab ,\alpha}} e^{A_{a , \alpha} (x,y) R^{a , \alpha}}
  e^{A_a (x) R^a} e^{\phi(x) R } e^{y m_i R^i} e^{\phi_i (x) R^i} \quad
  . \label{4.23}
  \end{equation}
The $y$ dependence of the fields in this last expression can be
derived using the relation
  \begin{equation}
  e^A e^B e^{-A} = e^{e^A B e^{-A}} \quad . \label{4.24}
  \end{equation}
This implies in particular that any $SL(2, \mathbb{R})$ doublet
acquires the same $y$ dependence. For instance for the 2-form this
is
  \begin{equation}
  A_{ab , \alpha} (x,y) = ( e^{y m_i D^i} )_\alpha{}^\beta A_{ab
  ,\beta} (x) \quad , \label{4.25}
  \end{equation}
and similarly for any doublet, including the Og fields, while the
$SL(2,\mathbb{R})$ singlets acquire no $y$ dependence. This is the
$y$ dependence that results in the Scherk-Schwarz dimensional
reduction, which consists in compactifying the ten-dimensional
theory to nine dimensions on a circle of coordinate $y$, while
performing a $y$-dependent $SL(2, \mathbb{R})$ transformation
\cite{27}.

From the group element in (\ref{4.23}) one obtains the Maurer-Cartan
form of eq. (\ref{4.18}). It is instructive to write this down to
show explicitly the $y$ dependence. The result is
  \begin{eqnarray}
  g^{-1} d g & = & dx^\mu [ P_\mu + A_\mu e^\phi Q +
  ( \partial_\mu A_{ab , \alpha} - \partial_\mu A_{a , \alpha} A_b -
  \Phi_{\mu , ab , \alpha} + \Phi_{(\mu a ) , \alpha} A_b -
  \Phi_{[\mu a ] , \alpha} A_b \nonumber \\
  & +& \Phi_{(\mu a)} A_{b , \alpha} + \Phi_\mu A_{a , \alpha} A_b
  )e^{- \phi_i R^i} e^{-y m_i R^i}  R^{ab , \alpha} e^{-y m_i R^i} e^{\phi_i
  R^i} \nonumber \\
  & + & (\partial_\mu A_{a , \alpha} - \Phi_{(\mu a ) , \alpha} +
  \Phi_{[\mu a ] , \alpha}
  - \Phi_\mu A_{a , \alpha}) e^{-\phi} e^{- \phi_i R^i}  e^{-y m_i R^i}  R^{a , \alpha}  e^{y m_i R^i}  e^{\phi_i
  R^i} \nonumber \\
  & + &  (\partial_\mu A_a - \Phi_{(\mu a )} + \Phi_\mu A_a  ) e^\phi R^a
  + ( \partial_\mu \phi - \Phi_\mu ) R
  +
  e^{-\phi_i R^i} \partial_\mu e^{\phi_i R^i} +  ...] \nonumber \\
  & + & d y [ e^\phi Q +
  (\partial_y A_{ab , \alpha}  -  \partial_y A_{a , \alpha} A_b
  -  \Phi_{[ab ] , \alpha}  +
  \Phi_{ a , \alpha} A_b \nonumber \\
  & + & \Phi A_{a ,\alpha} A_b )
  e^{- \phi_i R^i}  e^{-y m_i R^i}  R^{ab , \alpha}  e^{y m_i R^i} e^{\phi_i R^i} + ( \partial_y
  A_{a , \alpha}- \Phi_{a , \alpha} \nonumber \\
  &-& \Phi A_{a ,\alpha} ) e^{-\phi} e^{- \phi_i R^i}  e^{-y m_i R^i}  R^{a ,
  \alpha}  e^{y m_i R^i}
  e^{\phi_i R^i}
  + ( \partial_y A_a  - \Phi_a + \Phi A_a ) e^\phi R^a \nonumber \\
  &+&  ( \partial_y \phi - \Phi ) R
  + e^{-\phi_i R^i } m_i R^i e^{\phi_i R^i} + \dots ]
  \quad . \label{4.26}
  \end{eqnarray}
Alternatively, one can compute the Maurer-Cartan form with the group
element written as in eq. (\ref{4.22}). In this way of writing down
the group element, the fields have no $y$ dependence and the $dy$
part of eq. (\ref{4.26}) results from passing $m_i R^i$ through the
group element. Indeed it can be shown that the two ways of computing
the Maurer-Cartan form are identical using the $y$ dependence given
as in eq. (\ref{4.25}).

As for the massless case, we now use the inverse Higgs mechanism to
impose that all the terms in $dy$ proportional to positive level
generators vanish, and we get
  \begin{eqnarray}
  & & \Phi_{a , \alpha} (x) = ( m_i D^i )_\alpha{}^\beta A_{a , \beta}
  (x) \qquad   \Phi_{[ab] , \alpha} (x) = ( m_i D^i )_\alpha{}^\beta A_{ab ,
  \beta} (x) \nonumber \\
  & & \Phi = \Phi_a =0 \quad . \label{4.27}
  \end{eqnarray}
This does not apply to the scalars $\phi_i$, and indeed the $m_i
R^i$ term in the $dy$ part of eq. (\ref{4.26}) is not affected by
the inverse Higgs mechanism. This will be discussed later. We put
the relations of eq. (\ref{4.27}) back in the group element and so
the Cartan form, and we use the inverse Higgs mechanism on the
remaining Og fields such that the $d x^\mu$ part of the Cartan form
gives the field-strengths
  \begin{eqnarray}
  & & F_{abc ,\alpha} = \partial_{[a} A_{bc ] , \alpha} -
  \partial_{[a} A_{b , \alpha} A_{c ]} -
  ( m_i D^i )_\alpha{}^\beta A_{[ab , \beta} A_{c ]}
  \nonumber \\
  & & F_{ab , \alpha} = \partial_{[a} A_{b ] , \alpha}
  +( m_i D^i )_\alpha{}^\beta A_{ab , \beta} \quad , \nonumber \\
  & & F_{ab} = \partial_{[a} A_{b ]} \quad ,
  \label{4.28}
  \end{eqnarray}
which transform covariantly under the gauge transformations
  \begin{eqnarray}
  & & \delta A_{ab ,\alpha} = \partial_{[a}  \Lambda_{b ] , \alpha} -
  \partial_{[a} \Lambda A_{b ] , \alpha} + \Lambda
  ( m_i D^i )_\alpha{}^\beta A_{ab ,\beta} \nonumber \\
  & &  \delta A_{a , \alpha} = \partial_{a} \Lambda_{\alpha}
  + \Lambda ( m_i D^i )_\alpha{}^\beta A_{a , \beta} -
  ( m_i D^i )_\alpha{}^\beta \Lambda_{a , \beta} \quad ,\nonumber \\
  & & \delta A_a = \partial_a \Lambda \quad .
  \label{4.29}
  \end{eqnarray}
These are the field strengths and gauge transformations of the
1-forms and 2-forms of the nine-dimensional gauged maximal
supergravity that arises from Scherk-Schwarz reduction of the IIB
theory \cite{27}.

We now discuss the scalar sector. One obtains the correct covariant
derivative for the scalars observing that the metric that results
from the Maurer-Cartan form in eq. (\ref{4.26}) is
  \begin{equation}
  \left( \begin{array}{cc}
  e_\mu{}^a  & A_\mu e^\phi \\
  0 & e^\phi \end{array} \right) \label{4.30}
  \end{equation}
as the coefficients of the generators $P_a$ and $Q$ (for simplicity
we are actually not considering the gravity contribution in nine
dimensions and therefore the coefficient of $P_a$  in eq.
(\ref{4.26}) is the diagonal metric). The corresponding inverse
metric is
  \begin{equation}
  \left( \begin{array}{cc}
  e^\mu{}_a  & - A_a \\
  0 & e^{-\phi} \end{array} \right) \quad . \label{4.31}
  \end{equation}
Taking account of having applied the inverse Higgs mechanism, the
only other part of the Maurer-Cartan form along $dy$ is
  \begin{equation}
  G_{y i} R^i = m_i e^{-\phi_i R^i } R^i e^{\phi_i R^i} \quad ,
  \label{4.32}
  \end{equation}
while the $R^i$ term along $d x^\mu$ is
  \begin{equation}
  G_{\mu i} R^i = e^{-\phi_i R^i } \partial_\mu e^{\phi_i R^i} \quad
  . \label{4.33}
  \end{equation}
Therefore the covariant derivative for the scalar is given by
  \begin{equation}
  e^{\mu}{}_a G_{\mu ,i } - A_a G_{y , i} \quad , \label{4.34}
  \end{equation}
which reads
  \begin{equation}
  e^{-\phi_i R^i} \partial_a e^{\phi_i R^i} - A_a m_i e^{-\phi_i R^i} R^i e^{\phi_i
  R^i} \quad . \label{4.35}
  \end{equation}
This analysis therefore gives all the covariant quantities of the
nine-dimensional theory corresponding to the Scherk-Schwarz
reduction of IIB.

As we observed, eq. (\ref{4.27}) expresses some Og fields in terms
of $E_{11}$ fields. Similarly, requiring that the Og 1 operators
$K^{[ab], \alpha}$ and $K^{a , \alpha}$ have vanishing coefficients
in the $dy$ direction relates $\Phi_{[ab ], \alpha}$ and $\Phi_{a ,
\alpha}$ to Og 2 fields carrying the same spacetime and $SL(2,
\mathbb{R})$ representations. Iterating this one obtains for any $n$
an Og $n$ generator identified with $A_{ab ,\alpha}$ times the $n$th
power of the mass parameter, and similarly for $A_{a ,\alpha}$. This
generalises to all the fields in the theory. Putting these solutions
into the original group element of eq. (\ref{4.22}) we find that it
takes the form
  \begin{equation}
  g  = e^{x \cdot P} e^{y ( Q + m_i R^i )}  e^{\Phi_{\rm Og} (x) \tilde{K}^{\rm Og}}
  e^{A_{ab ,\alpha}(x) \tilde{R}^{ab ,\alpha}} e^{A_{a , \alpha} (x) \tilde{R}^{a , \alpha}}
  e^{A_a (x) R^a} e^{\phi(x) R } e^{\phi_i (x) R^i} \quad ,
  \label{4.36}
  \end{equation}
where
  \begin{eqnarray}
  & & \tilde{R}^{a , \alpha} = {R}^{a , \alpha} + m_i D^i_\beta{}^\alpha
  K^{a, \beta} + ... \nonumber \\
  & & \tilde{R}^{ab , \alpha} = {R}^{ab , \alpha} + m_i D^i_\beta{}^\alpha
  K^{[ab], \beta} + ... \quad , \label{4.37}
  \end{eqnarray}
where the dots correspond to higher powers in $m_i$ multiplying
higher grade Og generators, and $\tilde{K}$ denotes deformed Og
generators associated with nine-dimensional gauge transformations.
The group element of eq. (\ref{4.36}) resembles the group element
corresponding to the massless nine-dimensional theory, in the sense
that each generator in eq. (\ref{4.36}) corresponds to a generator
with identical index structure of the massless nine-dimensional
theory. As such we can interpret the $\tilde{R}$ generators as
deformed $E_{11}$ generators. In particular, we claim that although
the expansions in eq. (\ref{4.37}) are non-polynomial in $m_i$, all
the commutation relations involving these operators, or the
commutation relations between these operators and momentum, only
contain terms at most linear in $m_i$. In particular, the commutator
of $\tilde{R}^{ab ,\alpha}$ with $P_a$ is
  \begin{equation}
  [  \tilde{R}^{ab ,\alpha}, P_c ] = - (m_i D^i
  )_\beta{}^\alpha  \delta^{[a}_c \tilde{R}^{b ] , \beta} \quad ,
  \label{4.38}
  \end{equation}
while the commutator of $\tilde{R}^{a ,\alpha}$ with $P_a$ vanishes,
as can be seen from eq. (\ref{4.15}).

The deformed $E_{11}$ and indeed the deformed Og generators have a
simple algebraic classification. They are the operators that commute
with the operator $\tilde{Q}$ defined as
  \begin{equation}
  \tilde{Q} = Q + m_i R^i \quad . \label{4.39}
  \end{equation}
Indeed, the commutation relation of $R^a$, $R^{a, \alpha}$ and
$R^{ab, \alpha}$ with $\tilde{Q}$ is
  \begin{eqnarray}
  & & [ \tilde{Q} , R^a ] = 0 \nonumber \\
  & & [ \tilde{Q} , R^{a , \alpha} ] = m_i D^i_\beta{}^\alpha R^{a ,
  \beta} \nonumber \\
  & & [ \tilde{Q} , R^{ab , \alpha} ] = m_i D^i_\beta{}^\alpha R^{ab ,
  \beta} \quad . \label{4.40}
  \end{eqnarray}
We thus have to deform the operators $R^{a ,\alpha}$ and $R^{ab
,\alpha}$, and from eq. (\ref{4.15}) one gets that the deformed
operators given in eq. (\ref{4.37}) satisfy
  \begin{equation}
  [ \tilde{Q} , \tilde{R}^{a , \alpha} ] = [ \tilde{Q} , \tilde{R}^{ab , \alpha} ] = 0
  \quad . \label{4.41}
  \end{equation}
The Og generators of the nine-dimensional theory are also redefined
in order to commute with $\tilde{Q}$. One thus constructs the
generators $\tilde{K}^{a, bc , \alpha}$ and $\tilde{K}^{(ab) ,
\alpha}$, which are Og 1 generators followed by an expansion in
$m_i$ of higher grade Og generators. The Og 1 generator $K^{(ab)}$
of the singlet vector $R^a$ is not modified as it commutes with
$\tilde{Q}$.

Having introduced the operator $\tilde{Q}$, we can write down the
group element of eq. (\ref{4.36}) as
  \begin{equation}
  g  = e^{x \cdot P} e^{y \tilde{Q}}  e^{\Phi_{\rm Og} (x) \tilde{K}^{\rm Og}}
  e^{A_{ab ,\alpha}(x) \tilde{R}^{ab ,\alpha}} e^{A_{a , \alpha} (x) \tilde{R}^{a , \alpha}}
  e^{A_a (x) R^a} e^{\phi(x) R } e^{\phi_i (x) R^i} \quad .
  \label{4.42}
  \end{equation}
Indeed, we now show that using the operator $\tilde Q$ rather than
$Q$ one obtains the field strengths, including the covariant
derivative of the scalars, in a straightforward way. Calculating the
Cartan forms from the group element of eq. (\ref{4.42}) and using
eq. (\ref{4.38}) and the fact that all the positive level operators
commute with $\tilde{Q}$ we find
  \begin{eqnarray}
  g^{-1} d g & = & dx^a [ P_a + (\partial_a A_{bc , \alpha} -
  \partial_a A_{b ,\alpha} A_c - (m_i D^i )_\alpha{}^\beta A_{ab ,
  \beta} A_c + ...) e^{-\phi_i R^i} \tilde{R}^{bc , \alpha} e^{\phi_i
  R^i} \nonumber \\
  &+& (\partial_a A_{b ,\alpha} + (m_i D^i )_\alpha{}^\beta  A_{ab ,
  \beta} +...) e^{-\phi}  e^{-\phi_i R^i} \tilde{R}^{b , \alpha} e^{\phi_i
  R^i} + (\partial_a A_b + ...) e^\phi R^b \nonumber \\
  &+& \partial_a \phi R + A_a  e^{-\phi_i R^i} e^{-\phi R} \tilde{Q}
  e^{\phi R} e^{\phi_i R^i} + e^{-\phi_i R^i} (\partial_a - A_a m_i
  R^i ) e^{\phi_i R^i} ] \nonumber \\
  & + & dy e^{-\phi_i R^i} e^{-\phi R} \tilde{Q}
  e^{\phi R} e^{\phi_i R^i} \quad , \label{4.43}
  \end{eqnarray}
where the dots in each term denote the Og field contributions, whose
role is to cancel the non-antisymmetric terms in the Cartan form
using the inverse Higgs mechanism. As explained above $g^{-1}dg$ is
invariant under $g\to g_0 g$ and so all the coefficients of the
generators in the above equation are invariant. Hence, in particular
the two terms
  \begin{equation}
  dx^a P_a   \quad , \qquad
  dx^a e^{-\phi_i R^i}(\partial_a -A_a m_i R^i)e^{\phi_i R^i}
  \label{4.44}
  \end{equation}
are separately invariant under $g_0$ transformations.  Hence we can
identify the covariant derivative of the scalars as
  \begin{equation}
  e^{-\phi_i R^i}(\partial_a -A_a m_i R^i )e^{\phi_i R^i}
  \label{4.45}
  \end{equation}
which now only transforms under the local transformations. The
infinite number of rigid $g_0$ transformations constitute the gauge
transformations and so this covariant derivative is also covariant
in the conventional sense.

We note that the operator $\tilde Q$ and the variable $y$ although
important for the logic of the result did not appear explicitly in
the calculation of the terms that lead to this covariant derivative.
Indeed one could have  written down the group element without any
$\tilde Q$ or $y$ dependence and the Cartan forms would give the
correct covariant derivatives and so gauge invariant quantities.
Dropping the operator $\tilde{Q}$, one obtains in particular the
commutation relation
  \begin{equation}
  [ R^a , P_b ] =  \delta^a_b m_i R^i \quad
  . \label{4.46}
  \end{equation}
We now consider eq. (\ref{4.46}) as our starting point to define the
nine-dimensional algebra $\tilde{E}_{11,9}^{local}$. This is the
algebra that describes the deformed nine-dimensional theory
considered in this section, and contains the generators $m_i R^i$,
$P_a$ and all the positive level deformed generators, including the
deformed Og generators. In the remaining of this section we will
show that all the results obtained so far can be derived simply
requiring the closure of the Jacobi identities in
$\tilde{E}_{11,9}^{local}$ starting from eq. (\ref{4.46}). This
approach is entirely nine-dimensional, and one never makes use of
the fact that the theory has a ten-dimensional origin. As we will
see, this provides an extremely fast method of deriving the field
strengths of all gauged maximal supergravities.

We start considering the Jacobi identity involving $R^a$,
$\tilde{R}^{a , \alpha}$ and $P_a$. The commutator between $R^a$ and
$\tilde{R}^{b , \alpha}$ is a deformation of the commutator in eq.
(\ref{4.11}), and the most general expression we can write with the
generators at our disposal is
  \begin{equation}
  [ R^a , \tilde{R}^{b , \alpha}] = - \tilde{R}^{ab ,\alpha} + a m_i D^i_\beta{}^\alpha
  \tilde{K}^{(ab), \alpha} \quad
  , \label{4.47}
  \end{equation}
with $a$ to be determined, and where $\tilde{K}^{(ab), \alpha}$ is
the modified Og 1 generator satisfying
  \begin{equation}
  [ \tilde{K}^{(ab), \alpha} , P_c ] = \delta^{(a}_c \tilde{R}^{b),
  \alpha} \quad . \label{4.48}
  \end{equation}
We also demand that the commutator between $\tilde{R}^{ab, \alpha}$
and $P_c$ be of the form
  \begin{equation}
  [  \tilde{R}^{ab ,\alpha}, P_c ] = b (m_i D^i
  )_\beta{}^\alpha  \delta^{[a}_c \tilde{R}^{b ] , \beta} \quad ,
  \label{4.49}
  \end{equation}
with the parameter $b$ to be determined. The Jacobi identity
involving $R^a$, $\tilde{R}^{a , \alpha}$ and $P_a$ is satisfied
provided that the values of $a$ and $b$ are
  \begin{equation}
  a = 1 \qquad \qquad b = -1 \quad . \label{4.50}
  \end{equation}
To summarise, we have obtained the relations
  \begin{eqnarray}
  & & [ R^a , \tilde{R}^{b , \alpha}] = - \tilde{R}^{ab ,\alpha} + m_i D^i_\beta{}^\alpha
  \tilde{K}^{(ab), \alpha} \nonumber \\
  & & [  \tilde{R}^{ab ,\alpha}, P_c ] = - (m_i D^i
  )_\beta{}^\alpha  \delta^{[a}_c \tilde{R}^{b ] , \beta} \quad ,
  \label{4.51}
  \end{eqnarray}
and in particular the second relation coincides with eq.
(\ref{4.38}).

Proceeding this way, one can determine all the commutation relations
of the modified $E_{11}$ generators among themselves and with the
momentum operator $P_a$. For instance, the Jacobi identity involving
the operators $R^a$, $\tilde{R}^{ab , \alpha}$ and $P_a$ requires
the cancellation of terms linear in $m_i$ as well as terms quadratic
in $m_i$. The latter are cancelled by requiring that also the
commutator of $\tilde{K}^{a,bc ,\alpha}$ with $P_d$ receives a
correction at order $m_i$. The result is
  \begin{equation}
  [ R^a , \tilde{R}^{bc ,
  \alpha} ] = {3 \over 2} (m_i D^i
  )_\beta{}^\alpha  \tilde{K}^{a , bc , \beta}
  \label{4.52}
  \end{equation}
and
  \begin{equation}
  [ \tilde{K}^{a,bc ,\alpha} , P_d ] = \delta^a_d \tilde{R}^{bc ,
  \alpha} - \delta^{[a}_d \tilde{R}^{bc ] , \alpha} - { 1 \over 3} (m_i D^i
  )_\beta{}^\alpha ( \delta^b_d \tilde{K}^{ac , \beta} -  \delta^c_d \tilde{K}^{ab ,
  \beta} )
  \quad . \label{4.53}
  \end{equation}
Using the definition of the operator $\tilde{R}^{ab ,\alpha}$ in eq.
(\ref{4.37}), and eq. (\ref{4.14}), one can for instance recover eq.
(\ref{4.52}), that we have obtained requiring the closure of the
Jacobi identities, directly using the ten-dimensional commutation
relations. Indeed, at lowest order in the mass parameter, one gets
  \begin{equation}
  [ R^a , \tilde{R}^{bc ,
  \alpha} ] = (m_i D^i
  )_\beta{}^\alpha  [ K^a{}_y , K^{y,bc , \beta} - K^{[b,c]y ,
  \beta} ] = {3 \over 2}  (m_i D^i
  )_\beta{}^\alpha K^{a ,bc ,\beta} \quad . \label{4.54}
  \end{equation}

In order to show the power of this method, we now determine the
field-strengths for the 3-form and the 4-form of the
nine-dimensional massive theory without using its ten-dimensional
origin. We first write down the relevant commutators of the massless
theory. We thus add to the commutators of eqs. (\ref{4.11}) and
(\ref{4.12}) all the commutators that involve generators up to the
4-form included. We only write down the non-vanishing commutators,
that are
  \begin{eqnarray}
  & & [ R , R^{abc} ] = R^{abc} \qquad \qquad \qquad [ R^a , R^{bcd} ] = - R^{abcd}
  \nonumber \\
  & & [ R^{a , \alpha} , R^{bc , \beta} ] = \epsilon^{\alpha \beta}
  R^{abc} \quad \qquad \ [ R^{ab , \alpha} , R^{cd , \beta} ] = \epsilon^{\alpha \beta}
  R^{abcd} \quad , \label{4.55}
  \end{eqnarray}
where $\epsilon^{\alpha \beta}$ is the invariant antisymmetric
tensor of $SL(2, \mathbb{R})$. Starting from these relations and
using eq. (\ref{4.46}) we can determine all the commutation
relations involving such deformed generators by imposing the closure
of the Jacobi identities. Denoting with $\tilde{R}^{abc}$ and
$\tilde{R}^{abcd}$ the deformed generators, one can show that the
only commutation relation that needs to be modified with respect to
eq. (\ref{4.55}) is the commutator between two deformed 2-form
generators, which becomes
  \begin{equation}
  [ \tilde{R}^{ab ,\alpha} , \tilde{R}^{cd , \beta} ] =
  \epsilon^{\alpha \beta} \tilde{R}^{abcd} + 2 (m_i D^i )^{\alpha
  \beta} \tilde{K}^{[a , b] cd} \quad , \label{4.56}
  \end{equation}
where $\tilde{K}^{a , b cd}$ is the deformed Og 1 generator
associated to the deformed 3-form generator $\tilde{R}^{abc}$,
satisfying
  \begin{equation}
  [ \tilde{K}^{a , b_1 b_2 b_3} , P_c ] = \delta^a_c \tilde{R}^{b_1
  b_2 b_3} - \delta^{ [a}_c \tilde{R}^{b_1 b_2 b_3 ]} \label{4.57}
  \end{equation}
and we have used $\epsilon^{\alpha \beta}$ to raise the $SL(2,
\mathbb{R})$ index, that is
  \begin{equation}
  D^{i \ \alpha \beta} = \epsilon^{\alpha \gamma} D^i_\gamma{}^\beta
  \quad , \label{4.58}
  \end{equation}
and $D^{i \ \alpha \beta}$ is symmetric. The deformed Og 1 generator
for the 4-form is $\tilde{K}^{a , b_1 b_2 b_3}$, satisfying
  \begin{equation}
  [ \tilde{K}^{a , b_1 ...b_4} , P_c ] = \delta^a_c \tilde{R}^{b_1
  ... b_4} - \delta^{ [a}_c \tilde{R}^{b_1 ... b_4 ]} \quad .
  \label{4.59}
  \end{equation}
Both the deformed 3-form and the deformed 4-form commute with the
momentum operator (actually neither the 3-form nor the 4-form
generator are really deformed, but this is not relevant for this
analysis).

We now consider the group element
  \begin{equation}
  g  = e^{x \cdot P} e^{\Phi_{\rm Og}  \tilde{K}^{\rm Og}}
  e^{A_{abcd} \tilde{R}^{abcd}} e^{A_{abc} \tilde{R}^{abc}}
  e^{A_{ab ,\alpha} \tilde{R}^{ab ,\alpha}} e^{A_{a , \alpha}  \tilde{R}^{a , \alpha}}
  e^{A_a  R^a} e^{\phi R } e^{\phi_i  R^i} \quad ,
  \label{4.60}
  \end{equation}
which only depends on the nine-dimensional coordinates $x^a$.
Computing the Maurer-Cartan form and applying the inverse Higgs
mechanism, one can show that all the terms which are not
antisymmetric are set to zero by fixing the Og 1 fields in terms of
the $E_{11}$ fields, and one is left with completely antisymmetric
terms. These are the field-strengths of the 1-forms and 2-forms
given in eq. (\ref{4.28}), as well as the field-strengths
  \begin{eqnarray}
  & & F_{a_1 \dots a_4} = \partial_{[a_1} A_{a_2 a_3 a_4 ]} +
  \epsilon^{\alpha \beta} \partial_{[a_1} A_{a_2 a_3 , \alpha}
  A_{a_4 ] , \beta} - {1 \over 2} (m_i D^i )^{ \alpha \beta} A_{[
  a_1 a_2 , \alpha} A_{a_3 a_4 ] , \beta} \nonumber \\
  & & F_{a_1 \dots a_5} = \partial_{[a_1} A_{a_2 ... a_5 ]} -
  \partial_{[a_1} A_{a_2 a_3 a_4} A_{a_5 ]}+ {1 \over 2} \epsilon^{\alpha \beta}
  \partial_{[a_1} A_{a_2 a_3 , \alpha } A_{a_4 a_5 ] , \beta} \nonumber \\
  & & \qquad \quad -
  \epsilon^{\alpha \beta} \partial_{[a_1} A_{a_2 a_3 , \alpha}
  A_{a_4  , \beta} A_{a_5 ]} + {1 \over 2} (m_i D^i )^{ \alpha \beta} A_{[
  a_1 a_2 , \alpha} A_{a_3 a_4  , \beta} A_{a_5 ]} \label{4.61}
  \end{eqnarray}
for the 3-form and the 4-form. These are indeed the field-strengths
of the 3-form and its dual 4-form of the massive nine-dimensional
supergravity. The gauge transformations of the fields arise in the
non-linear realisation as rigid transformations of the group element
with the Og generators included. One obtains the transformations of
the 2-forms and 1-forms given in eq. (\ref{4.29}) as well as the
transformations
  \begin{eqnarray}
  & & \delta A_{abc} = \partial_{[a} \Lambda_{bd ]} + \epsilon^{\alpha
  \beta} \partial_{[a} \Lambda_{\alpha} A_{bc ] , \beta} + {1 \over
  2} \epsilon^{\alpha \beta} \partial_{[a} \Lambda A_{b , \alpha}
  A_{c] , \beta} + (m_i D^i )^{ \alpha \beta} \Lambda_{[a , \alpha}
  A_{bc ], \beta} \nonumber \\
  & & \delta A_{a_1 ... a_4} = \partial_{[a_1} \Lambda_{a_2 a_3 a_4 ]}
  + {1 \over 2}\epsilon^{\alpha
  \beta} \partial_{[a_1} \Lambda_{a_2 , \alpha} A_{a_3 a_4 ] , \beta}
  - \partial_{[a_1} \Lambda A_{a_2 a_3 a_4 ]} \nonumber \\
  & & \qquad \qquad
  + {1 \over
  2} \epsilon^{\alpha \beta} \partial_{[a_1} \Lambda A_{a_2 , \alpha}
  A_{a_3 a_4 ] , \beta} \label{4.62}
  \end{eqnarray}
of the 3-form and the 4-form.

Observe that although the operators $R^i$ other than $m_i R^i$ do
not belong to the algebra $\tilde{E}_{11,9}^{local}$, one can
nonetheless use the group element of eq. (\ref{4.60}). Indeed, the
covariant derivative for the scalars is also obtained from the
nine-dimensional group element of eq. (\ref{4.60}). Indeed, the
Maurer-Cartan form contains the terms
  \begin{equation}
  e^{-\phi_i R^i }
  \partial_\mu e^{\phi_i R^i }
  - A_\mu e^{-\phi_i R^i } m_i R^i  e^{\phi_i R^i }  \quad , \label{4.63}
  \end{equation}
which we recognise as the covariant derivative of the scalars.

To summarise, we have found a general pattern for carrying out
dimensional reduction to obtain gauged supergravities. The higher
dimensional coordinates have a generator ($Q$ in the massless case
and $\tilde{Q}$ in the Scherk-Schwarz case) which is associated with
the space being reduced on. From the set on $E_{11}$ and Og
generators, we can find a set of deformed $E_{11}$ generators which
are just those that commute with the preferred generator associated
with the reduction ($Q$ or $\tilde{Q}$). The field strengths can
then just be deduced from this deformed $E_{11}$ algebra. We will
see that this method transcends dimensional reduction and in fact
applies to all gauged maximal supergravities. This will be the focus
on the next two sections.

\section{$E_{11}$ and massive IIA}
In the last section we have analysed the massless and Scherk-Schwarz
reductions to nine dimensions of ten-dimensional IIB supergravity
from an $E_{11}$ perspective. Starting from the algebra
$E_{11,10B}^{local}$ of $E_{11}$ plus the Og generators that encodes
all the gauge symmetries of the ten-dimensional theory, the massless
dimensional reduction corresponds to taking the $E_{11}$ generators
together with the subset of Og generators that commute with the
momentum operator in the internal direction. On the other hand, the
Scherk-Schwarz dimensional reduction corresponds to choosing
operators that commute with a twisted internal momentum operator,
and the twist is such that these operators are combinations of the
ten-dimensional $E_{11}$ and Og generators. It is important to
stress that the content of the sets of generators in the massless
and the Scherk-Schwarz theory are exactly the same, and the two
theories differ because the commutation relations are different. In
particular, the set of non-negative level $E_{11}$ generators in the
massless theory is the same as the set of operators in the
Scherk-Schwarz reduction case that are obtained by adding to the
$E_{11}$ generators suitable Og  generators of the ten-dimensional
theory multiplied by powers of the mass deformation parameter $m_i$,
where $i$ is an $SL(2, \mathbb{R})$ triplet index. From the
nine-dimensional perspective, these operators look like $E_{11}$
generators, but their commutation relation receives a correction at
order $m_i$. Therefore, the algebra appears from the
nine-dimensional perspective as a deformation of the original
$E_{11}$ algebra. This deformation is such that the commutator of
two positive level generators gives the standard $E_{11}$ result at
zero order in $m_i$ together with an order $m_i$ deformation
proportional to the Og generators of the nine-dimensional theory.
Correspondingly, the commutator of the deformed positive level
$E_{11}$ generators with the nine-dimensional momentum is
proportional to the deformed $E_{11}$ generators times the mass
parameter. Starting from the commutation relation (\ref{4.46}), the
entire algebra of the nine-dimensional deformed theory can be
determined by requiring that the Jacobi identities close.

In this section we consider the case of the massive deformation of
the IIA theory, discovered by Romans in \cite{8}. In this case the
theory does not arise as a dimensional reduction of
eleven-dimensional supergravity, and therefore one cannot deform the
$E_{11}$ generators adding eleven-dimensional Og generators.
Nonetheless, we will show that from the ten-dimensional perspective
one can still consider deformed $E_{11}$ and Og generators, and the
corresponding algebra $\tilde{E}_{11,10A}^{local}$, which appears as
a deformation of the massless ten-dimensional algebra, determines
all the field-strengths of the theory. In \cite{11} it was shown
that the massive IIA theory can be recovered from an $E_{11}$
perspective by adopting a non-trivial commutation relation between
the momentum operator and the positive level generators. In
particular the commutator of the 2-form generator with momentum
gives the 1-form generator multiplied by the mass deformation
parameter. The resulting algebra \cite{11} though has a problem of
consistency because the corresponding Jacobi identities do not
close. In \cite{28} it was shown that if one insists on requiring
the consistency of the algebra for the lower-rank forms, the
commutator of two 2-forms cannot vanish in the massive theory, but
instead is proportional to an operator in the (3,1) representation
of $GL(10, \mathbb{R})$. This operator is indeed the Og 1 operator
for the 3-form. We show that the whole algebra corresponding to the
massive IIA theory is determined starting from the deformed
commutation relation of the 2-form with momentum and requiring the
closure of all the Jacobi identities. A different approach, based on
the Kac-Moody algebra $E_{10}$ \cite{29}, has recently been given in
\cite{30}.

We start by writing down the algebra associated to the massless IIA
theory. The massless IIA theory arises from the dimensional
reduction of eleven-dimensional supergravity. The corresponding
algebra arises from a decomposition of the $E_{11}$ algebra in terms
of $GL(10,\mathbb{R})$ as relevant for the IIA theory, which
corresponds to deleting nodes 10 and 11 in the Dynkin diagram in
fig. \ref{fig1}. In this section we denote with $a,b,...$ the
tangent spacetime indices in ten dimensions. In deriving the
$E_{11}$ generators in terms of their $GL(10,\mathbb{R})$ IIA
representations it is useful to consider the eleven-dimensional
generators and denote with $y$ the internal 11th coordinate. One
then obtains that the theory contains a scalar $R$, which is the
$GL(11,\mathbb{R})$ generator $K^y{}_y$, a vector $R^a$
corresponding to the eleven-dimensional $K^a{}_y$, a 2-form $R^{ab}$
which arises from the eleven-dimensional 3-form with one index in
the internal direction $R^{aby}$, and then a 3-form $R^{abc}$, a
5-form $R^{a_1 \dots a_5}$, a 6-form $R^{a_1 \dots a_6}$, a 7-form
$R^{a_1 \dots a_7}$, an 8-form $R^{a_1 \dots a_8}$, a 9-form $R^{a_1
\dots a_9}$ and two 10-forms $R^{a_1 \dots a_{10}}$ and $R^{\prime
a_1 \dots a_{10}}$, together with an infinite set of generators with
mixed, {\it i.e.} not completely antisymmetric, indices \cite{12}.

We now write down the part of the $E_{11}$ algebra that involves
these completely antisymmetric generators. This was first derived in
\cite{11}, but we use different normalisations for the generators,
that make the eleven-dimensional origin of the algebra more
transparent. For simplicity, we will neglect the contribution from
the 10-form generators, that is we will consider a level truncation
only involving generators up to the 9-form included. The algebra is
  \begin{eqnarray}
  & & [ R, R^a ] = - R^a \qquad \qquad \quad  \ \ \  [ R , R^{ab} ] = R^{ab}\nonumber \\
  & &   [ R ,
  R^{a_1 \dots a_5} ] = R^{a_1 \dots a_5} \qquad \  \quad [ R ,  R^{a_1 \dots a_7} ]
  = 2  R^{a_1 \dots a_7} \nonumber \\
  & & [ R , R^{a_1 \dots
  a_8} ] = R^{a_1 \dots a_8} \qquad \quad  \  [ R, R^{a_1 \dots a_9} ] = 3 R^{a_1 \dots
  a_9} \nonumber \\
  & & [ R^a , R^{bc} ] = R^{abc}  \qquad \qquad \quad  \ \   [ R^{a_1} , R^{a_2 \dots
  a_6} ] = - R^{a_1 \dots a_6} \nonumber \\
  & &  [ R^{a_1} , R^{a_2 \dots
  a_8} ] = 3 R^{a_1 \dots a_8} \qquad \  [ R^{a_1 a_2} , R^{a_3 \dots
  a_5} ] = - 2R^{a_1 \dots a_5} \nonumber \\
  & &   [ R^{a_1 a_2} , R^{a_3 \dots
  a_7} ] = R^{a_1 \dots a_7} \qquad [ R^{a_1 a_2} , R^{a_3 \dots a_8} ] = - 2 R^{a_1 \dots a_8} \nonumber
  \\
  & &  [ R^{a_1 a_2} , R^{a_3 \dots a_9} ] =  R^{a_1 \dots a_9} \qquad
 [ R^{a_1 \dots a_3} , R^{a_4 \dots
  a_6} ] = 2 R^{a_1 \dots a_6}\nonumber \\
  & &  [ R^{a_1 \dots a_3} , R^{a_4 \dots
  a_8} ] = R^{a_1 \dots a_8} \quad , \label{5.1}
  \end{eqnarray}
with all the other commutators vanishing or giving generators with
mixed symmetries. One can show that all the Jacobi identities
involving these operators are satisfied. Following the results of
section 4, one can then obtain the Og generators of the
ten-dimensional theory by decomposing the Og generators of the
eleven-dimensional theory in terms of representations of
$GL(10,\mathbb{R})$. The subset of such generators that commute with
the momentum operator along the 11th direction are the Og generators
of the massless IIA theory.  These are exactly the operators that
are needed to encode all the gauge symmetries of the fields of the
massless IIA theory. This corresponds to the fact that the massless
IIA theory arises as a circle dimensional reduction of
eleven-dimensional supergravity. In particular, for each $n$-form
$E_{11}$ generator the corresponding Og 1 operator is $K^{a, b_1
\dots b_n}$ satisfying $K^{[ a, b_1 \dots b_n ]}=0$.

We now consider the deformation of the massless IIA algebra giving
rise to massive IIA. This theory was constructed by Romans in
\cite{8}, and it corresponds to a Higgs mechanism in which the
2-form acquires a mass by absorbing the vector. In \cite{11} this
mechanism was recovered from an $E_{11}$ perspective by adopting a
non-trivial commutation relation between the 2-form generator and
momentum. Following the results of section 4, we interpret this as a
redefinition of the $E_{11}$ generators. We thus denote all the
generators of the massive theory with a tilde. These generators,
although forming a set identical to the one corresponding to the
$E_{11}$ generators of the massless theory, have different
commutation relations. These commutation relations make the
corresponding algebra look like a deformation of the massless
algebra involving the mass parameter. We thus write down the
commutation relation between the 2-form and momentum as
  \begin{equation}
  [ \tilde{R}^{ab} , P_c ] = - m \delta^{[a}_c \tilde{R}^{b]} \quad ,
  \label{5.2}
  \end{equation}
where $m$ is the Romans mass parameter. Our strategy is to use eq.
(\ref{5.2}) as our starting point, and to derive all the commutation
relations of the deformed theory from it imposing the closure of the
Jacobi identities. We will show that this will fix all the
field-strengths and gauge transformations of the forms in the
theory. In \cite{31} it was shown that the supersymmetry algebra of
IIA closes on all the forms predicted by $E_{11}$, and the
field-strengths and gauge transformations of the form fields were
derived imposing the closure of the supersymmetry algebra. We will
show that the field-strengths and gauge transformations as obtained
using supersymmetry exactly coincide, up to field redefinitions,
with the ones obtained here from $E_{11}$.

The Jacobi identity involving the operators $\tilde{R}$,
$\tilde{R}^{ab}$ and $P_a$ imposes that
  \begin{equation}
  [ \tilde{R} , P_a ] =-2 P_a \quad . \label{5.3}
  \end{equation}
Introducing the Og 1 operator for the deformed 3-form, defined as
  \begin{equation}
  [ \tilde{K}^{a , b_1 b_2 b_3} , P_c ] = \delta^a_c \tilde{R}^{b_1 b_2 b_3}
  - \delta^{[a}_c \tilde{R}^{b_1 b_2 b_3 ]} \label{5.4}
  \end{equation}
one can then show that the Jacobi identity between two 2-forms and
momentum imposes \cite{28}
  \begin{equation}
  [ \tilde{R}^{ab} , \tilde{R}^{cd} ] = - 2m \tilde{K}^{[a,b]cd}
  \label{5.5}
  \quad .
  \end{equation}
One can then show that the Jacobi identities involving the scalar
operator $\tilde{R}$ require a non-trivial commutation relation
between the deformed 7-form generator and the momentum operator,
while the commutator between the 2-form and the 5-form generator has
to be modified by a term proportional to the Og 1 $\tilde{K}^{a ,
b_1 \dots b_6}$ the 6-form, which satisfies
  \begin{equation}
  [ \tilde{K}^{a , b_1 \dots b_6} , P_c ] = \delta^a_c \tilde{R}^{b_1 \dots b_6}
  - \delta^{[a}_c \tilde{R}^{b_1 \dots  b_6 ]} \quad . \label{5.6}
  \end{equation}
The result is
  \begin{equation}
  [ \tilde{R}^{a_1 \dots a_7} , P_b ] =  m \delta^{[a_1}_b \tilde{R}^{a_2 \dots a_7
  ]} \label{5.7}
  \end{equation}
and
  \begin{equation}
  [ \tilde{R}^{a_1 a_2} , \tilde{R}^{b_1\dots b_5} ] = \tilde{R}^{a_1 a_2 b_1 \dots b_5}
  +  m \tilde{K}^{[ a_1 , a_2 ] b_1 \dots b_5}
  \label{5.8}
  \quad .
  \end{equation}
Finally, the Jacobi identities also impose that the commutator
between the 9-form and momentum, as well as the commutator between
the 2-form and the 7-form, must be modified. The result is
  \begin{equation}
  [ \tilde{R}^{a_1 \dots a_9} , P_b ] = - 5 m \delta^{[a_1}_b \tilde{R}^{a_2 \dots a_9
  ]} \label{5.9}
  \end{equation}
and
  \begin{equation}
  [ \tilde{R}^{a_1 a_2} , \tilde{R}^{b_1\dots b_7} ] = \tilde{R}^{a_1 a_2 b_1 \dots b_7}
  - {17 \over 7}
  m \tilde{K}^{[ a_1 , a_2 ] b_1 \dots b_7}
  \label{5.10}
  \quad ,
  \end{equation}
where $\tilde{K}^{a, b_1 \dots b_8}$ is the Og 1 operator for the
8-form, satisfying
  \begin{equation}
  [ \tilde{K}^{a , b_1 \dots b_8} , P_c ] = \delta^a_c \tilde{R}^{b_1 \dots b_8}
  - \delta^{[a}_c \tilde{R}^{b_1 \dots  b_8 ]} \quad . \label{5.11}
  \end{equation}
All the other commutators are not modified, and they are as in eq.
(\ref{5.1}) with all operators replaced by deformed operators.

To summarise, we have shown that starting from the $E_{11}$ algebra
of eq. (\ref{5.1}) and introducing the deformed 2-form generator
which satisfies the commutation relation of eq. (\ref{5.2}), the
Jacobi identities determine completely the rest of the algebra. In
particular, once the algebra is expressed in terms of the tilde
generators, the only commutators that are modified with respect to
eq. (\ref{5.1}) are those of eqs. (\ref{5.5}), (\ref{5.8}) and
(\ref{5.10}), while the additional non-trivial commutation relations
with $P_a$ are given is eqs. (\ref{5.3}), (\ref{5.7}) and
(\ref{5.9}).

We now consider the group element
  \begin{equation}
  g  = e^{x \cdot P} e^{\Phi_{\rm Og} \tilde{K}^{\rm Og}}
  e^{A_{a_1 \dots a_9} \tilde{R}^{a_1 \dots a_9}} \dots
  e^{A_{a_1 \dots a_5} \tilde{R}^{a_1 \dots a_5}}
  e^{A_{a_1 \dots a_3} \tilde{R}^{a_1 \dots a_3}}  e^{A_{a_1 a_2} \tilde{R}^{a_1
  a_2}}
  e^{A_{a} \tilde{R}^{a}} e^{\phi \tilde{R}}
  \label{5.12}
  \end{equation}
where we denote with $\Phi_{\rm Og}$ the whole set of Og 1 field of
the ten-dimensional massive IIA theory. Similarly we denote with
$\tilde{K}^{\rm Og}$ the whole set of deformed ten-dimensional Og 1
operators, which we treat as commuting because we are ignoring the
contribution of Og 2 generators for simplicity. One can compute the
Maurer-Cartan form that results from the group element in eq.
(\ref{5.12}). The result is
  \begin{eqnarray}
  g^{-1} \partial_\mu g & = & e^{2\phi} P_\mu + \partial_\mu \phi \tilde{R} + (\partial_\mu A_a +
  m A_{\mu a} + ...) e^{\phi} \tilde{R}^a + (\partial_\mu A_{a_1 a_2} + ...)
  e^{- \phi}
  \tilde{R}^{a_1 a_2}\nonumber \\
  & + & (\partial_\mu A_{a_1 a_2 a_3} - \partial_\mu A_{a_1
  a_2} A_{a_3} + {m \over 2} A_{\mu a_1} A_{a_2 a_3} +...)  \tilde{R}^{a_1
  ... a_3} \nonumber \\
  & +& (\partial_\mu A_{a_1 ...a_5} +2  \partial_\mu A_{a_1
  a_2 a_3} A_{a_4 a_5} + {m \over 3} A_{\mu a_1} A_{a_2 a_3} A_{a_4 a_5}+...)
  e^{- \phi} \tilde{R}^{a_1
  ... a_5} \nonumber \\
  & + & (\partial_\mu A_{a_1 ...a_6} -  \partial_\mu A_{a_1 ...a_5}
  A_{a_6} -2 \partial_\mu A_{a_1 a_2 a_3} A_{a_4 a_5} A_{a_6} +
  \partial_\mu A_{a_1 a_2 a_3 } A_{a_4 a_5 a_6}\nonumber \\
  & -& {m \over 3} A_{\mu a_1} A_{a_2 a_3} A_{a_4 a_5} A_{a_6} - m A_{\mu a_1 ...a_6} +...)
  \tilde{R}^{a_1
  ... a_6} \nonumber \\
  & +& (\partial_\mu A_{a_1 ...a_7} -  \partial_\mu A_{a_1 ...a_5} A_{a_6
  a_7} - \partial_\mu A_{a_1 a_2 a_3} A_{a_4 a_5 } A_{a_6 a_7}
  - {m \over 12} A_{\mu a_1} A_{a_2 a_3} A_{a_4 a_5} A_{a_6 a_7} \nonumber \\
  &+& ...)
  e^{- 2 \phi} \tilde{R}^{a_1
  ... a_7} + ( \partial_\mu A_{a_1 ...a_8} + 3 \partial_\mu A_{a_1
  ...a_7} A_{a_8} +2 \partial_\mu A_{a_1 ...a_6} A_{a_7 a_8} \nonumber \\
  &+&
  \partial_\mu A_{a_1 ...a_5} A_{a_6 ...a_8} -3 \partial_\mu A_{a_1
  ...a_5} A_{a_6 a_7} A_{a_8} -3 \partial_\mu A_{a_1 a_2 a_3} A_{a_4
  a_5 }A_{a_6 a_7} A_{ a_8} \nonumber \\
  &+& 2 \partial_\mu A_{a_1 ..a_3} A_{a_4 ...a_6} A_{a_7 a_8} -{m
  \over 4} A_{\mu a_1} A_{a_2 a_3} A_{a_4 a_5} A_{a_6 a_7} A_{a_8} -
  2m A_{\mu a_1 ...a_6} A_{a_7 a_8} \nonumber \\
  &+& 5m A_{\mu a_1 ...a_8}+ ...) e^{-  \phi} \tilde{R}^{a_1 ... a_8} + (
  \partial_\mu A_{a_1 ...a_9} - \partial_\mu A_{a_1 ...a_7} A_{a_8
  a_9}\nonumber \\
  & + &{1 \over 2} \partial_\mu A_{a_1 ...a_5} A_{a_6 a_7} A_{a_8
  a_9} + {1 \over 3} \partial_\mu A_{a_1 a_2 a_3} A_{a_4 a_5} A_{a_6
  a_7} A_{a_8 a_9} \nonumber \\
  &+& {m \over 60} A_{\mu a_1 } A_{a_2 a_3 } A_{a_4 a_5} A_{a_6 a_7}
  A_{a_8 a_9} +...) e^{- 3 \phi} \tilde{R}^{a_1 ...a_9} + ... \quad .
  \label{5.13}
  \end{eqnarray}
The dots at the end denote terms proportional to the higher level
deformed $E_{11}$ generators as well as all the Og generators, while
the dots in each bracket denote the contributions from the Og 1
fields, which we did not write down explicitly because their
contribution vanishes after antisymmetrisation of the $\mu$ index
with the other indices. Indeed, the Og 1 fields in the group element
of eq. (\ref{5.12}) are $\Phi_{a, b_1 \dots b_n}$, for
$n=1,2,3,5,...$, satisfying $\Phi_{[ a, b_1 \dots b_n ]}=0$. The
inverse Higgs mechanism relates these Og fields to the deformed
$E_{11}$ fields in such a way that only the completely antisymmetric
terms in (\ref{5.13}) survive. These terms are
   \begin{eqnarray}
   & & F_{a_1 a_2}  =   \partial_{[a_1} A_{a_2]} +
   m A_{a_1 a_2} \nonumber \\
   & & F_{a_1 a_2 a_3}  = \partial_{[a_1} A_{a_2 a_3 ] }\nonumber \\
   & & F_{a_1 ... a_4}  =    \partial_{[a_1} A_{a_2 a_3 a_4 ]} - \partial_{[a_1}
   A_{a_2
   a_3} A_{a_4 ]} + {m \over 2} A_{[ a_1 a_2} A_{a_3 a_4 ]} \nonumber
   \\
   & & F_{a_1 ... a_6}  =  \partial_{[a_1} A_{a_2 ...a_6 ]} +2  \partial_{[a_1}
   A_{a_2
   a_3 a_4} A_{a_5 a_6 ]} + {m \over 3} A_{[ a_1 a_2} A_{a_3 a_4} A_{a_5 a_6
   ]} \nonumber \\
   & & F_{a_1 ... a_7}  =
  \partial_{[ a_1} A_{a_2 ...a_7 ]} -  \partial_{[ a_1} A_{a_2 ...a_6} A_{a_7 ]}
  -2 \partial_{[ a_1}  A_{a_2 a_3 a_4} A_{a_5 a_6} A_{a_7 ]} +
  \partial_{[a_1} A_{a_2 a_3 a_4 } A_{a_5 a_6 a_7 ]}\nonumber \\
  & & \qquad \quad -{m \over 3} A_{[ a_1 a_2} A_{a_3 a_4} A_{a_5 a_6} A_{a_7 ]} - m A_{a_1
  ...a_7} \nonumber \\
  & & F_{a_1 ...a_8} = \partial_{[a_1} A_{a_2 ...a_8 ]} -
  \partial_{[ a_1 } A_{a_2 ...a_6} A_{a_7 a_8 ]}
  - \partial_{[a_1} A_{a_2 a_3 a_4} A_{a_5 a_6 } A_{a_7 a_8
  ]}\nonumber \\
  & & \qquad \quad
  - {m \over 12} A_{[ a_1 a_2} A_{a_3 a_4} A_{a_5 a_6} A_{a_7 a_8 ]}
  \nonumber \\
  & & F_{a_1 ...a_9} = \partial_{[a_1} A_{a_2 ...a_9 ]} +
  3 \partial_{[a_1} A_{a_2
  ...a_8} A_{a_9 ]} +2 \partial_{[a_1} A_{a_2 ...a_7} A_{a_8 a_9 ]}
  +
  \partial_{[a_1} A_{a_2 ...a_6} A_{a_7 ...a_9 ]} \nonumber \\
  & & \qquad \quad  -3 \partial_{[a_1}
  A_{a_2
  ...a_6} A_{a_7 a_8} A_{a_9 ]} -3 \partial_{[a_1} A_{a_2 a_3 a_4}
  A_{a_5
  a_6 }A_{a_7 a_8} A_{ a_9 ]}
  + 2 \partial_{[a_1} A_{a_2 ..a_4} A_{a_5 ...a_7} A_{a_8 a_9 ]}\nonumber \\
  & & \qquad \quad  -{m
  \over 4} A_{[ a_1 a_2} A_{a_3 a_4} A_{a_5 a_6} A_{a_7 a_8} A_{a_9 ]} -
  2m A_{[ a_1 ...a_7} A_{a_8 a_9 ]}
  + 5m A_{a_1 ...a_9} \nonumber \\
  & & F_{a_1 ... a_{10}} =
  \partial_{[a_1} A_{a_2 ...a_{10}]} - \partial_{[a_1} A_{a_2 ...a_8}
  A_{a_9
  a_{10}]}
  + {1 \over 2} \partial_{[a_1} A_{a_2 ...a_6} A_{a_7 a_8} A_{a_9
  a_{10}]} \nonumber \\
  & & \qquad \quad + {1 \over 3} \partial_{[a_1} A_{a_2 a_3 a_4} A_{a_5 a_6}
  A_{a_7
  a_8} A_{a_9 a_{10}]}
  + {m \over 60} A_{[ a_1 a_2} A_{a_3 a_4 } A_{a_5 a_6} A_{a_7 a_8}
  A_{a_9 a_{10}]} \quad . \label{5.14}
  \end{eqnarray}
These are the field-strengths of the fields of the massive IIA
theory. Out of these field-strengths one can construct the field
equations, which are duality relations between the various
field-strengths. In particular, the 2-form $F_{a_1 a_2}$  is dual to
the 8-form $F_{a_1 ... a_8}$, the 3-form $F_{a_1 a_2 a_3}$ is dual
to the 7-form $F_{a_1 ... a_7}$ and the 4-form $F_{a_1 ... a_4}$ is
dual to the 6-form $F_{a_1 ... a_6}$, while the 9-form $F_{a_1 ...
a_9}$ is dual to the derivative of the scalar and the 10-form
$F_{a_1 ... a_{10}}$ is dual to the mass parameter $m$. All these
relations are covariant under the local subalgebra of the non-linear
realisation, which is $SO(9,1)$.

The gauge transformations of the fields arise in the non-linear
realisation as rigid transformations of the group element, $g
\rightarrow g_0 g$, as long as one includes the Og generators. One
obtains
   \begin{eqnarray}
   & & \delta A_{a} = \partial_a \Lambda - m \Lambda_{a} \nonumber
   \\
   & & \delta A_{a_1 a_2} = \partial_{[a_1} \Lambda_{a_2 ]}
   \nonumber \\
   & & \delta A_{a_1 a_2 a_3} = \partial_{[a_1} \Lambda_{a_2 a_3 ]}
   + \partial_{[ a_1} \Lambda A_{a_2 a_3 ]} - m \Lambda_{[a_1} A_{a_2 a_3
   ]} \nonumber \\
   & & \delta A_{a_1 ... a_5} = \partial_{[a_1} \Lambda_{a_2 ... a_5
   ]} - \partial_{[a_1} \Lambda A_{a_2 a_3 } A_{a_4 a_5 ]} -2
   \partial_{[a_1} \Lambda_{a_2} A_{a_3 a_4 a_5 ]} + m \Lambda_{[a_1}
   A_{a_2 a_3 } A_{a_4 a_5 ]} \nonumber \\
   & & \delta A_{a_1 ... a_6} = \partial_{[a_1} \Lambda_{a_2 ... a_6
   ]} + \partial_{[a_1} \Lambda_{a_2 a_3} A_{a_4 a_5 a_6 ]} -
   \partial_{[a_1} \Lambda A_{a_2 ...a_6] } - \partial_{[a_1}
   \Lambda A_{a_2 a_3} A_{a_4 a_5 a_6 ]} \nonumber \\
   & & \qquad \quad + m \Lambda_{a_1 ... a_6} + m \Lambda_{[a_1} A_{a_2
   ... a_6 ]} + m \Lambda_{[a_1} A_{a_2 a_3} A_{a_4 a_5 a_6 ]}
   \nonumber \\
   & & \delta A_{a_1 ... a_7} = \partial_{[a_1} \Lambda_{a_2 ... a_7
   ]} - {1 \over 3} \partial_{[a_1} \Lambda A_{a_2 a_3} A_{a_4 a_5}
   A_{a_6 a_7 ]} + \partial_{[a_1} \Lambda_{a_2} A_{a_3 ... a_7 ]} \nonumber \\
   & & \qquad \quad +
   {1 \over 3} m  \Lambda_{[a_1} A_{a_2 a_3} A_{a_4 a_5}
   A_{a_6 a_7 ]}\nonumber \\
   & & \delta A_{a_1 ... a_8} = \partial_{[a_1} \Lambda_{a_2 ... a_8
   ]} + \partial_{[a_1} \Lambda_{a_2 a_3} A_{a_4 ...a_8]} + 3 \partial_{[a_1}
   \Lambda A_{a_2 ...a_8 ]} + \partial_{[a_1} \Lambda A_{a_2 a_3 a_4}
   A_{a_5 a_6} A_{a_7 a_8 ]} \nonumber \\
   & & \qquad \quad - 2 \partial_{[a_1} \Lambda_{a_2} A_{a_3 ...
   a_8 ]} - 5m \Lambda_{a_1 ... a_8} - 3 m \Lambda_{[a_1}
   A_{a_2 ...a_8 ]} - m \Lambda_{[a_1} A_{a_2 a_3 a_4}
   A_{a_5 a_6} A_{a_7 a_8 ]}\nonumber \\
   & & \delta A_{a_1 ... a_9} = \partial_{[a_1} \Lambda_{a_2 ... a_9
   ]} - {1 \over 12} \partial_{[a_1} \Lambda A_{a_2 a_3} A_{a_4 a_5}
   A_{a_6 a_7 } A_{a_8 a_9]}+ \partial_{[a_1} \Lambda_{a_2} A_{a_3 ... a_9 ]} \nonumber \\
   & & \qquad \quad +
   {1 \over 12} m  \Lambda_{[a_1} A_{a_2 a_3} A_{a_4 a_5}
   A_{a_6 a_7 }A_{a_8 a_9 ]}\quad . \label{5.15}
   \end{eqnarray}

In \cite{31} the supersymmetry transformations of all the forms and
dual forms of the massive IIA theory where determined. The
supersymmetry algebra closes on all the local symmetries of the
theory, and this was used to determine all the gauge transformations
and the field-strengths of the various forms, as well as their
duality relations. These forms are exactly those predicted by
$E_{11}$. One can show that the field strengths and gauge
transformations of \cite{31} coincide with those given in eqs.
(\ref{5.14}) and (\ref{5.15}) up to field redefinitions. The fact
that using simple algebraic techniques one can easily determine
these quantities proves the power of the $E_{11}$ formulation of
maximal supergravities and of the methods explained in this paper.
In the next section we will apply these methods to the case of
maximal gauged supergravity in five dimensions, deriving again the
results of \cite{20}.

\section{$E_{11}$ and gauged five-dimensional supergravity\label{D=5section}}
In section 4 we derived the algebra $\tilde{E}_{11,9}^{local}$
associated to the Scherk-Schwarz dimensional reduction of the IIB
theory to nine dimensions. After deriving the field-strengths of the
theory from a ten-dimensional group element with a given dependence
on the 10th coordinate $y$, we have shown that the same results can
be obtained directly in nine dimensions. Indeed from the
nine-dimensional perspective the fact that the ten-dimensional group
element has a non-trivial $y$ dependence translates in having
generators of the nine-dimensional theory that are deformed with
respect to the massless case. We have shown that the algebra of
these deformed generators is uniquely fixed by the Jacobi
identities, and in deriving the deformed algebra in this way one
never makes use of the fact that the theory has a ten-dimensional
origin. This approach was indeed taken in the previous section,
where we derived the algebra $\tilde{E}_{11,10A}^{local}$
corresponding to the massive IIA theory by requiring the closure of
the Jacobi identities. From this algebra we have then derived the
field-strengths of all the forms in the theory.

In this section we will perform precisely the same analysis for the
case of maximal gauged supergravity in five dimensions. We will
derive the algebra $\tilde{E}_{11,5}^{local}$ and from it we will
determine the field-strengths of the forms in the theory. The
analysis follows exactly the same steps as we have shown in the
previous section for the case of the massive IIA theory in ten
dimensions. We will first review the $E_{11}$ algebra as decomposed
with respect to its $GL(5, \mathbb{R}) \otimes E_6$ subalgebra
\cite{14} which is relevant for the five-dimensional analysis. This
corresponds to deleting node 5 in the Dynkin diagram of fig.
\ref{fig1}. We will then consider the algebra of the deformed
generators which occur in the description of the gauged theory from
the $E_{11}$ perspective. The commutation relations of these
generators are completely fixed by imposing Jacobi identities. The
resulting algebra is such that the non-linear realisation determines
completely all the field-strengths of gauged maximal supergravity in
five dimensions. A different approach to gauge supergravities, based
on $E_{10}$, was presented in \cite{3dime10} for the
three-dimensional case.

We now review the $E_{11}$ commutation relations of the form
generators up to the 4-form included that occur in the decomposition
of $E_{11}$ with respect to $GL(5, \mathbb{R}) \otimes E_6$
\cite{14}. These generators are
  \begin{equation}
  R^\alpha  \qquad R^{a , M} \qquad R^{ab}{}_M \qquad R^{abc , \alpha}  \qquad  R^{abcd}{}_{[MN]}
  \quad , \label{6.1}
  \end{equation}
where  $R^\alpha$, $\alpha = 1 ,\dots ,78$ are the $E_6$ generators,
and an upstairs $M$ index, $M= 1 ,\dots , 27$, corresponds to the
${\bf \overline{27}}$ representation of $E_6$, a downstairs  $M$
index to the ${\bf 27}$ of $E_6$ and a pair of antisymmetric
downstairs indices $[MN]$ correspond to the ${\bf  \overline{351}}$.
The commutation relations for the $E_6$ generators is
  \begin{equation}
  [R^\alpha , R^\beta ]= f^{\alpha\beta}{}_{\gamma} R^\gamma \quad ,
  \label{6.2}
  \end{equation}
where $f^{\alpha\beta}{}_{\gamma}$ are the structure constants of
$E_6$. The commutation relations of $R^\alpha$ with all the other
generators is determined by the $E_6$ representations that they
carry. This gives
  \begin{eqnarray}
  & & [R^\alpha , R^{a,M} ]= (D^\alpha )_N{}^M R^{a, N} \nonumber \\
  & & [R^\alpha , R^{ab}{}_M ]= -(D^\alpha )_M{}^N R^{ab}{}_N
  \nonumber \\
  & &  [R^\alpha , R^{abc ,\beta} ]= f^{\alpha\beta}{}_{\gamma}
  R^{abc,\gamma} \nonumber \\
  & & [R^\alpha , R^{abcd}{}_{[MN]} ]= -(D^\alpha )_M{}^P
  R^{abcd}{}_{[PN]}-(D^\alpha )_N{}^P  R^{abcd}{}_{[MP]}\quad ,
  \label{6.3}
  \end{eqnarray}
where $(D^\alpha )_N{}^M $ obey
  \begin{equation}
  [D^\alpha , D^\beta ]_M{}^N = f^{\alpha\beta}{}_\gamma (D^\gamma )_M{}^N \quad
  . \label{6.4}
  \end{equation}
The commutation relations of all the other generators are
  \begin{eqnarray}
  & & [R^{a,M} , R^{b,N} ]= d^{MNP} R^{ab}{}_P \nonumber \\
  & & [R^{a,N} , R^{bc}{}_M ]= g_{\alpha\beta} (D^\alpha )_M{}^N R^{abc, \beta}
  \nonumber \\
  & & [R^{ab}{}_M , R^{cd}{}_N ]= R^{abcd}{}_{[MN]} \nonumber \\
  & & [R^{a, P}, R^{bcd , \alpha} ]= S^{\alpha P[MN]} R^{abcd}{}_{[MN]} \quad
  , \label{6.5}
  \end{eqnarray}
where $d^{MNP}$ is the symmetric invariant tensor of $E_6$ and
$g_{\alpha\beta}$ is the Cartan-Killing metric of $E_6$. $S^{\alpha
P[MN]}$ is also an invariant tensor, which the Jacobi identities fix
to be
  \begin{equation}
  S^{\alpha P[MN]} = -{1 \over 2} D^\alpha_Q{}^{[M} d^{N]QP } \quad , \label{6.6}
  \end{equation}
and which satisfies the further identity
  \begin{equation}
  g_{\alpha\beta} D^\alpha_Q{}^{(P} S^{\beta R)[MN]} =  -{1 \over 2} \delta_{Q}^{[M} d^{N]PR}
  \quad . \label{6.7}
  \end{equation}
One can show that all the Jacobi identities involving the generators
in eq. (\ref{6.1}) are satisfied using the commutators listed above.

To obtain the field-strengths of the massless theory, one introduces
the Og generators, that encode the gauge transformations of all the
fields. We focus in particular on the Og 1 generators for the
$E_{11}$ generators listed in eq. (\ref{6.1}), that are
  \begin{equation}
  K^{a,b , M} \qquad K^{a , b_1 b_2}{}_M \qquad K^{a, b_1 b_2 b_3 ,
  \alpha} \qquad K^{a , b_1 ... b_4}{}_{[MN]} \quad ,
  \label{6.8}
  \end{equation}
and whose commutators with the momentum operator are
  \begin{eqnarray}
  & & [  K^{a,b , M} , P_c ] = \delta^{(a}_c R^{b),M} \nonumber \\
  & & [ K^{a , b_1 b_2}{}_M  , P_c ] = \delta^a_c R^{b_1 b_2 }{}_M -
  \delta^{[a}_c R^{b_1 b_2 ] }{}_M \nonumber \\
  & & [ K^{a , b_1 b_2 b_3 , \alpha}  , P_c ] = \delta^a_c R^{b_1 b_2 b_3 , \alpha}-
  \delta^{[a}_c R^{b_1 b_2 b_3] , \alpha}\nonumber \\
  & & [ K^{a , b_1 ... b_4}{}_{[MN]}  , P_c ] = \delta^a_c R^{b_1 ... b_4}{}_{[MN]} -
  \delta^{[a}_c R^{b_1 ... b_4 ] }{}_{[MN]} \quad . \label{6.9}
  \end{eqnarray}
We then write down the group element
  \begin{equation}
  g = e^{x \cdot P} e^{ \Phi_{\rm Og} K^{\rm Og}} e^{A_{a_1
  ...a_4}^{[MN]} R^{a_1 ... a_4}_{[MN]} } e^{A_{a_1 a_2 a_3, \alpha}
  R^{a_1 a_2 a_3, \alpha}} e^{A_{a_1 a_2}^M R^{a_1 a_2}_M } e^{A_{a
  ,M} R^{a , M}} e^{\phi_\alpha R^\alpha} \quad , \label{6.10}
  \end{equation}
where we denote with $K^{\rm Og}$ all the Og 1 generators listed in
eq. (\ref{6.8}) and with $\Phi_{\rm Og}$ their corresponding fields.
One can then compute the Maurer-Cartan form, and use the inverse
Higgs mechanism to fix all the Og 1 fields in terms of derivatives
of the $E_{11}$ fields, in such a way that only the completely
antisymmetric terms in the Maurer-Cartan form survive. These
quantities are the gauge-invariant field-strengths of the massless
theory obtained in \cite{20}, which we list here
  \begin{eqnarray}
  & & F_{a_1 a_2 ,M} = \partial_{[a_1} A_{a_2 ] ,M} \nonumber \\
  & & F^{M}_{a_1 a_2 a_3} = \partial_{[a_1} A^{M}_{a_2 a_3 ]} + {1 \over 2} \partial_{[a_1} A_{a_2 ,N}
  A_{a_3 ] ,P} d^{MNP} \nonumber \\
  & & F^{\alpha}_{a_1 a_2 a_3 a_4} = \partial_{[a_1} A^{\alpha}_{a_2 a_3 a_4 ]} - {1 \over 6}
  \partial_{[a_1}
  A_{a_2 ,M} A_{a_3 ,N} A_{a_4 ] ,P} d^{MNQ} D^\alpha_Q{}^P - \partial_{[a_1}
  A^{M}_{a_2
  a_3} A_{a_4 ],N} D^\alpha_M{}^N \nonumber \\
  & & F^{MN}_{a_1 \dots a_5} = \partial_{[a_1} A^{MN}_{a_2 \dots a_5 ]} - {1 \over 24}
  \partial_{[a_1}
  A_{a_2 , P}  A_{a_3 , Q} A_{a_4 , R} A_{a_5 ] , S} d^{PQT} D^\alpha_T{}^R
  S^{\alpha S [MN]} \nonumber \\
  & & \qquad \qquad - {1 \over 2} \partial_{[a_1} A^{P}_{a_2 a_3 } A_{a_4 , Q} A_{a_5 ], R}
  D^\alpha_P{}^Q S^{\alpha R[MN]} + {1 \over 2} \partial_{[a_1} A^{[M}_{a_2 a_3 } A^{N]}_{a_4 a_5 ]
  }\nonumber \\
  & & \qquad \qquad  +
  \partial_{[a_1} A^{\alpha}_{a_2 a_3 a_4} A_{a_5 ] , P} S^{\alpha P[MN]}
  \label{6.11}
  \end{eqnarray}
for completeness.

We now consider the deformed case. We take as our set of generators
that of eqs. (\ref{6.1}) and (\ref{6.8}), but to indicate that these
generators themselves have been deformed, we denote them with a
tilde. The commutation relations receive order $g$ corrections with
respect to the massless ones, where $g$ is the deformation
parameter. We start from the commutation relation between the
deformed vector generator and the momentum operator. We impose this
to be
  \begin{equation}
  [ \tilde{R}^{a , M} , P_b ] = - g \delta^a_b \Theta^M_\alpha
  {R}^\alpha \quad . \label{6.12}
  \end{equation}
The quantity $\Theta^M_\alpha$ turns out to be the embedding tensor
\cite{7}, and the generators $\Theta^M_\alpha R^\alpha$ are the
generators of the subgroup $G$ of $E_6$ that is gauged. These
generators belong to the algebra $\tilde{E}_{11,5}^{local}$. We now
show that all the commutation relations of the deformed operators
among themselves and with the momentum operators are uniquely fixed
by Jacobi identities. From the resulting algebra we construct the
non-linear realisation whose Maurer-Cartan form gives the
field-strengths of the gauged theory.

We first consider the Jacobi identity between $\Theta^M_\alpha
{R}^\alpha$, $\tilde{R}^{a , N}$ and $P_b$. Defining
  \begin{equation}
  X^{MN}_P = \Theta^M_\alpha D^\alpha_P{}^N \label{6.13}
  \end{equation}
one gets
   \begin{equation}
   \Theta^M_\alpha \Theta^N_\beta f^{\alpha \beta}{}_\gamma -
   \Theta^P_\gamma X^{MN}_P =0 \quad , \label{6.14}
   \end{equation}
which turns out to be the condition that the embedding tensor is
invariant under the gauge group. We then write the commutator of the
2-form $\tilde{R}^{ab}{}_M$ with $P_c$ as
  \begin{equation}
  [\tilde{R}^{ab}{}_M , P_c ] = -2 g W_{MN} \delta^{[a}_c
  \tilde{R}^{ b ] , N} \quad , \label{6.15}
  \end{equation}
which defines the antisymmetric tensor $W_{MN}$. The Jacobi identity
involving $\tilde{R}^{ab}{}_M$, $P_c$ and $P_d$ gives
  \begin{equation}
  W_{MN} \Theta^N_\alpha =0 \quad . \label{6.16}
  \end{equation}
The Jacobi identity involving the operators $\tilde{R}^{a , M}$,
$\tilde{R}^{b , N}$ and $P_c$ gives
  \begin{equation}
  [ \tilde{R}^{a ,M} , \tilde{R}^{ b ,N}] = d^{MNP}
  \tilde{R}^{ab}{}_P - 2 g X^{[MN]}_P \tilde{K}^{a,b,P} \quad ,
  \label{6.17}
  \end{equation}
using the fact that the Og 1 for the vector $\tilde{K}^{a,b,M}$ is
symmetric in $ab$ and satisfies
  \begin{equation}
  [ \tilde{K}^{a,b,M}, P_c ] = \delta^{(a}_c \tilde{R}^{b ) , M}
  \quad . \label{6.18}
  \end{equation}
To get eq. (\ref{6.17}) one has also to impose
  \begin{equation}
  X^{(MN)}_P = - W_{PQ} d^{QMN} \quad . \label{6.19}
  \end{equation}

The Jacobi identity between $\tilde{R}^{ab}{}_M$, $\Theta^N_\alpha
R^\alpha$ and $P_c$ gives
  \begin{equation}
  X^{MN}_{[P} W_{Q]N} =0 \quad , \label{6.20}
  \end{equation}
which is the condition that the tensor $W_{MN}$ is invariant under
the gauge subgroup $G$. The Jacobi identity involving $\tilde{R}^{a,
M}$, $\tilde{R}^{bc}{}_N$ and $P_d$ gives
  \begin{equation}
  [ \tilde{R}^{a ,M} , \tilde{R}^{bc}{}_N ] = ( D^\alpha )_N{}^M
  \tilde{R}^{abc}{}_\alpha + {3 \over 2} g (X^{MP}_N + {1\over 3}
  X^{PM}_N ) \tilde{K}^{a ,bc}{}_P \label{6.21}
  \end{equation}
and
  \begin{equation}
  [ \tilde{R}^{abc}{}_\alpha , P_d ] = - g \Theta^M_\alpha
  \delta^{[a}_d \tilde{R}^{bc ]}{}_M \quad , \label{6.22}
  \end{equation}
where the Og 1 operator $\tilde{K}^{a, bc}{}_M$ satisfies
  \begin{equation}
  [ \tilde{K}^{a, bc}{}_M , P_d ] = \delta^a_d \tilde{R}^{bc}{}_M -
  \delta^{[a}_d \tilde{R}^{bc ]}{}_{M} - {2 \over 3} g W_{MN} (
  \delta^b_d \tilde{K}^{a ,c , N} - \delta^c_d \tilde{K}^{a,b,N} )
  \quad . \label{6.23}
  \end{equation}
Proceeding this way, one can determine all the commutators requiring
the closure of the Jacobi identities. This gives
  \begin{equation}
  [ \tilde{R}^{ab}{}_M , \tilde{R}^{cd}{}_N ] =
  \tilde{R}^{abcd}{}_{MN} - 4 g W_{(M \vert P \vert }
  D^\alpha_{N)}{}^P \tilde{K}^{[a,b]cd}{}_\alpha \quad ,
  \label{6.24}
  \end{equation}
  \begin{equation}
  [ \tilde{R}^{a , M} , \tilde{R}^{bcd}{}_\alpha ] =
  S^{M[NP]}_\alpha  \tilde{R}^{abcd}{}_{MN} - g ( f^{\beta
  \gamma}{}_\alpha \Theta^M_\gamma + {1 \over 3} D^\beta_P{}^M
  \Theta^P_\alpha ) \tilde{K}^{a , bcd}{}_\beta \label{6.25}
  \end{equation}
and
  \begin{equation}
  [ \tilde{R}^{a_1 ... a_4}{}_{MN} , P_b ] = - 4 g W_{[M\vert P
  \vert } D^\alpha_{N]}{}^P \delta^{[ a_1}_b \tilde{R}^{a_2 a_3 a_4
  ]}{}_\alpha \quad , \label{6.26}
  \end{equation}
where we introduce the Og 1 generator for the 3-form $\tilde{K}^{a,
b_1 b_2 b_3}{}_\alpha$, satisfying
  \begin{equation}
  [ \tilde{K}^{a, b_1 b_2 b_3}{}_\alpha , P_c ] = \delta^a_c
  \tilde{R}^{b_1 b_2 b_3}{}_\alpha - \delta^{[a}_c
  \tilde{R}^{b_1 b_2 b_3 ]}{}_\alpha - {3 \over 4} g \Theta^M_\alpha
  \delta^{[ b_1}_c \tilde{K}^{\vert a \vert , b_2 b_3 ]}
  \quad . \label{6.27}
  \end{equation}
In order to get these results one has to impose an additional
constraint
  \begin{equation}
  f^{\alpha\beta}{}_\gamma \Theta^Q_\beta - D^\alpha_P{}^Q \Theta^P_\gamma = 4 D^\alpha_M{}^P W_{PN} g_{\beta
  \gamma} S^{\beta Q[MN]} \quad , \label{6.28}
  \end{equation}
which shows that the embedding tensor and $W_{MN}$ are related by
the invariant tensor $S^{\alpha P[MN]}$, and thus belong to the same
representation of $E_6$, which is the ${\bf \overline{351}}$.

To summarise, we have determined the commutation relations satisfied
by the deformed $E_{11}$ $p$-form generators, the corresponding
deformed Og 1 generators and the momentum operator of the
five-dimensional massive theory starting from eq. (\ref{6.12}) and
imposing the closure of the Jacobi identities. We will now determine
the field-strengths of the 1-forms, 2-forms and 3-forms of the
theory using these results. We consider the group element
  \begin{equation}
  g = e^{x \cdot P} e^{ \Phi_{\rm Og} \tilde{K}^{\rm Og}} e^{A_{a_1
  ...a_4}^{[MN]} \tilde{R}^{a_1 ... a_4}_{[MN]} } e^{A_{a_1 a_2 a_3, \alpha}
  \tilde{R}^{a_1 a_2 a_3, \alpha}} e^{A_{a_1 a_2}^M \tilde{R}^{a_1 a_2}_M } e^{A_{a
  ,M} \tilde{R}^{a , M}} e^{\phi_\alpha R^\alpha} \quad ,
  \label{6.29}
  \end{equation}
where as in the massless case we denote with $\tilde{K}^{\rm Og}$
all the deformed Og 1 generators and with $\Phi_{\rm Og}$ the
corresponding fields. One can then compute the Maurer-Cartan form,
and use the inverse Higgs mechanism to fix all the Og 1 fields in
such a way that only the completely antisymmetric terms in the
Maurer-Cartan form survive. To compute the field-strengths, it is
thus sufficient to consider only the $\tilde{R}$ operators and $P_a$
in the group element above. The final result is
  \begin{eqnarray}
  & & \! \! \! \! \! \! \!   F_{a_1 a_2 ,M} = \partial_{[a_1} A_{a_2 ] ,M} + {1 \over 2} g
  X^{[NP]}_M A_{[a_1 ,N} A_{a_2], P} - 2 g W_{MN}A_{a_1
  a_2}^N \nonumber \\
  & &  \! \! \! \! \! \! \!  F^{M}_{a_1 a_2 a_3} = \partial_{[a_1} A^{M}_{a_2 a_3 ]} + {1 \over 2} \partial_{[a_1} A_{a_2 ,N}
  A_{a_3 ] ,P} d^{MNP} - 2 g X^{(MN)}_P A_{[a_1 a_2}^P A_{a_3 ], N}
  \nonumber \\
  & &  +{1 \over 6} g X^{[NP]}_R d^{RQM} A_{[a_1 , N} A_{a_2 , P} A_{[a_3 ] , Q} + g \Theta^M_\alpha A_{a_1 a_2
  a_3 }^\alpha \nonumber \\
  &  & \! \! \! \! \! \! \!
  F^{\alpha}_{a_1 \dots  a_4} = \partial_{[a_1} A^{\alpha}_{a_2 \dots a_4 ]} - {1 \over 6} \partial_{[a_1}
  A_{a_2 ,M} A_{a_3 ,N} A_{a_4 ] ,P} d^{MNQ} D^\alpha_Q{}^P - \partial_{[a_1} A^{M}_{a_2
  a_3} A_{a_4 ],N} D^\alpha_M{}^N \nonumber \\
  & &   +  g D^\alpha_M{}^P \Theta^M_\beta A_{[a_1 , P} A^\beta_{a_2 \dots a_4 ]} + 4 g  D^\alpha_M{}^P  W_{PN}
  A^{MN}_{a_1 \dots a_4} - g  D^\alpha_M{}^P W_{PN} A^M_{[a_1 a_2} A^N_{a_3 a_4
  ]} \nonumber \\
  & &   -  g  D^\alpha_M{}^P X^{(MR)}_Q A_{[a_1 , P} A_{a_2 , R} A_{a_3 a_4 ]}^Q - {1 \over 24} g X^{[MN]}_R d^{RPS}
  D^\alpha_S{}^Q A_{[a_1 , M} A_{a_2 , N} A_{a_3 , P} A_{a_4 ] , Q}
  \quad \label{6.30}
  \end{eqnarray}
These are the field-strengths of the five-dimensional gauged maximal
supergravity \cite{20}. One can also derive the gauge
transformations of the fields from the non-linear realisation as
they arise as rigid transformations of the group element, $g
\rightarrow g_0 g$, as long as one includes the Og generators. The
result is
  \begin{eqnarray}
  & &   \delta A_{a , N}= \partial_a \Lambda_N -g \Lambda_S X^{SM}_N A_{aM}+ 2 gW_{MP}
  \Lambda^P_a \nonumber \\
  & &   \delta A^N_{a_1a_2}=  \partial_{[a_1}\Lambda^N_{a_2]}  +{1\over 2} \partial_{[a_1} \Lambda_S  A_{a_2 ]T}
  d^{STN} +g \Lambda_S X^{SN}_M A_{a_1a_2}^M  +2W_{SP} \Lambda^P_{[a_1} A_{a_2
  ]T}d^{STN}\nonumber \\
  & & \quad - g \Lambda_{a_1a_2}^\alpha \Theta _\alpha^N \nonumber
  \\
  & &   \delta A_{a_1a_2a_3}^\alpha=  \partial_{[a_1} \Lambda_{a_2 a_3 ]}^\alpha +  \partial_{[a_1} \Lambda_M
  A^{N}_{a_2 a_3 ]} D^\alpha_N{}^M + {1 \over 6}
  \partial_{[a_1} \Lambda_M  A_{a_2 , N}  A_{a_3 ], P} d^{MNQ}
  D^\alpha_Q{}^P \nonumber \\
  & & \quad
  - g \Lambda_P  \Theta^P_\beta f^{\alpha\beta}{}_\gamma A^\gamma_{a_1 a_2
  a_3}   + 2 g W_{MP} \Lambda_{[a_1}^P  D^\alpha_N{}^M A_{a_2 a_3 ]}^N
  + {1 \over 3} g W_{MR} \Lambda_{[a_1}^R   d^{MNQ}
  D^\alpha_Q{}^P A_{a_2 , N} A_{a_3 ] }  \nonumber \\
  & & \quad - 4 g D^\alpha_M{}^P W_{PN} \Lambda_{a_1 a_2 a_3}^{MN}
  \quad . \label{6.31}
  \end{eqnarray}
In \cite{20} these transformations were derived both from $E_{11}$
and from requiring the closure of the supersymmetry algebra. Indeed,
the commutator of two supersymmetry transformations on these fields
gives the gauge transformations above provided that the fields are
related by dualities. In particular the 1-forms are dual to 2-forms
while the 3-forms are dual to scalars in five dimensions. The
field-strength of the 4-form is dual to the mass parameter. The
field strengths and gauge transformations obtained here precisely
agree with those obtained by supersymmetry.

In the above we have taken Jacobi identities with all the deformed
generators of eqs. (\ref{6.1}) and (\ref{6.8}) with the exception of
$R^\alpha$, but we have instead restricted our use to
$\Theta^M_\alpha R^\alpha$. Using the Jacobi identities for
$R^\alpha$ would lead to results that are too strong. A solution to
this problem, at least at low levels, requires adding to spacetime
the scalar charges in the $l$ multiplet, as was done in \cite{20}.
In this case the Jacobi identities are automatically satisfied.
Adding the higher charges in the $l$ multiplet may also resolve this
problem at higher level.

To summarise, we have thus shown that the methods explained in this
paper give an extremely fast way of computing the field strengths of
all the forms and dual forms of five-dimensional gauged maximal
supergravity. These methods can be easily generalised to any
dimension, providing a remarkably efficient way of determining the
gauge algebra of any massless or massive theory with maximal
supersymmetry.

\section{The dual graviton}
Any very-extended Kac-Moody algebra, when decomposed in terms of a
$GL(D, \mathbb{R})$ subalgebra which one associates to the
non-linear realisation of gravity, contains a generator with $D-2$
indices in the hook Young Tableaux irreducible representation with
$D-3$ completely antisymmetric indices, that is $R^{a, b_1 \dots
b_{D-3}}$ with $R^{[a, b_1 \dots b_{D-3}]}=0$ (in the case of
$E_{11}$ decomposed in terms of $GL(11, \mathbb{R})$, this generator
is $R^{a, b_1 \dots b_8}$). The field associated to this generator
in the non-linear realisation has the degrees of freedom of the dual
graviton. The Kac-Moody algebra therefore describes together the
graviton and the dual graviton. In this section we will consider the
Og operators for the dual graviton. We will focus on the
four-dimensional case, in which the dual graviton generator $R^{ab}$
is symmetric in its two indices.

In subsection 7.1 we will first consider the case of the dual
graviton in flat space. This corresponds to considering the dual
graviton generator by itself, together with its corresponding Og
generators. This does not arise from any very-extended Kac-Moody
algebra. A field theory description of a linearised dual graviton is
known to exist, and its field equations in four dimensions were
first obtained by Curtright \cite{32}. For subsequent developments
see \cite{1,33}.

We then consider the dual graviton coupled to gravity. The simplest
very-extended Kac-Moody algebra whose non-linear realisation gives
rise to a four-dimensional theory is the algebra $A_1^{+++}$, whose
Dynkin diagram is shown in fig. \ref{fig5}. The corresponding
spectrum does not contain any form. We will show that there is no
consistent solution of the inverse Higgs mechanism that leaves a
propagating dual graviton. We will also consider the case of
$E_{11}$ in four dimensions, which corresponds to deleting node 4 in
the diagram of fig. \ref{fig1}, leading to the internal symmetry
algebra $E_7$. In this case we will show that even considering
linearised gravity, that is neglecting the $GL(4,\mathbb{R})$
generators and the corresponding Og generators, and only considering
interactions of the dual graviton with matter, one is left with no
consistent field strength for the dual graviton. This result is
consistent with \cite{34}, where it was shown that it is impossible
to write down a dual Riemann tensor in the presence of matter even
when gravity is treated at the linearised level.

In the first subsection we will only consider the algebra of
$R^{ab}$ and all the corresponding Og generators. The Maurer-Cartan
form, that can be thought as the Maurer-Cartan form of $A_1^{+++}$
or $E_{11}$ truncated to this sector, leads to invariant quantities
that can be constrained by means of the inverse Higgs mechanism to
generate the Riemann tensor for the linearised dual graviton. In the
second subsection we will then consider the case of dual graviton
coupled to gravity, which corresponds to the algebra $A_1^{+++}$,
and in the third subsection we will consider the $E_{11}$ case of
the dual graviton coupled to vectors with linearised gravity.

\subsection{The dual graviton in four dimensions}
In this subsection we want to consider the dual graviton alone, that
is without introducing the generator associated to the graviton or
any other matter generator. We want to show that one can introduce
suitable Og generators for the dual graviton in such a way that
gives rise to a consistent field strength and consistent gauge
transformations. The dual graviton generator in four dimensions is a
generator with 2 symmetric indices $R^{ab}$. Following the notation
of the previous section, we define two ${\rm Og} \ 1$ generators
$K_1^{a,bc}$ and $\tilde{K}_1^{abc}$ in the irreducible
$GL(4,\mathbb{R})$ representations defined as
  \begin{eqnarray}
  & & K_1^{a,bc}= K_1^{a,(bc)} \qquad K_1^{(a,bc)}=0 \nonumber \\
  & & \tilde{K}_1^{abc}= \tilde{K}_1^{(abc)} \quad , \label{7.1}
  \end{eqnarray}
and whose corresponding Young Tableaux are shown in fig. \ref{fig3}.
Note that the sum of these two representations corresponds to an
object with three indices, symmetric under the exchange of two of
them and with no further constraint. These operators satisfy the
commutation relations
 \begin{eqnarray}
 & & [ K_1^{a,bc} , P_d ]= \delta^a_d R^{bc} - \delta^{(a}_d R^{bc)}
 \nonumber \\
 & & [ \tilde{K}_1^{abc} , P_d ]= \delta^{(a}_d R^{bc)} \quad ,
 \label{7.2}
 \end{eqnarray}
while we take $R^{ab}$ as commuting with $P_a$. We also take
$R^{ab}$ as commuting with itself because we are considering the
dual graviton alone (for instance in $A_1^{+++}$ the dual graviton
is a generator at level 1, and therefore the commutator of two dual
graviton generators leads to an operator at level 2).   Also all the
dual graviton Og generators are taken to commute with each other,
and to commute with $R^{ab}$ as well.
\begin{figure}[h]
\centering
\begin{picture}(340,140)

\put(-10,115){dual graviton} \put(100,115){Og 1} \put(190,115){Og 2}
\put(290,115){${\overline{\rm Og}}\ 2$}

\put(-10,100){\line(1,0){360}}

\put(0,50){$R^{ab}$:}

\put(40,50){\line(1,0){20}} \put(40,60){\line(1,0){20}}
\put(40,60){\line(0,-1){10}} \put(50,60){\line(0,-1){10}}
\put(60,60){\line(0,-1){10}}

\put(70,10){\line(0,1){120}}

\put(80,80){$K_1^{a,bc}$:}

\put(120,90){\line(1,0){20}} \put(120,80){\line(1,0){20}}
\put(120,70){\line(1,0){10}}\put(120,90){\line(0,-1){20}}
\put(130,90){\line(0,-1){20}} \put(140,90){\line(0,-1){10}}

\put(80,20){$\tilde{K}_1^{abc}$:}

\put(120,30){\line(1,0){30}} \put(120,20){\line(1,0){30}}
\put(120,30){\line(0,-1){10}} \put(130,30){\line(0,-1){10}}
\put(140,30){\line(0,-1){10}} \put(150,30){\line(0,-1){10}}

\put(160,10){\line(0,1){120}}

\put(170,80){$K_2^{a,bcd}$:}

\put(210,90){\line(1,0){30}} \put(210,80){\line(1,0){30}}
\put(210,70){\line(1,0){10}}\put(210,90){\line(0,-1){20}}
\put(220,90){\line(0,-1){20}} \put(230,90){\line(0,-1){10}}
\put(240,90){\line(0,-1){10}}

\put(170,20){$\tilde{K}_2^{abcd}$:}

\put(210,30){\line(1,0){40}} \put(210,20){\line(1,0){40}}
\put(210,30){\line(0,-1){10}} \put(220,30){\line(0,-1){10}}
\put(230,30){\line(0,-1){10}} \put(240,30){\line(0,-1){10}}
\put(250,30){\line(0,-1){10}}

\put(260,10){\line(0,1){120}}

\put(270,50){$\bar{K}_2^{ab,cd}$:}

\put(310,60){\line(1,0){20}} \put(310,50){\line(1,0){20}}
\put(310,40){\line(1,0){20}} \put(310,60){\line(0,-1){20}}
\put(320,60){\line(0,-1){20}} \put(330,60){\line(0,-1){20}}


\end{picture}
\caption{\label{fig3}\sl The Young Tableaux of the first ${\rm Og}$
and ${\overline{\rm Og}}$ generators for the dual graviton in four
dimensions. }
\end{figure}

We now consider the ${\rm Og} \ 2$ operators. These are
$K_2^{a,bcd}$ and $\tilde{K}_2^{abcd}$ in the $GL(4,\mathbb{R})$
representations
  \begin{eqnarray}
  & & K_2^{a,bcd}= K_2^{a,(bcd)} \qquad K_2^{(a,bcd)}=0 \nonumber \\
  & & \tilde{K}_2^{abcd}= \tilde{K}_2^{(abcd)} \quad ,\label{7.3}
  \end{eqnarray}
whose Young Tableaux are shown in fig. \ref{fig3}, and their
commutation relation with $P_a$ is
 \begin{eqnarray}
 & & [ K_2^{a,bcd} , P_e ]= \delta^{(b}_e K_1^{\vert a \vert ,cd)} +
 {2 \over 3} (\delta^a_e \tilde{K}_1^{bcd} - \delta^{(a}_e
 \tilde{K}_1^{bcd)} )
 \nonumber \\
 & & [ \tilde{K}_2^{abcd} , P_e ]= \delta^{(a}_e \tilde{K}_1^{bcd)} \quad
 . \label{7.4}
 \end{eqnarray}
The coefficient ${2 \over 3}$ in the first commutator can be
obtained from the Jacobi identity involving $K_2^{a,bcd}$ and two
$P$'s.

We will now compute the Maurer-Cartan form, and we will first
consider only the contribution from dual graviton and the Og 1
fields, while the Og 2 fields will be included later. We thus
consider the group element
  \begin{equation}
  g = e^{x \cdot P}  e^{\Phi^1_{a,bc} K_1^{a,bc}}
  e^{\tilde{\Phi}^1_{abc} \tilde{K}_1^{abc}} e^{A_{ab}R^{ab}} \quad
  , \label{7.5}
  \end{equation}

from which one computes the Maurer-Cartan form
 \begin{equation}
 g^{-1} \partial_\mu g = P_\mu + (\partial_\mu A_{ab} - \Phi^1_{\mu,ab}
 - \tilde{\Phi}^1_{\mu ab} ) R^{ab} + ... \quad , \label{7.6}
 \end{equation}
which is invariant under
  \begin{eqnarray}
  & & \delta A_{ab} = a_{ab} + x^c b_{c,ab} + x^c
  \tilde{b}_{abc} \nonumber\\
  & & \delta \Phi^1_{a,bc} = b_{a,bc} \nonumber\\
  & & \delta \tilde{\Phi}^1_{a,bc} = \tilde{b}_{a,bc} \quad .
  \label{7.7}
  \end{eqnarray}
The first of eqs. (\ref{7.7}) is reproducing the gauge
transformation for the dual graviton in flat space,
  \begin{equation}
  \delta A_{ab} = \partial_{(a} \Lambda_{b)} \label{7.8}
  \end{equation}
at linear order in $x$, that is quadratic order in $x$ for the gauge
parameter $\Lambda_a$,
  \begin{equation}
  \Lambda_a = a_{ab} x^b - b_{a, bc} x^b x^c + {1 \over 2}
  \tilde{b}_{abc} x^b x^c \quad . \label{7.9}
  \end{equation}

One can solve for inverse Higgs in such a way that the whole
Maurer-Cartan form proportional to $R^{ab}$ vanishes compatibly with
the symmetries. This corresponds to fixing
 \begin{equation}
 \Phi^1_{\mu,ab} = \partial_\mu A_{ab} - \partial_{ (\mu} A_{ab)}
 \label{7.10}
 \end{equation}
and
 \begin{equation}
 \tilde{\Phi}^1_{\mu,ab} = \partial_{ (\mu} A_{ab)} \quad .
 \label{7.11}
 \end{equation}
The fact that reproducing the gauge transformations for $A_{ab}$ at
linear order in $x$ allows one to eliminate completely the
Maurer-Cartan form proportional to $R^{ab}$ by means of the inverse
Higgs mechanism corresponds to the fact that one cannot write a
gauge invariant quantity at linear order in the derivatives. Note
that there is a crucial difference here with respect to the
non-linear realisation of gravity discussed in section 2. In that
case the part of the Maurer-Cartan form proportional to the
generators of the local subalgebra $SO(D)$ gives the $SO(D)$
connection, which becomes the spin connection once the inverse Higgs
mechanism is applied. In this case the dual graviton field is
already symmetric, and thus there is no corresponding local Lorentz
symmetry. It is for this reason that at this level the Maurer-Cartan
form vanishes once the inverse Higgs mechanism is applied.

We now consider the contribution from the Og 2 fields. We write the
group element as
  \begin{equation}
  g = e^{x \cdot P} e^{\Phi^2_{a,bcd} K_2^{a,bcd}}
  e^{\tilde{\Phi}^2_{abcd} \tilde{K}_2^{abcd}}  e^{\Phi^1_{a,bc} K_1^{a,bc}}
  e^{\tilde{\Phi}^1_{abc} \tilde{K}_1^{abc}} e^{A_{ab}R^{ab}} \quad ,
  \label{7.12}
  \end{equation}
and obtain the corresponding Maurer-Cartan form
 \begin{eqnarray}
 g^{-1} \partial_\mu g &= & P_\mu + (\partial_\mu A_{ab} - \Phi^1_{\mu,ab}
 - \tilde{\Phi}^1_{\mu ab} ) R^{ab} \nonumber \\
 & +& (\partial_\mu \Phi^1_{a,bc} - \Phi^2_{a, \mu bc} ) K_1^{a,bc}
 \nonumber \\
 & +& (\partial_\mu \tilde{\Phi}^1_{abc} - {2 \over 3} \Phi^2_{\mu , abc}
 - \tilde{\Phi}^2_{\mu abc } )
 \tilde{K}_1^{abc} + ... \quad . \label{7.13}
 \end{eqnarray}
Having introduced the ${\rm Og} \ 2$ operators, the transformations
of the field that leave the Maurer-Cartan form invariant acquire
additional contributions, and in particular there is a term in the
variation of $A_{ab}$ which is quadratic in $x$. The result is
  \begin{eqnarray}
  & & \delta A_{ab} = a_{ab} + x^c b_{c,ab} + x^c
  \tilde{b}_{abc} +{5 \over 6} x^c x^d c_{c, abd} - {1 \over 2}x^c x^d c_{(c, ab)d}
  + {1 \over 2} x^c x^d \tilde{c}_{abcd} \nonumber\\
  & & \delta \Phi^1_{a,bc} = b_{a,bc} + x^d c_{a, bcd} - x^d c_{(a ,bc)d}\nonumber\\
  & & \delta \tilde{\Phi}^1_{a,bc} = \tilde{b}_{a,bc} +{2 \over 3} x^d c_{d,abc} + x^d
  \tilde{c}_{abcd}\nonumber \\
  & & \delta \Phi^2_{a,bcd} = c_{a,bcd}\nonumber \\
  & & \delta \tilde{\Phi}^2_{a,bcd} = \tilde{c}_{a,bcd} \quad .
  \label{7.14}
  \end{eqnarray}
In particular the first of these variations is the most general
gauge transformation for the field $A_{ab}$ of the form (\ref{7.8})
up to terms cubic in $x$.

We now apply the inverse Higgs mechanism, solving for the fields
$\Phi^2_{a,bcd}$ and $\tilde{\Phi}^2_{abcd}$ in terms of $A_{ab}$.
The result is
  \begin{eqnarray}
  \Phi^2_{a,bcd} & =& {1 \over 4} [ \partial_a \partial_b A_{cd} +  \partial_a \partial_c
  A_{db} +  \partial_a \partial_d A_{bc} -  \partial_b \partial_c
  A_{ab} - \partial_b \partial_d A_{ac} -  \partial_c \partial_d
  A_{ab} ] \nonumber \\
  \tilde{\Phi}^2_{abcd} & =& \partial_{( a} \partial_b A_{cd )} \quad
  . \label{7.15}
  \end{eqnarray}
Plugging this into the Maurer-Cartan form, one notices that there is
a non-vanishing term proportional to $K_1^{a,bc}$, that is
  \begin{equation}
  g^{-1} \partial_\mu g =  P_\mu + ( {1 \over 3} \partial_\mu
  \partial_a A_{bc} - {1 \over 3} \partial_\mu \partial_b A_{ac} -
  {1\over 3} \partial_a \partial_b A_{\mu c} + {1 \over 3} \partial_b
  \partial_c A_{\mu a} ) K_1^{a,bc} + ... \quad . \label{7.16}
  \end{equation}
This is indeed the Riemann tensor of linearised gravity $D_{ab,cd}$,
which is a tensor in the window-like Young Tableaux representation.

One can introduce in the same way the higher ${\rm Og}$ generators,
constructing in this way gauge invariant quantities which are
derivatives of the Riemann tensor. The end result is thus
  \begin{equation}
  g^{-1} \partial_\mu g =  P_\mu + D_{\mu a ,bc}  K_1^{a,bc} + ...
  \quad , \label{7.17}
  \end{equation}
where the dots correspond to derivatives of the dual graviton
Riemann tensor
  \begin{equation}
  D_{ab ,cd} = {1 \over 3} \partial_a
  \partial_b A_{cd} - {1 \over 6}  \partial_a \partial_c A_{bd} - {1
  \over 6} \partial_a \partial_d A_{bc} - {1 \over 6}
  \partial_b \partial_c A_{ad} - {1 \over 6} \partial_b \partial_d A_{ac}+ {1 \over 3}
  \partial_c \partial_d A_{ab} \label{7.18}
  \end{equation}
contracted with higher order Og generators. This shows that the
linearised dual graviton admits a description in terms of $R^{ab}$
and Og generators, and the corresponding Maurer-Cartan form contains
the correct Riemann tensor, which can be used to construct the
dynamics.

\begin{figure}[h]
\centering
\begin{picture}(260,300)

\put(0,295){field} \put(80,295){Og 1} \put(190,295){Og 2}

\put(-10,280){\line(1,0){280}}

\put(40,305){\line(0,-1){305}}

\put(150,305){\line(0,-1){305}}

\put(0,260){\line(1,0){20}} \put(0,270){\line(1,0){20}}
\put(0,270){\line(0,-1){10}} \put(10,270){\line(0,-1){10}}
\put(20,270){\line(0,-1){10}}

\put(60,270){\line(1,0){20}} \put(60,260){\line(1,0){20}}
\put(60,250){\line(1,0){10}}\put(60,270){\line(0,-1){20}}
\put(70,270){\line(0,-1){20}} \put(80,270){\line(0,-1){10}}

\put(85,260){${\bf \oplus}$}

\put(100,270){\line(1,0){30}} \put(100,260){\line(1,0){30}}
\put(100,270){\line(0,-1){10}} \put(110,270){\line(0,-1){10}}
\put(120,270){\line(0,-1){10}} \put(130,270){\line(0,-1){10}}

\put(170,270){\line(1,0){30}} \put(170,260){\line(1,0){30}}
\put(170,250){\line(1,0){10}}\put(170,270){\line(0,-1){20}}
\put(180,270){\line(0,-1){20}} \put(190,270){\line(0,-1){10}}
\put(200,270){\line(0,-1){10}}

\put(205,260){${\bf \oplus}$}

\put(220,270){\line(1,0){40}} \put(220,260){\line(1,0){40}}
\put(220,270){\line(0,-1){10}} \put(230,270){\line(0,-1){10}}
\put(240,270){\line(0,-1){10}} \put(250,270){\line(0,-1){10}}
\put(260,270){\line(0,-1){10}}

\put(10,240){\vector(0,-1){20}}

\put(90,240){\vector(0,-1){20}}

\put(210,240){\vector(0,-1){20}}

\put(0,200){\line(1,0){20}} \put(0,210){\line(1,0){20}}
\put(0,210){\line(0,-1){10}} \put(10,210){\line(0,-1){10}}
\put(20,210){\line(0,-1){10}}

\put(0,180){\line(1,0){10}} \put(0,170){\line(1,0){10}}
\put(0,180){\line(0,-1){10}} \put(10,180){\line(0,-1){10}}

\put(0,150){${\bf 1}$}

\put(60,210){\line(1,0){20}} \put(60,200){\line(1,0){20}}
\put(60,190){\line(1,0){10}}\put(60,210){\line(0,-1){20}}
\put(70,210){\line(0,-1){20}} \put(80,210){\line(0,-1){10}}

\put(85,200){${\bf \oplus}$}

\put(100,210){\line(1,0){30}} \put(100,200){\line(1,0){30}}
\put(100,210){\line(0,-1){10}} \put(110,210){\line(0,-1){10}}
\put(120,210){\line(0,-1){10}} \put(130,210){\line(0,-1){10}}

\put(60,180){\line(1,0){20}} \put(60,170){\line(1,0){20}}
\put(60,180){\line(0,-1){10}} \put(70,180){\line(0,-1){10}}
\put(80,180){\line(0,-1){10}}

\put(170,210){\line(1,0){30}} \put(170,200){\line(1,0){30}}
\put(170,190){\line(1,0){10}}\put(170,210){\line(0,-1){20}}
\put(180,210){\line(0,-1){20}} \put(190,210){\line(0,-1){10}}
\put(200,210){\line(0,-1){10}}

\put(205,200){${\bf \oplus}$}

\put(220,210){\line(1,0){40}} \put(220,200){\line(1,0){40}}
\put(220,210){\line(0,-1){10}} \put(230,210){\line(0,-1){10}}
\put(240,210){\line(0,-1){10}} \put(250,210){\line(0,-1){10}}
\put(260,210){\line(0,-1){10}}

\put(170,180){\line(1,0){30}} \put(170,170){\line(1,0){30}}
\put(170,180){\line(0,-1){10}} \put(180,180){\line(0,-1){10}}
\put(190,180){\line(0,-1){10}} \put(200,180){\line(0,-1){10}}

\put(80,120){$\partial_y ({\rm field} )$}

\put(60,90){$\overbrace{\phantom{allthisfillsuuu}}$}

\put(60,90){\line(1,0){20}} \put(60,80){\line(1,0){20}}
\put(60,90){\line(0,-1){10}} \put(70,90){\line(0,-1){10}}
\put(80,90){\line(0,-1){10}}

\put(85,80){${\bf \oplus}$}

\put(100,90){\line(1,0){10}} \put(100,80){\line(1,0){10}}
\put(100,90){\line(0,-1){10}} \put(110,90){\line(0,-1){10}}

\put(115,80){${\bf \oplus}$}

\put(130,80){${\bf 1}$}

\put(75,45){$\overline{\rm Og}$}

\put(60,20){$\overbrace{\phantom{allthisf}}$}

\put(60,20){\line(1,0){10}} \put(60,10){\line(1,0){10}}
\put(60,0){\line(1,0){10}} \put(60,20){\line(0,-1){20}}
\put(70,20){\line(0,-1){20}}

\put(75,10){${\bf \oplus}$}

\put(90,20){\line(1,0){10}} \put(90,10){\line(1,0){10}}
\put(90,20){\line(0,-1){10}} \put(100,20){\line(0,-1){10}}

\put(190,120){$\partial_y ({\rm Og}  \ 1)$}

\put(170,90){$\overbrace{\phantom{allthivvvvvnnsf}}$}

\put(170,90){\line(1,0){20}} \put(170,80){\line(1,0){20}}
\put(170,70){\line(1,0){10}}\put(170,90){\line(0,-1){20}}
\put(180,90){\line(0,-1){20}} \put(190,90){\line(0,-1){10}}

\put(195,80){${\bf \oplus}$}

\put(210,90){\line(1,0){30}} \put(210,80){\line(1,0){30}}
\put(210,90){\line(0,-1){10}} \put(220,90){\line(0,-1){10}}
\put(230,90){\line(0,-1){10}} \put(240,90){\line(0,-1){10}}

\put(245,80){${\bf \oplus}$}

\put(170,60){\line(1,0){20}} \put(170,50){\line(1,0){20}}
\put(170,60){\line(0,-1){10}} \put(180,60){\line(0,-1){10}}
\put(190,60){\line(0,-1){10}}

\put(195,50){${\bf \oplus}$}

\put(210,60){\line(1,0){20}} \put(210,50){\line(1,0){20}}
\put(210,60){\line(0,-1){10}} \put(220,60){\line(0,-1){10}}
\put(230,60){\line(0,-1){10}}

\put(235,50){${\bf \oplus}$}

\put(170,40){\line(1,0){10}} \put(170,30){\line(1,0){10}}
\put(170,40){\line(0,-1){10}} \put(180,40){\line(0,-1){10}}

\put(185,30){${\bf \oplus}$}

\put(197,30){${\bf 1}$}

\put(205,30){${\bf \oplus}$}

\put(220,40){\line(1,0){10}} \put(220,30){\line(1,0){10}}
\put(220,40){\line(0,-1){10}} \put(230,40){\line(0,-1){10}}

\put(235,30){${\bf \oplus}$}

\put(170,20){\line(1,0){10}} \put(170,10){\line(1,0){10}}
\put(170,0){\line(1,0){10}} \put(170,20){\line(0,-1){20}}
\put(180,20){\line(0,-1){20}}

\end{picture}
\caption{\label{fig4}\sl The dimensional reduction of the dual
graviton and its ${\rm Og}$ fields. Each field is aligned
horizontally with its corresponding Og fields. The rest of the ${\rm
Og}$ $n$ fields are associated to the $y$ derivative of the ${\rm
Og}$ $(n-1)$ fields. The dimensional reduction also produces
$\overline{\rm Og}$ 1 fields for the vector and the scalar.}
\end{figure}
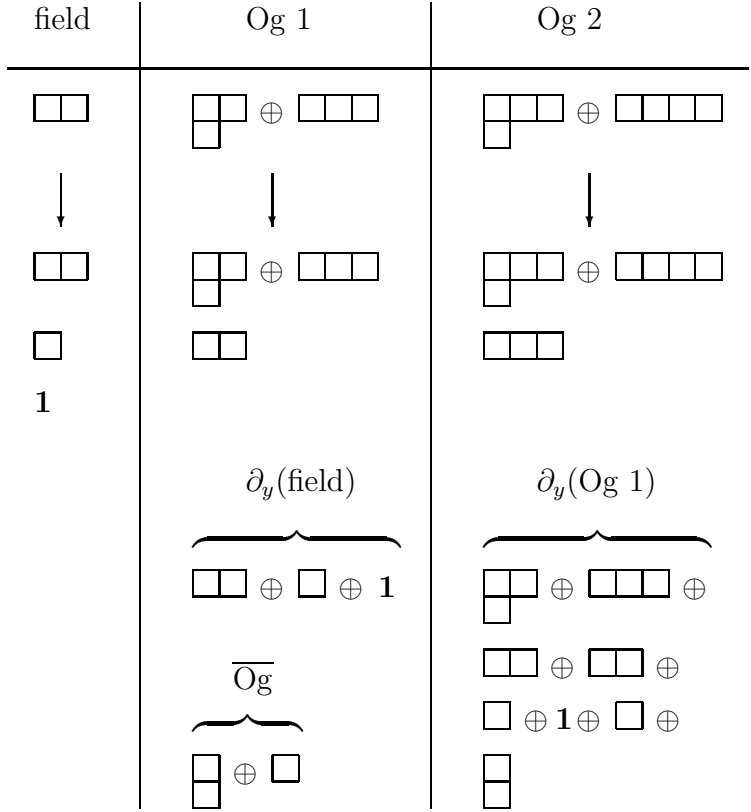

We now want to perform a dimensional reduction on a circle of
coordinate $y$. We thus take the dual graviton and all the Og fields
to be $y$ independent. The representations of $GL(3,\mathbb{R})$
that arise in the three-dimensional compactified theory are shown in
fig. \ref{fig4} for the dual graviton field and the first two Og
fields. The circle dimensional reduction corresponds to the
assumption that neither the dual graviton field nor the Og fields
depend on $y$.

The dual graviton $A_{ab}$ in four dimensions has two symmetric
indices, and after dimensional reduction it leads to an object with
two symmetric indices $A_{ab}$, a vector $A_a = A_{a y}$ and a
scalar $A = A_{yy}$. As the figure shows, the Og 1 fields can be
divided in three sets. The first one contains the Og fields for the
field with two symmetric indices $A_{ab}$ and the vector $A_{a}$.
This is precisely what we want in order to obtain the correct gauge
transformations for the three-dimensional fields. The second set
contains the same representations as the dimensionally reduced
fields. The $dy$ part of the Maurer-Cartan form contains these
fields summed to the $y$ derivative of the dimensionally reduced
fields. Thus, from the requirement that the fields do not depend on
$y$ it follows that these Og fields can be put to zero using the
inverse Higgs mechanism. Finally, the third set contains a field
with two antisymmetric indices and a vector, and we call these
fields the $\overline{\rm Og}$ 1 fields for the vector and the
scalars. These fields are the ones of interest to us in the
following. The dimensional reduction of the Og 2 fields gives the Og
2 fields for the field with two symmetric indices and the vector,
together with a set of fields in the same representations as the
dimensionally reduced Og 1 fields, and again the $dy$ part of the
Maurer-Cartan form contains these Og 2 fields summed to the $y$
derivative of all the Og 1 fields. Using $y$ independence and the
inverse Higgs mechanism one thus sets to zero these Og 2 fields.

We now explain the occurrence of the $\overline{\rm Og}$ 1 fields
and generators in the dimensional reduction. In the four-dimensional
theory, once the inverse Higgs mechanism is applied the
Maurer-Cartan form is given in eq. (\ref{7.17}), and the first
non-vanishing term is the dual graviton Riemann tensor, which is at
second order in derivatives. Using the $y$-independence of the
fields, the dimensional reduction of the Riemann tensor leads to the
Riemann tensor for $A_{ab}$ in three dimensions together with
  \begin{eqnarray}
  & & D_{ab,c y} = {1 \over 2} \partial_a F_{bc} + {1 \over 3}
  \partial_{(b} F_{\vert a \vert c )} \nonumber \\
  & & D_{ab, yy} = {1 \over 3} \partial_a \partial_b A \quad ,
  \label{7.19}
  \end{eqnarray}
while $D_{ay,yy}$ vanishes. Here we have denoted with $F_{ab} =
\partial_{[a} A_{b ]}$ the field-strength of the vector. As eq.
(\ref{7.19}) shows, the Maurer-Cartan form in three dimensions thus
contains the Riemann tensor of $A_{ab}$ together with the derivative
of the field-strength of the vector and the double derivative of the
scalar. This implies that among the rest, the Maurer-Cartan form is
invariant under the transformations
  \begin{equation}
  \delta A_a = x^b b_{[ba]}  \qquad \delta A = b_a x^a \quad ,
  \label{7.20}
  \end{equation}
which indeed lead to
  \begin{equation}
  \delta F_{ab} = b_{[ab]} \qquad \delta ( \partial_a A ) = b_a \quad
  . \label{7.21}
  \end{equation}
Such transformations cannot be written as standard gauge
transformations for the corresponding fields, and indeed they do not
leave the field strength invariant, although they are symmetries of
the dimensionally reduced Riemann tensor. They are generated by the
operators associated to the $\overline{\rm Og}$ 1 fields in fig.
\ref{fig4}, and in general we define the $\overline{\rm Og}$
generators as those producing transformations that cannot be written
as gauge transformations. The $\overline{\rm Og}$ 1 fields in fig.
\ref{fig4}, together with the standard Og 1 fields for $A_{ab}$ and
$A_a$, are such that all the terms with one derivative of the fields
in the Maurer-Cartan form vanish once the inverse Higgs mechanism is
applied. The standard gauge transformations of the fields are
obtained by performing a truncation that projects out the
$\overline{\rm Og}$ 1 generators, and once this truncation is
performed one can no longer use the inverse Higgs mechanism to
cancel the one derivative terms completely, which indeed give
$F_{ab}$ and the derivative of the scalar.

\subsection{The dual graviton in $A_1^{+++}$ in four dimensions}

\begin{figure}[h]
\begin{center}
\begin{picture}(140,30)
 \multiput(10,10)(40,0){4}{\circle{10}}
 \multiput(15,10)(40,0){2}{\line(1,0){30}}
 \put(95,12){\line(1,0){30}}
  \put(95,8){\line(1,0){30}}
    \put(8,-8){$1$} \put(48,-8){$2$} \put(88,-8){$3$} \put(128,-8){$4$}
\end{picture}
\end{center}
\caption{\sl The $A_{1}^{+++}$ Dynkin diagram.} \label{fig5}
\end{figure}
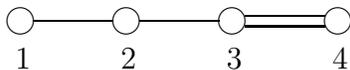

The non-linear realisation based on the algebra $A_1^{+++}$, whose
Dynkin diagram is shown in fig. \ref{fig5}, has the particular
feature of only containing in four dimensions the graviton and its
duals, which are fields with two symmetric indices together with an
arbitrary number of blocks of two antisymmetric indices, as well as
generators with sets of 3 or 4 antisymmetric indices. This in
particular means that the spectrum does not contain any forms, that
is fields with completely antisymmetric indices.

Decomposing the adjoint representation of $A_1^{+++}$ in
representations of $GL(4,\mathbb{R})$ one gets $K^a{}_b$ at level
zero, which are the generators of $GL(4,\mathbb{R})$, and $R^{ab}$
at level one. The generators at higher level can be obtained as
multiple commutators of $R^{ab}$ subject to the Serre relations, and
the number of indices of a generator at level $l$ is $2l$. We will
ignore all the generators of level higher than 1 in this subsection.
The commutation relation between $K^a{}_b$ and $R^{ab}$ is
  \begin{equation}
  [ K^a{}_b , R^{cd} ] = \delta^c_b R^{ad} + \delta^d_b R^{ca}
  \quad . \label{7.22}
  \end{equation}

We now want to introduce the Og generators for both the graviton and
the dual graviton, in order to reproduce the correct general
coordinate transformation for the fields, as well as the expected
gauge transformation for the dual graviton. As in the previous
subsection, we have
  \begin{equation}
  [R^{ab} , P_c ] =0 \quad . \label{7.23}
  \end{equation}
We thus obtain the commutator between the gravity Og 1 operator
$K^{ab}{}_c$ and $R^{ab}$ by the Jacobi identity with $P_a$. The
result is
  \begin{equation}
  [ K^{ab}{}_c , R^{de}] = \delta^d_c K_1^{e,ab} + \delta^e_c
  K_1^{d,ab} -2 (\delta^d_c \tilde{K}_1^{abe} +\delta^e_c
  \tilde{K}_1^{abd} ) \quad . \label{7.24}
  \end{equation}
Similarly, imposing the Jacobi identity between $P_a$, $K^a{}_b$ and
the Og 1 dual graviton operators gives
  \begin{eqnarray}
  & & [ K^a{}_b , K_1^{c,de} ] = \delta^c_b K_1^{a,de} + \delta^d_b
  K_1^{c,ae} + \delta^e_b K_1^{c,da} \nonumber \\
  & & [ K^a{}_b , \tilde{K}_1^{cde} ] = 3 \delta^{(c}_b
  \tilde{K}_1^{de)a} \label{7.25}
  \end{eqnarray}
as expected from the index structure of the operators.

We  would expect that the commutator between the Og 1 gravity
operator $K^{ab}{}_c$ and the dual graviton Og 1 operators gave the
Og 2 dual graviton operators in fig. \ref{fig3} if a description of
both gravity and dual gravity were possible. This turns out to be
impossible, {\it i.e.} one can show that the Jacobi identity between
$K^{ab}{}_c$, $P_a$ and either $K_1^{a,bc}$ or $\tilde{K}_1^{abc}$
is not satisfied if the commutator between $K^{ab}{}_c$ and the dual
graviton Og 1 operators gives dual graviton Og 2 operators. This is
indeed the problem that one typically encounters when trying to
construct a dual Riemann tensor.

One can define an operator $\bar{K}_2^{ab,cd}$ satisfying
  \begin{equation}
  \bar{K}_2^{ab,cd} = \bar{K}_2^{(ab),cd} = \bar{K}_2^{ab,(cd)} =
  \bar{K}_2^{cd,ab} \qquad \bar{K}_2^{a(b,cd)} = 0 \quad ,
  \label{7.26}
  \end{equation}
whose corresponding Young Tableaux is shown in the last column in
fig. \ref{fig3}. We define the commutation relation of this operator
with $P_a$ to be
  \begin{equation}
  [ \bar{K}_2^{ab,cd} , P_e ] = {1 \over 2} \delta^{(a}_e K_1^{b
  ),cd} +  {1 \over 2} \delta^{(c}_e K_1^{d ),ab} \quad .
  \label{7.27}
  \end{equation}
This is indeed the most general result compatible with the
symmetries, and one can show that the Jacobi identity with a further
$P_a$ operator is satisfied.

If one adds the term ${\rm exp}(\bar{\Phi}^2_{ab,cd}
\bar{K}_2^{ab,cd})$ to the group element of eq. (\ref{7.12}), one
obtains that a transformation
  \begin{equation}
  \delta \bar{\Phi}^2_{ab,cd} = \bar{c}_{ab,cd} \label{7.28}
  \end{equation}
implies an $x^2$ transformation for $A_{ab}$ of the form
  \begin{equation}
  \delta A_{ab} = {1 \over 2} \bar{c}_{ab,cd} x^c x^d \quad .
  \label{7.29}
  \end{equation}
This transformation cannot be written as a gauge transformation of
eq. (\ref{7.8}) for the linearised graviton. Following the arguments
of the previous subsection, we refer to $\bar{K}_2^{ab,cd}$ as an
${\overline{\rm Og}}$ operator. The inverse Higgs mechanism allows
to gauge away completely all the terms at most quadratic in $x$ in
the field, with this still being compatible with all the symmetries.

Having introduced the operator $\bar{K}_2^{ab,cd}$, one obtains that
the commutator between $K^{ab}{}_c$ and the dual graviton Og 1
operators can now be made compatible with the Jacobi identity with
$P_a$. The result is
  \begin{eqnarray}
  & & [ K^{ab}{}_c , K_1^{d,ef}] = -6 \delta^{(e}_c K_2^{\vert d
  \vert  , f) a b} + 6 \delta^{(d}_c K_2^{e , f) a b} +2 \delta^d_c
  \tilde{K}_2^{efab} - 2 \delta^{(d}_c \tilde{K}_2^{ef) ab} \nonumber \\
  & &\qquad \qquad \qquad  \ \  - 4 \delta^d_c
  \bar{K}_2^{ef,ab} + 4 \delta^{(d}_c \bar{K}_2^{ef), ab}
  \nonumber\\
  & & [ K^{ab}{}_c , \tilde{K}_1^{def} ]   = 3
  \delta^{(d}_c K_2^{e,f)ab} - 10 \delta^{(d}_c \tilde{K}_2^{ef)ab} +
  2 \delta^{(d}_c \bar{K}_2^{ef),ab} \quad . \label{7.30}
  \end{eqnarray}
The fact that $\bar{K}_2^{ab,cd}$ must appear on the right hand side
of this commutation relation is the main result of this section.
This shows that the only inverse Higgs mechanism compatible with the
symmetries is the one that gauges away the dual graviton completely.
We expect that once all the ${\overline{\rm Og}}$ operators for the
dual graviton are introduced together with the Og operators for both
the graviton and the dual graviton, the resulting algebra is well
defined. We conjecture that the same applies to all the generators
of $A_1^{+++}$ with positive level. As a consequence of this, after
having applied the inverse Higgs mechanism, the Maurer-Cartan form
will contain only the graviton Riemann tensor and its derivatives.

It is interesting to discuss the dimensional reduction to three
dimensions in this case as we have done in the previous subsection.
Following arguments similar to that case, one can show that the
dimensional reduction of all the generators in fig. \ref{fig3}
contains the ${\overline{\rm Og}}$ 1 and ${\overline{\rm Og}}$ 2
generators for the scalar and the vector, and more generally the
dimensional reduction of all the Og and ${\overline{\rm Og}}$ dual
graviton generators leads to all the Og and ${\overline{\rm Og}}$
generators for the dimensionally reduced fields. The algebra of the
dimensionally reduced theory can be truncated in such a way that the
${\overline{\rm Og}}$ generators for the scalar and the vector that
arise from the reduction of the dual graviton can be consistently
projected out, so that the corresponding Maurer-Cartan form results
in the field-strengths for this fields, as well as their
derivatives, once the inverse Higgs mechanism is applied.

\subsection{The dual graviton in $E_8^{+++}$ in four dimensions}
In this subsection we want to discuss the case in which the dual
graviton couples to matter. We will discuss the case of the
non-linear realisation of $E_8^{+++}$, {\it i.e.} $E_{11}$, in four
dimensions, which corresponds to the bosonic sector of
four-dimensional maximal supergravity. The Dynkin diagram of
$E_{11}$ is shown in fig. \ref{fig1}, and the four dimensional
theory is obtained deleting node 4 in the diagram. The internal
symmetry is $E_{7}$, and the spectrum contains among the rest
vectors in the ${\bf 56}$ of $E_7$. We will show that even
neglecting couplings to gravity, it is impossible to make the gauge
transformation of the dual graviton compatible with that of the
vector. The situation is exactly as in the previous subsection: the
commutator of two Og 1 operators for the vector generate the
operator $\bar{K}_2^{ab,cd}$, which is an ${\overline{\rm Og}}\ 2$
operator for the dual graviton.

Decomposing the adjoint representation of $E_{11}$ in
representations of $GL(4, \mathbb{R})$ one gets at level zero the
gravity generators $K^a{}_b$ and the $E_7$ generators $R^\alpha$,
while at level 1 one gets $R^{a, M}$, where $M$ denotes the ${\bf
56}$ of $E_7$. The higher level generators can be obtained as
multiple commutators of $R^{a, M}$. In particular at level 2 one
gets
  \begin{equation}
  [ R^{a, M} , R^{b, N} ] = D_{\alpha}^{MN} R^{[ab] , \alpha} +
  \Omega^{MN} R^{ab} \quad , \label{7.31}
  \end{equation}
where $R^{ab , \alpha}$ is the 2-form generator in the adjoint of
$E_7$ and $R^{ab}$ is the dual graviton generator. We have also
introduced
  \begin{equation}
  D^{\alpha MN} = \Omega^{MP} D^\alpha_P{}^N \label{7.32}
  \end{equation}
which is symmetric in $MN$, and $D^\alpha_M{}^N$ are the generators
in the ${\bf 56}$. Finally $\Omega^{MN}$ is the antisymmetric
invariant tensor of $E_7$. The field associated to the generator
$R^{[ab] , \alpha}$ is related to the scalars by duality. In the
rest of this section we will ignore the 2-form contribution to the
commutator of eq. (\ref{7.31}), and we will only consider the dual
graviton contribution,
  \begin{equation}
  [ R^{a, M} , R^{b, N} ] =
  \Omega^{MN} R^{ab} \quad . \label{7.33}
  \end{equation}
This truncation of the algebra is consistent because the Jacobi
identities close independently on the 2-form generators and on the
dual graviton generators.

The Og 1 generator for the vector $R^{a, M}$ is a generator $K^{ab ,
M}$ symmetric in $ab$, whose commutation relation with $P_a$ is
  \begin{equation}
  [ K^{ab , M} , P_c ] = \delta^{(a}_c R^{b), M} \quad .
  \label{7.34}
  \end{equation}
The commutation relation of $K^{ab , M}$ with $R^{a, M}$ can be
obtained by imposing the Jacobi identity of these operators with
$P_a$ and using eqs. (\ref{7.2}), eq. (\ref{7.33}) and eq.
(\ref{7.34}), as well as the fact that $R^{a, M}$ commutes with
$P_a$. The result is
  \begin{equation}
  [ R^{a, M} , K^{bc , N} ] = - {1 \over 2 } \Omega^{MN} K_1^{a, bc}
  + \Omega^{MN} \tilde{K}_1^{abc} \quad . \label{7.35}
  \end{equation}

We can now write the group element up to Og 2 generators,
  \begin{equation}
  g = e^{x \cdot P}   e^{\Phi^1_{a,bc} K_1^{a,bc}}
  e^{\tilde{\Phi}^1_{abc} \tilde{K}_1^{abc}} e^{\Phi_{ab ,M} K^{ab ,M}}
  e^{A_{ab}R^{ab}} e^{A_{a ,M} R^{a, M}}
\quad , \label{7.36}
  \end{equation}
which leads to the Maurer-Cartan form
 \begin{eqnarray}
 g^{-1} \partial_\mu g & = & P_\mu + (\partial_\mu A_{a , M} - \Phi_{\mu
 a , M} ) R^{a , M}
 + (\partial_\mu A_{ab}
 +  {1 \over 2} \Omega^{MN} \partial_\mu
 A_{a, M} A_{b ,N} \nonumber \\
 &-& \Phi^1_{\mu,ab}
 - \tilde{\Phi}^1_{\mu ab} - \Phi_{\mu a , M} A_{b ,N} \Omega^{MN} ) R^{ab} + ... \quad
 . \label{7.37}
 \end{eqnarray}
The inverse Higgs mechanism then leaves the field strength for the
vector, while the term contracting $R^{ab}$ is put to zero by
solving for $\Phi^1_{\mu,ab}$ and $\tilde{\Phi}^1_{\mu ab}$ in terms
of $A_{ab}$ and $A_{a ,M}$.

We now compute the commutator of two Og 1 operators $K^{ab ,M}$ for
the vector, and we determine which Og 2 generators are needed to
satisfy the Jacobi identities. It turns out that because of the
symmetry of the commutator, it is not possible to generate the Og 2
dual graviton operator $K_2^{a,bcd}$, and the Jacobi identity with
$P_a$ imposes that this actually closes on $\tilde{K}_2^{abcd}$ and
$\bar{K}_2^{ab,cd}$. The result is
  \begin{equation}
  [ K^{ab ,M} , K^{cd ,N} ] = 2 \Omega^{MN} \tilde{K}_2^{abcd} -
  \Omega^{MN} \bar{K}_2^{ab,cd} \quad . \label{7.38}
  \end{equation}
Thus exactly as in the case of the dual graviton coupled to gravity
of the previous subsection we have found here that the commutator of
two Og operators generates an ${\overline{\rm Og}}$ operator for the
dual graviton, which means that a gauge invariant field strength for
the dual graviton is not compatible with vector gauge invariance.

We claim that this is a generic feature of $E_{11}$ positive level
generators with spacetime indices with mixed symmetry. The algebra
of their Og generators does not close, and one is forced to
introduce ${\overline{\rm Og}}$ generators for all these mixed
symmetry generators. Only for the gravity generator, which has level
zero, and for the generators with completely antisymmetric indices
the Og algebra closes. As a consequence only for these fields can
one use the inverse Higgs mechanism and be left with a non-vanishing
field-strength. The fact that the positive level mixed symmetry
generators require the introduction of the Og and ${\overline{\rm
Og}}$ generators implies instead that the corresponding fields do
not allow a gauge invariant field strength and the inverse Higgs
mechanism gauges away these fields completely. To show this one
computes Jacobi identities involving positive level $E_{11}$
generators, Og generators and the momentum operator $P_a$. Thus this
result deeply relies on the structure of the $E_{11}$ algebra. The
dimensional reduction allows a further truncation of the algebra in
the case in which a mixed symmetry generator gives rise to a
generator with completely antisymmetric indices. Indeed in this
case, as was shown in the previous subsections, the  ${\overline{\rm
Og}}$ generators can be consistently projected out.

It is important to stress that the dynamics is compatible with this
construction. The field strengths of the antisymmetric fields are
first order in derivatives, and therefore one needs fields and dual
fields to construct duality relations which are first order
equations for these fields. The gravity Riemann tensor instead is at
second order in derivatives and thus there is no need of a dual
field to construct its equation of motion.

\section{Conclusions}
In this paper we have given a method of obtaining field strengths
and gauge transformations of all the massless and massive maximal
supergravity theories starting from $E_{11}$. The global $E_{11}$
transformations of the fields are promoted to gauge transformations
by the inclusion in the algebra of additional generators.

We have first shown how this mechanism works for pure gravity. We
have constructed Einstein's theory of gravity using a non-linear
realisation which takes as its underlying algebra one that consists
of  $IGL(D,\mathbb{R} )$ and an infinite set of additional
generators whose effect is to promote the rigid $IGL(D,\mathbb{R})$
to be local. This infinite number of additional generators lead to
local translations, that is general coordinate transformations, but
to no new fields in the final theory as their Goldstone fields are
solved in terms of the graviton field using a set of invariant
constraints placed on the Cartan forms. This is an example of what
has been called the inverse Higgs effect \cite{23}.

We have then generalised this procedure to $E_{11}$ at low levels.
We have taken the algebra, called $E_{11}^{local}$  consisting of
non-negative level $E_{11}$ generators, the generators $P_a$  and an
infinite number of additional generators, whose role is to promote
all the low level $E_{11}$ symmetries to gauge symmetries. Again, as
in the gravity case these generators do not lead to new Goldstone
fields. We have shown that the non-linear realisation of the algebra
$E_{11}^{local}$ describes at low levels in eleven dimensions the
3-form and the 6-form of the eleven dimensional supergravity theory
with all their gauge symmetries.

We have then  considered in general the formulation of
$D$-dimensional maximal gauged supergravity theories from the
viewpoint of the enlarged algebra $\tilde{E}_{11,D}^{local}$. We
have first considered as a toy model the Scherk-Schwarz dimensional
reduction of the IIB supergravity theory from this viewpoint. One
starts from the algebra $E_{11,10B}^{local}$ corresponding to the
IIB theory and take the ten dimensional space-time to arise from an
operator $\tilde Q$ which is constructed from $Q=P_{9}$ and part of
the $SL(2,\mathbb{R})$ symmetry of the theory. This means that the
10th direction of space-time is twisted to contain a part in the
$SL(2,\mathbb{R})$ coset symmetry of the theory. This non-linear
realisation gives a nine dimensional gauged supergravity. We have
observed that not all of the algebra $E_{11,10B}^{local}$ is
essential for the construction of the gauged supergravity in nine
dimensions, but only an algebra which we call $\tilde E_{11,
9}^{local}$ which is the subalgebra of $E_{11,10B}^{local}$ that
commutes with $\tilde Q$. Its generators are non-trivial
combinations of $E_{11}$ generators and the additional generators
and in general the generators of $\tilde E_{11, 9}^{local}$ have
non-trivial commutation relations with nine dimensional  space-time
translations. Although the subalgebra $\tilde E_{11, 9}^{local}$
appears to be a deformation of the original $E_{11}$ algebra and the
space-time translations we have not changed the original
commutators,  but rather the new algebra arises due to the presence
of the additional generators which are added to the $E_{11}$
generators.

However, we have then shown that one can find the algebra $\tilde
E_{11, 9}^{local}$ without carrying out all the above steps. Given
the non-trivial relation between the lowest non-trivial positive
level generator of $\tilde E_{11,9}^{local}$ and the nine
dimensional space-time translations one can derive the rest of the
algebra $\tilde E_{11, 9}^{local}$ simply using Jacobi identities.
This algebra determines uniquely all the field strengths of the
theory, and thus one finds a very quick way of deriving the gauged
supergravity theory.

This picture applies to all gauged supergravity theories, as one can
easily find the algebra  $\tilde E_{11, D}^{local}$  without using
its derivation from $E_{11}^{local}$ and this provides a very
efficient method of constructing all gauged supergravities. We have
illustrated how this works by constructing the massive IIA theory as
well as all the gauged maximal supergravities in five dimensions.

Finally, we have considered how this construction generalises to the
fields with mixed symmetry, {\it i.e.} not completely antisymmetric,
of $E_{11}$ and in general of any non-linear realisation of a
very-extended Kac-Moody algebra. We have considered as a prototype
of such fields the dual graviton in four dimensions. If one tries to
promote the global shift symmetry of the dual graviton field to a
gauge symmetry, one finds that this is not compatible with the
$E_{11}$ algebra. The solution of this problem is that actually
$E_{11}$ forces to include additional generators, whose role is to
enlarge the gauge symmetry of the dual graviton so that one can
gauge away the field completely. This also applies if one only
restricts his attention to the compatibility of the dual graviton
with matter fields, that is if one neglects the gravity generators.
This result agrees with the field theory analysis of \cite{34}. More
generally, this agrees with the no-go theorems of \cite{35} on the
consistency of self-interactions for the dual graviton. Recently, an
alternative approach to the construction of an action for the dual
graviton has been taken \cite{36}, in which the metric only appears
via topological couplings, and an additional shift gauge field is
included.

As we have mentioned in the introduction it is not obvious  how to
to implement the conformal group, or equivalently, add the Og fields
in the presence of the generators of the $l$ multiplet. The rational
for introducing the $l$ multiplet was that it would allow an
$E_{11}$ way of encoding space-time. However, in this paper we have
chosen to take only the first component of the $l$ multiplet, namely
the space-time translations and we have taken this to commute with
the positive level $E_{11}$ generators. As a result we have had to
discard the negative level $E_{11}$ generators. This is
unsatisfactory as $E_{11}$ is defined from its Chevalley generators
and relations and there is no definition that uses only the positive
levels. For this reason the content of the adjoint representation
and the $l$ multiplet also rely on the negative root generators.
However, we know that many of the generators, and so fields in the
non-linear realisation,  in the former  and brane charges in the
latter are in very convincing agreement with what one might expect
in M theory. One example being the classification of all gauged
supergravities using  the $D-1$ forms found in the adjoint
representation of $E_{11}$. How to reconcile local symmetries,
space-time and the full $E_{11}$ algebra is for future work.

\vskip 2cm

\section*{Acknowledgments}
This work is supported by the PPARC rolling grant PP/C5071745/1, the
EU Marie Curie research training network grant MRTN-CT-2004-512194
and the STFC rolling grant ST/G000/395/1.

\end{document}